\documentclass[prd,eqsecnum,aps,showpacs]{revtex4}

\usepackage{amsmath,amssymb}
\usepackage[dvips]{graphics}
\usepackage{calc}

\newcommand{\bra}[1]{{\!}\left\langle{}{ #1 }{}\right|{}}
\newcommand{\ket}[1]{{}\left|{}{ #1}{}\right\rangle{\!}}
\newcommand{\bracket}[2]{{\!}\left\langle{}{#1}\left.\!\!\vphantom{#2}%
\vphantom{#1}{}\right|{}\!{#2}{}\right\rangle{}\!}
\newcommand{\ev}[1]{\left\langle{}{#1}\right\rangle{}}

\newcommand{\Z}{\hspace{0.2zw}}
\renewcommand{\vec}[1]{\mathbf #1}
\newcommand{\nn}{\nonumber}
\newcommand{\sla}[1]{{\not \! \! \: {#1}}}
\newcommand{\del}[2]{\frac{\partial #1}{\partial #2}}

\newcommand{\olra}[1]{\overleftrightarrow #1}
\newcommand{\udl}[1]{\underline{#1}}
\newcommand{\vl}[1]{\underline{\vec{#1}}}

\newcommand{\Zc}{Z^0_{\text{clas}}}
\newcommand{\Zd}{\udl{Z}^0}
\newcommand{\Zdp}{(\Zd)_{\text{para}}}
\newcommand{\Zdc}{(\Zd)_{\text{clas}}}
\newcommand{\Yd}{\udl{Y}^0}

\newcommand{\vev}[1]{\ev{\vec{v}_{#1}}}
\newcommand{\xa}{\vec{X}_A}
\newcommand{\xb}{\vec{X}_B}

\newcommand{\tb}{X^0_B}
\newcommand{\sg}{\sigma^2}
\newcommand{\f}{\vec{F}}
\newcommand{\fac}{\vec{F}^{AC}}

\newcommand{\fbc}{\vec{F}^{BC}}
\newcommand{\Tac}{T^{AC}}

\newcommand{\tacp}[1]{(\Tac_{#1})_{\text{para}}}
\newcommand{\Tbc}{T^{BC}}

\newcommand{\tbcp}[1]{(\Tbc_{#1})_{\text{para}}}

\newcommand{\life}{\ev{\Gamma_A}}
\newcommand{\plt}{\ev{\Gamma_\pi}}
\newcommand{\vL}{\vec{L}}
\newcommand{\vv}{\vec{v}}
\newcommand{\im}{\text{Im}}

\begin{document}
\title{Expectation values of flavor-neutrino currents in field theoretical
approach to oscillation problem ------ formulation
}
\author{Kanji Fujii and Takashi Shimomura}

\address{Department of Physics, Faculty of Science,Hokkaido University, Sapporo 060-0810, Japan}
\date{\today}
\begin{abstract}
As a possible approach to the neutrino oscillation on the basis of quantum
field theory, the expectation values of the flavor-neutrino currents are
investigated by employing the finite-time transition matrix in the interaction
representation.
Such expectation values give us in the simplest form a possible way of treating the 
neutrino oscillation without recourse to any one flavor-neutrino states. The present 
paper is devoted to presenting the formulation and the main structures of the relevant 
expectation values.
\end{abstract}
\draft
\maketitle
%\input{1sec.tex}
%\input{2sec.tex}
%\input{3sec.tex}
%\input{4sec.tex}

%% section 1 %%
%%%%%%%%%%%%%%%%%%%%%%%%%%%%%%%%%%%%%%%%%%%%%%%%%%%%%%%%%%%%%%%%%%%%%%%%%%%%%%%%%%%%%%
\section{Introduction}
Many papers have been published concerning the theoretical basis of the standard
formulas of the neutrino oscillations. As to the works appeared till the
middle of the year 2001, we can find the list of references in the review article written 
by Buete\cite{r1}; in this article he gave his view by arranging the controvercial
points in the neutrino- and meson-oscillations.

%%%%%%%%%%%%%%%%%%%%%%%%%%%%%%%%%%%%%%%%%%%%%%%%%%%%%%%%%%%%%%%%%%%%%%%%%%%%%%%%%%%%%%
\subsection{Proposal of problems}

The analyses of the neutrino experimental data have been usually done on the basis
of the standard formula of the neutrino oscillation. In the present work we concentrate
on the following two problems included in the ordinary derivation of the standard formula 
of neutrino oscillations in vacuum \cite{r2,r3}. 

First we summarize the ordinary derivation of the standard formula in vacuum\cite{r2,r3}.
Let's consider the neutrino with momentum $\vec{k}$ and helicity $r = -1$ , produced through 
the charged-current weak interaction. Such a flavor-neutrino state is assumedto be a
superposition of the mass eigenstates, expressed as 
\begin{align}
\bra{\nu_\sigma(\vec{k})}=\sum_j
 \bracket{\nu_\sigma(\vec{k})}{\nu_j(\vec{k})}\bra{\nu_j(\vec{k})},
\hspace{0.5cm} \sigma=(e,\mu,\cdots),j=(1,2,\cdots) ;\label{1.1}
\end{align}
the matrix $U=\left[U_{\sigma j}\right]$ with $U_{\sigma j}=\bracket{\nu_\sigma(\vec{k})}{\nu_j(\vec{k})}$
is unitary. 

By employing the time evolution of $\bra{\nu_j(\vec{k})}$,
\begin{align}
\bra{\nu_j(\vec{k};t)}=e^{-i\omega_j(\vec{k})t}\bra{\nu_j(\vec{k})},
\hspace{0.5cm} \omega_j(\vec{k})=\sqrt{\vec{k}^2+m_j^2}, \label{1.2}
\end{align}
the flavor-neutrino state is seen to evolve in time as 
\begin{align}
\bra{\nu_\rho(\vec{k};t)}
&=\sum_{\lambda,j}U_{\rho j}e^{-i\omega_j(\vec{k})t}U^\ast_{\lambda j}\bra{\nu_\lambda(\vec{k})} \nn \\
&=\sum_{\lambda}\mathcal{A}_{\lambda
 \rho}(\vec{k};t)\bra{\nu_\lambda(\vec{k})}, \label{1.3}
\end{align}
where
\begin{align}
\mathcal{A}_{\sigma \rho}(\vec{k};t)
&=\left( Ue^{-i\omega_{\text{diag}}(\vec{k})t}U^\dagger \right)_{\rho \sigma}, \nn \\
\omega_{\text{diag}} &=
\begin{pmatrix}
\omega_1(\vec{k}) & &   \\
& \omega_2(\vec{k}) &   \\
& & \ddots              \\
\end{pmatrix}.
\label{1.4}
\end{align}
Then, the transition probability is
\begin{align}
P_{\sigma \rightarrow \rho}(t) 
&= \left| \mathcal{A}_{\sigma \rho}(\vec{k};t) \right|^2 \nn \\
&=\sum_{j,l}U_{\sigma j}U^\ast_{\rho j}U^\ast_{\sigma l}U_{\rho l}
e^{i( \omega_j(\vec{k})-\omega_l(\vec{k}) )t}. \label{1.5}
\end{align}
Under the condition of the neutrino velocities 
\begin{align}
v_j(\vec{k}) \equiv \frac{|\vec{k}|}{\omega_j(\vec{k})} \simeq 1
\qquad\text{(the light velocity in vacuum)}, \label{1.6}
\end{align}
the time t which is regarded as the time of neutrino 
propagation is replaced by the distance L from source to detector approximately.    
Then we have 
\begin{align}
P_{\sigma \rightarrow \rho}(L) 
&=\sum_j |U_{\sigma j}|^2 |U_{\rho j}|^2 
+ 2 \text{Re}\sum_{j>l}U_{\sigma j}U^\ast_{\rho j}U^\ast_{\sigma l}U_{\rho l}
\ e^{i\frac{\Delta m^2_{jl}}{2 |\vec{k}|}L}.\label{1.7}
\end{align}
here we used
%\begin{align}
$ \omega_j(\vec{k})-\omega_l(\vec{k}) \simeq \frac{(m^2_j-m^2_l)}{2|\vec{k}|} \equiv \frac{\Delta m^2_{jl}}{2 |\vec{k}|},$
%\end{align}
which holds for extremely relativistic neutrinos.

Now we mention the two problems included in the derivation described above.

Problem I:  The matrix $U = [U_{\sigma \rho}]$ is defined as the matrix which
diagonalizes the mass matrix in the flavor bases M ;
\begin{align}
M&=
\begin{pmatrix}
m_{ee} & m_{e \mu} & \cdots    \\
m_{\mu e} & m_{\mu \mu} &\cdots \\
\vdots & \vdots & \ddots        
\end{pmatrix}
,\hspace{0.5cm} M=M^\dagger \\
U^\dagger M U &=
\begin{pmatrix}
m_1 &    &     \\
    &m_2 &     \\
    &    &\ddots 
\end{pmatrix}
\equiv M_{\text{diag}} \label{Mdiag}
\end{align}
(\ref{Mdiag}) leads to
\begin{align}
U^\dagger \left( |\vec{k}| + \frac{M M^\dagger}{2 |\vec{k}|} \right) U
=|\vec{k}| + \frac{m^2_{\text{diag}}}{2 |\vec{k}|}
=\omega_{\text{diag}}(\vec{k}),
\end{align}
in the extremely relativistic case.

The important point is that the matrix U is set equal to the mixing matrix
$Z^{1/2} = [z^{1/2}_{\rho j}]$ between the two kinds of neutrino fields, i.e.
\begin{align}
\nu_F(x)=Z^{1/2}\nu_M(x),\hspace{0.5cm} {Z^{1/2}}^\dagger Z^{1/2}=1 \label{1.11}
\end{align}
for
\begin{align}
\nu_F(x)=
\begin{pmatrix}
\nu_e(x) \\
\nu_\mu(x) \\
\vdots
\end{pmatrix},
\hspace{0.5cm}
\nu_M(x)=
\begin{pmatrix}
\nu_1(x) \\
\nu_2(x) \\
\vdots
\end{pmatrix},
\end{align}
where $Z^{1/2}$ is defined by
\begin{align}
{Z^{1/2}}^\dagger M Z^{1/2} = M_{\text{diag}} \label{1.13};
\end{align}
M  is the mass matrix for the "free part" 
$\mathcal{L}_0(x)=-\bar{\nu}_F(x)(\sla{\partial}+M)\nu_F(x)$ in the total Lagrangin
\begin{align}
\mathcal{L}(x)= \mathcal{L}_0(x)+\mathcal{L}_{\text{int}}(x). \label{1.14}
\end{align}
Here, $\mathcal{L}_{\text{int}}(x)$ is assumed to have no bilinear term and no 
derivative of the neutrino fields. The Hamiltonian $H_0(x)$ obtained from $\mathcal{L}_0(x)$ 
is
\begin{align}
%\mathcal{H}_0(x)&=\bar{\nu}_F(x)(\vec{\gamma}\dot\vec{\partial}+M)\nu_F(x),\\
\mathcal{H}_0(x)&=\bar{\nu}_F(x)(\overrightarrow{\gamma} \cdot \overrightarrow{\partial}+M)\nu_F(x) \nn \\
&=\sum_j \bar{\nu}_j(x)(\overrightarrow{\gamma} \cdot \overrightarrow{\partial}+m_j)\nu_j(x), \label{1.15}
\end{align}
which means the energy is diagonal with respect to $\nu_j(x)$'s, irrespectively of
the momentum. Thus, in spite of the situation that , apart from a trivial phase
factor, we can take
\begin{align}
Z^{1/2}=U,
\end{align}
there is a certain inconsistency. It is often asserted \cite{r4} that such an
inconsistency disappears in extremely relativistic neutrinos as in existing
experiments ;  therefore, the state (\ref{1.1}) with $U = Z^{1/2}$ reflects
correctly the present experimental situation.  It is , however, not so simple 
conceptually for us to solve such an inconsistency from the field theoretical 
viewpoint as noted in Refs.\cite{r5}-\cite{r8}.
The problem with which we are now confronted is what relation the state has with 
the field-operator relation; in other words, what definition should be
given to the one-particle state of flavor neutrino properly under the
condition $\nu_F(x)=Z^{1/2}\nu_M(x)$.

\begin{figure}[h]
 \begin{center}
  \scalebox{0.9}[1]{%WinTpicVersion3.08
\unitlength 0.1in
\begin{picture}( 76.1500, 17.5000)(6,-29)
% DOT 1 0 3 0
% 3 1620 1990 1620 1990 1620 1990
% 
\special{pn 13}%
\special{sh 1}%
\special{ar 1620 1990 10 10 0  6.28318530717959E+0000}%
\special{sh 1}%
\special{ar 1620 1990 10 10 0  6.28318530717959E+0000}%
\special{sh 1}%
\special{ar 1620 1990 10 10 0  6.28318530717959E+0000}%
% BOX 1 0 3 1
% 2 1000 1610 2000 2410
% 
\special{pn 13}%
\special{pa 1000 1610}%
\special{pa 2000 1610}%
\special{pa 2000 2410}%
\special{pa 1000 2410}%
\special{pa 1000 1610}%
\special{fp}%
% LINE 1 0 3 2
% 2 1010 1610 1810 1270
% 
\special{pn 13}%
\special{pa 1010 1610}%
\special{pa 1810 1270}%
\special{fp}%
% LINE 1 0 3 3
% 2 2010 1610 2810 1270
% 
\special{pn 13}%
\special{pa 2010 1610}%
\special{pa 2810 1270}%
\special{fp}%
% LINE 1 0 3 4
% 2 2010 2410 2810 2070
% 
\special{pn 13}%
\special{pa 2010 2410}%
\special{pa 2810 2070}%
\special{fp}%
% LINE 1 0 3 5
% 4 2810 1270 2810 2070 1810 1270 2810 1270
% 
\special{pn 13}%
\special{pa 2810 1270}%
\special{pa 2810 2070}%
\special{fp}%
\special{pa 1810 1270}%
\special{pa 2810 1270}%
\special{fp}%
% DOT 1 0 3 6
% 3 6620 1990 6620 1990 6620 1990
% 
\special{pn 13}%
\special{sh 1}%
\special{ar 6620 1990 10 10 0  6.28318530717959E+0000}%
\special{sh 1}%
\special{ar 6620 1990 10 10 0  6.28318530717959E+0000}%
\special{sh 1}%
\special{ar 6620 1990 10 10 0  6.28318530717959E+0000}%
% BOX 1 0 3 7
% 2 6000 1610 7000 2410
% 
\special{pn 13}%
\special{pa 6000 1610}%
\special{pa 7000 1610}%
\special{pa 7000 2410}%
\special{pa 6000 2410}%
\special{pa 6000 1610}%
\special{fp}%
% LINE 1 0 3 8
% 2 6010 1610 6810 1270
% 
\special{pn 13}%
\special{pa 6010 1610}%
\special{pa 6810 1270}%
\special{fp}%
% LINE 1 0 3 9
% 2 7010 1610 7810 1270
% 
\special{pn 13}%
\special{pa 7010 1610}%
\special{pa 7810 1270}%
\special{fp}%
% LINE 1 0 3 10
% 2 7010 2410 7810 2070
% 
\special{pn 13}%
\special{pa 7010 2410}%
\special{pa 7810 2070}%
\special{fp}%
% LINE 1 0 3 11
% 4 7810 1270 7810 2070 6810 1270 7810 1270
% 
\special{pn 13}%
\special{pa 7810 1270}%
\special{pa 7810 2070}%
\special{fp}%
\special{pa 6810 1270}%
\special{pa 7810 1270}%
\special{fp}%
% VECTOR 1 0 3 12
% 2 1120 1990 1560 1990
% 
\special{pn 13}%
\special{pa 1120 1990}%
\special{pa 1560 1990}%
\special{fp}%
\special{sh 1}%
\special{pa 1560 1990}%
\special{pa 1494 1970}%
\special{pa 1508 1990}%
\special{pa 1494 2010}%
\special{pa 1560 1990}%
\special{fp}%
% VECTOR 1 0 3 13
% 2 1650 1960 2150 1330
% 
\special{pn 13}%
\special{pa 1650 1960}%
\special{pa 2150 1330}%
\special{fp}%
\special{sh 1}%
\special{pa 2150 1330}%
\special{pa 2094 1370}%
\special{pa 2118 1372}%
\special{pa 2124 1396}%
\special{pa 2150 1330}%
\special{fp}%
% VECTOR 1 0 3 14
% 2 1660 2030 1870 2350
% 
\special{pn 13}%
\special{pa 1660 2030}%
\special{pa 1870 2350}%
\special{fp}%
\special{sh 1}%
\special{pa 1870 2350}%
\special{pa 1850 2284}%
\special{pa 1842 2306}%
\special{pa 1818 2306}%
\special{pa 1870 2350}%
\special{fp}%
% VECTOR 1 0 3 15
% 2 1680 1990 3070 1980
% 
\special{pn 13}%
\special{pa 1680 1990}%
\special{pa 3070 1980}%
\special{fp}%
\special{sh 1}%
\special{pa 3070 1980}%
\special{pa 3004 1960}%
\special{pa 3018 1980}%
\special{pa 3004 2000}%
\special{pa 3070 1980}%
\special{fp}%
% LINE 1 1 3 16
% 2 3100 1990 5700 1990
% 
\special{pn 13}%
\special{pa 3100 1990}%
\special{pa 5700 1990}%
\special{da 0.070}%
% VECTOR 1 0 3 17
% 2 5710 1990 6590 1990
% 
\special{pn 13}%
\special{pa 5710 1990}%
\special{pa 6590 1990}%
\special{fp}%
\special{sh 1}%
\special{pa 6590 1990}%
\special{pa 6524 1970}%
\special{pa 6538 1990}%
\special{pa 6524 2010}%
\special{pa 6590 1990}%
\special{fp}%
% VECTOR 1 0 3 18
% 4 6650 1970 7290 1800 6670 2020 6940 2140
% 
\special{pn 13}%
\special{pa 6650 1970}%
\special{pa 7290 1800}%
\special{fp}%
\special{sh 1}%
\special{pa 7290 1800}%
\special{pa 7220 1798}%
\special{pa 7238 1814}%
\special{pa 7232 1836}%
\special{pa 7290 1800}%
\special{fp}%
\special{pa 6670 2020}%
\special{pa 6940 2140}%
\special{fp}%
\special{sh 1}%
\special{pa 6940 2140}%
\special{pa 6888 2096}%
\special{pa 6892 2118}%
\special{pa 6872 2132}%
\special{pa 6940 2140}%
\special{fp}%
% STR 2 0 3 19
% 3 1120 1790 1120 1890 5 0
% $P_S$
\put(13,-18.9000){\makebox(0,0){$P_S$}}%
% STR 2 0 3 20
% 3 2040 2030 2040 2130 5 0
% $P_S'$
\put(19,-21.5){\makebox(0,0){$P_S'$}}%
% STR 2 0 3 21
% 3 2180 1230 2180 1330 5 0
% $\bar{l}_\sigma$
\put(19.5,-13.7){\makebox(0,0){$\bar{l}_\sigma$}}%
% STR 2 0 3 22
% 3 2880 1720 2880 1820 5 0
% $\nu_\sigma$
\put(25,-19){\makebox(0,0){$\nu_\sigma$}}%
% STR 2 0 3 23
% 3 6050 1750 6050 1850 5 0
% $\nu_\rho$
\put(62,-19){\makebox(0,0){$\nu_\rho$}}%
% STR 2 0 3 24
% 3 6470 2110 6470 2210 5 0
% $P_D$
\put(63,-22.1000){\makebox(0,0){$P_D$}}%
% STR 2 0 3 25
% 3 7040 2050 7040 2150 5 0
% $P_D'$
\put(69,-22.5){\makebox(0,0){$P_D'$}}%
% STR 2 0 3 26
% 3 7300 1610 7300 1710 5 0
% $l_\rho$
\put(73.0000,-17.1000){\makebox(0,0){$l_\rho$}}%
% LINE 1 1 3 27
% 2 1620 2520 1620 2920
% 
\special{pn 13}%
\special{pa 1620 2520}%
\special{pa 1620 2920}%
\special{da 0.070}%
% LINE 1 1 3 28
% 2 6630 2520 6630 2920
% 
\special{pn 13}%
\special{pa 6630 2520}%
\special{pa 6630 2920}%
\special{da 0.070}%
% STR 2 0 3 29
% 3 4010 2810 4010 2910 5 0
% $L$
\put(40.1000,-29){\makebox(0,0){$L$}}%
% VECTOR 2 0 3 30
% 2 4200 2880 6580 2880
% 
\special{pn 8}%
\special{pa 4200 2880}%
\special{pa 6580 2880}%
\special{fp}%
\special{sh 1}%
\special{pa 6580 2880}%
\special{pa 6514 2860}%
\special{pa 6528 2880}%
\special{pa 6514 2900}%
\special{pa 6580 2880}%
\special{fp}%
% VECTOR 2 0 3 31
% 2 3820 2880 1640 2880
% 
\special{pn 8}%
\special{pa 3820 2880}%
\special{pa 1640 2880}%
\special{fp}%
\special{sh 1}%
\special{pa 1640 2880}%
\special{pa 1708 2900}%
\special{pa 1694 2880}%
\special{pa 1708 2860}%
\special{pa 1640 2880}%
\special{fp}%
% STR 2 0 3 32
% 3 960 2470 960 2570 5 0
% $(\vec{X}_S,t_S)$
\put(13,-25.7000){\makebox(0,0){$(\vec{X}_S,t_S)$}}%
% STR 2 0 3 33
% 3 6920 2490 6920 2590 5 0
% $(\vec{X}_D,t_D)$
\put(70,-25.9000){\makebox(0,0){$(\vec{X}_D,t_D)$}}%
% VECTOR 1 0 3 34
% 2 6380 2300 6590 2050
% 
\special{pn 13}%
\special{pa 6380 2300}%
\special{pa 6590 2050}%
\special{fp}%
\special{sh 1}%
\special{pa 6590 2050}%
\special{pa 6532 2088}%
\special{pa 6556 2092}%
\special{pa 6562 2114}%
\special{pa 6590 2050}%
\special{fp}%
\end{picture}%
}
 \end{center}
\caption{Real experimental situation}
\label{fig1}
\end{figure}
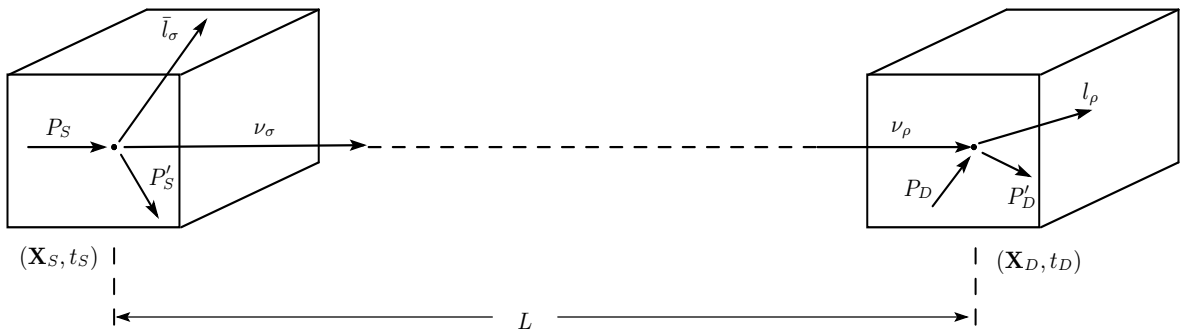

In connection with this problem, various authors have proposed \cite{r9}-\cite{r11} that
the neutrino oscillation shoud be investigated on the basis of the quantum
field theory by examining transition amplitudes, in which flavor neutrinos
appear only intermediate states between source-  and detection-interactions,
as schematically shown in FIG.\ref{fig1}.    In such approaches, it is unnecessary
for us to prepare any one-particle state of flavor neutrino  from the beginning ;
therefore, on the contrary to the standard formulation, we are free from the
trouble concerning Problem I.

Problem II :   The replacement of ct by the source-detector distance L for
obtaining (1.7) reflects such a situation that $P_{\sigma \rightarrow \rho}(L)$ 
mimics the usual experimental circumstances more closely than 
$P_{\sigma \rightarrow \rho}(t)$, because the times of reactions are not known 
and we measure distances rather than times. This simple replacement causes some 
doubts concerning the consistence of this procedure with the approximation
$\omega_j(\vec{k})-\omega_l(\vec{k}) \simeq \Delta m^2_{jl}/(2|\vec{k}|)$ .

Concerning a proper definition of space-dependent probability $P_{\sigma \rightarrow \rho}(L)$,
an interesting approach called "the current-density approach" has been proposed 
by Ancochea et al.\cite{r12} ; some modification is found in Ref.\cite{r13}. In this
quantum mechanical approach, the oscillation formulas $P_{\sigma \rightarrow \rho}(L)$, as a function
of the source-detector distance L, are obtained after integrating the
probability current density over a surface $\partial A$ arround a detector as 
well as the time duration of measurements ;
\begin{align}
P_{\sigma \rightarrow \rho}(L)=\int^{t_F}_{t_I}\!\!dt \int_{\partial A}\!\!d\vec{S}
\cdot\vec{j}(\vec{x},t). \label{1.17}
\end{align}
By employing the appropriate probability current densities of $K^0$ and $\bar{K}^0$
constructed with the use of wave-packet functions of $K_s$ and $K_L$, the usual 
formulas have been shown to be derived.

Blasone et al. \cite{r14} considered the neutrino oscillation along the line of the
current density approach. The current density $\vec{j}_{\rho}(\vec{x},t)$ is 
replaced by the expectation values
\begin{align}
\bra{\nu_\sigma(\vec{x}_I,t_I)}\vec{j}_{\rho}(\vec{x},t)\ket{\nu_\sigma(\vec{x}_I,t_I)},
\end{align}
where $\ket{\nu_\sigma(\vec{x}_I,t_I)}$ is an initial wave-packet state of the flavor
constructed on the
flavor vacuum in accordance with the formalism given by Blasone and Vitiello
\cite{r5,r6}.
It is pointed out that in low-energy region of neutrinos certain significant
deviations
from the standard formula may appear.

\subsection{Purpose and starting formula}

The purpose of the present work is to investigate the neutrino oscillation by
taking into consideration the two kinds of approaches explained in connection with 
Problems I and II.  We consider the formula (\ref{1.17}) with $\vec{j}_\rho(\vec{x},x^0)$ 
replaced by a certain kind of expectation value in the quantum field theory.
In order to do this, we notice that in the interaction representation the expectation 
value of a physical observable $F(x)$ at a space-time point $x=(\vec{x},x^0)$ with respect 
to a state $\ket{\Psi (x^0)}$ is given, in accordance with e.g. Umezawa's textbook \cite{r15}, 
by
\begin{align}
\bra{\Psi (x^0)} & F(x)\ket{\Psi (x^0)}=\bra{\Psi (x^0_I)}S^{-1}(x^0,x^0_I)F(x)S(x^0,x^0_I)\ket{\Psi (x^0_I)},\label{1.19}\\
\intertext{where}
\ket{\Psi (x^0)}&=S(x^0,x^0_I)\ket{\Psi (x^0_I)}, \nonumber \\
S(x^0,x^0_I)&=\sum_{m=0}(-i)^m \int^{x^0}_{x^0_I}\!\!d^4y_1\int^{y^0_1}_{x^0_I}\!\!d^4y_2
\cdots\int^{y^0_{m-1}}_{x^0_I}\!\!d^4y_m 
H_{\text{int}}(y_1) H_{\text{int}}(y_2) \cdots H_{\text{int}}(y_m). \label{1.20}
\end{align}
We consider the simplest case, where one $\pi^+$ decays into $\nu_\mu$ and
$\bar{\mu}$; for convenience we write $\pi^+ \rightarrow \nu_\sigma+\bar{l}_\sigma$.
As the initial state $\ket{\Psi_\pi(x^0)}$, we adopt a wave-packet state which is to 
be explained later in Sec.III. So we examine the expectation value of the flavor-neutrino 
current $\it j^{a}_{(\rho)}(x)$, which is to be defined below, with respect to the 
state $\ket{\Psi_\pi(x^0)}$. 
\begin{figure}[h]
 \begin{center}
  \scalebox{0.9}[0.9]{%WinTpicVersion3.08
\unitlength 0.1in
\begin{picture}( 75.4000, 19.4000)(5,-31)
% DOT 1 0 3 0
% 3 1620 1990 1620 1990 1620 1990
% 
\special{pn 13}%
\special{sh 1}%
\special{ar 1620 1990 10 10 0  6.28318530717959E+0000}%
\special{sh 1}%
\special{ar 1620 1990 10 10 0  6.28318530717959E+0000}%
\special{sh 1}%
\special{ar 1620 1990 10 10 0  6.28318530717959E+0000}%
% STR 2 0 3 1
% 3 6880 1450 6880 1550 5 0
% $j_{(\rho)}^b(x)$
\put(68.8000,-15.5000){\makebox(0,0){$j_{(\rho)}^b(x)$}}%
% BOX 1 0 3 2
% 2 6230 1270 7590 2480
% 
\special{pn 13}%
\special{pa 6230 1270}%
\special{pa 7590 1270}%
\special{pa 7590 2480}%
\special{pa 6230 2480}%
\special{pa 6230 1270}%
\special{fp}%
% LINE 2 0 3 3
% 4 2510 2160 6540 2240 2530 1790 6510 1580
% 
\special{pn 8}%
\special{pa 2510 2160}%
\special{pa 6540 2240}%
\special{fp}%
\special{pa 2530 1790}%
\special{pa 6510 1580}%
\special{fp}%
% ELLIPSE 2 0 3 4
% 4 6510 1910 6670 2237 6719 1890 6830 1929
% 
\special{pn 8}%
\special{ar 6510 1910 160 328  0.0281176 6.2353749}%
% LINE 1 1 3 5
% 2 6740 2510 6740 2910
% 
\special{pn 13}%
\special{pa 6740 2510}%
\special{pa 6740 2910}%
\special{da 0.070}%
% ELLIPSE 2 0 3 6
% 4 2470 1980 2570 2160 2600 1970 2670 1990
% 
\special{pn 8}%
\special{ar 2470 1980 100 180  0.0299910 6.2370642}%
% ELLIPSE 2 0 3 7
% 4 3660 1430 3740 1580 3750 1550 3770 1560
% 
\special{pn 8}%
\special{ar 3660 1430 80 150  0.5602320 0.6181442}%
% STR 2 0 3 8
% 3 3900 3000 3900 3100 5 0
% $(L \gg \vev{\pi}/\plt)$
\put(42,-31.0000){\makebox(0,0){$(L \gg \vev{\pi}/\plt)$}}%
% BOX 1 0 3 9
% 2 1000 1610 2000 2410
% 
\special{pn 13}%
\special{pa 1000 1610}%
\special{pa 2000 1610}%
\special{pa 2000 2410}%
\special{pa 1000 2410}%
\special{pa 1000 1610}%
\special{fp}%
% LINE 1 0 3 10
% 2 1010 1610 1810 1270
% 
\special{pn 13}%
\special{pa 1010 1610}%
\special{pa 1810 1270}%
\special{fp}%
% LINE 1 0 3 11
% 2 2010 1610 2810 1270
% 
\special{pn 13}%
\special{pa 2010 1610}%
\special{pa 2810 1270}%
\special{fp}%
% LINE 1 0 3 12
% 2 2010 2410 2810 2070
% 
\special{pn 13}%
\special{pa 2010 2410}%
\special{pa 2810 2070}%
\special{fp}%
% LINE 1 0 3 13
% 4 2810 1270 2810 2070 1810 1270 2810 1270
% 
\special{pn 13}%
\special{pa 2810 1270}%
\special{pa 2810 2070}%
\special{fp}%
\special{pa 1810 1270}%
\special{pa 2810 1270}%
\special{fp}%
% VECTOR 1 0 3 14
% 2 1120 1990 1560 1990
% 
\special{pn 13}%
\special{pa 1120 1990}%
\special{pa 1560 1990}%
\special{fp}%
\special{sh 1}%
\special{pa 1560 1990}%
\special{pa 1494 1970}%
\special{pa 1508 1990}%
\special{pa 1494 2010}%
\special{pa 1560 1990}%
\special{fp}%
% VECTOR 1 0 3 15
% 2 1650 1960 2150 1330
% 
\special{pn 13}%
\special{pa 1650 1960}%
\special{pa 2150 1330}%
\special{fp}%
\special{sh 1}%
\special{pa 2150 1330}%
\special{pa 2094 1370}%
\special{pa 2118 1372}%
\special{pa 2124 1396}%
\special{pa 2150 1330}%
\special{fp}%
% VECTOR 1 0 3 16
% 2 1680 1990 2295 2068
% 
\special{pn 13}%
\special{pa 1680 1990}%
\special{pa 2296 2068}%
\special{fp}%
\special{sh 1}%
\special{pa 2296 2068}%
\special{pa 2232 2040}%
\special{pa 2242 2062}%
\special{pa 2226 2080}%
\special{pa 2296 2068}%
\special{fp}%
% STR 2 0 3 17
% 3 1120 1790 1120 1890 5 0
% $\pi^+$
\put(13,-18.9000){\makebox(0,0){$\pi^+$}}%
% STR 2 0 3 18
% 3 2180 1230 2180 1330 5 0
% $l^+_\sigma$
\put(20,-13.8){\makebox(0,0){$l^+_\sigma$}}%
% STR 2 0 3 19
% 3 2280 1720 2280 1820 5 0
% $\nu_\sigma$
\put(22,-19.5){\makebox(0,0){$\nu_\sigma$}}%
% LINE 1 1 3 20
% 2 1620 2520 1620 2920
% 
\special{pn 13}%
\special{pa 1620 2520}%
\special{pa 1620 2920}%
\special{da 0.070}%
% STR 2 0 3 21
% 3 4060 2810 4060 2910 5 0
% $L$
\put(40.6000,-29.1000){\makebox(0,0){$L$}}%
% VECTOR 2 0 3 22
% 2 3840 2880 1660 2880
% 
\special{pn 8}%
\special{pa 3840 2880}%
\special{pa 1660 2880}%
\special{fp}%
\special{sh 1}%
\special{pa 1660 2880}%
\special{pa 1728 2900}%
\special{pa 1714 2880}%
\special{pa 1728 2860}%
\special{pa 1660 2880}%
\special{fp}%
% STR 2 0 3 23
% 3 960 2470 960 2570 5 0
% $(\vec{X}_I,x_I^0)$
\put(13,-25.7000){\makebox(0,0){$(\vec{X}_I,x_I^0)$}}%
% VECTOR 2 0 3 24
% 2 4280 2880 6660 2880
% 
\special{pn 8}%
\special{pa 4280 2880}%
\special{pa 6660 2880}%
\special{fp}%
\special{sh 1}%
\special{pa 6660 2880}%
\special{pa 6594 2860}%
\special{pa 6608 2880}%
\special{pa 6594 2900}%
\special{pa 6660 2880}%
\special{fp}%
\end{picture}%
}
 \end{center}
\caption{}
\label{fig2}
\end{figure}

The content of the first-step investigation is schematically shown in FIG.\ref{fig2}. 
Thus this is seen to give the simplest case, in which the oscillation probabilities 
are calculated in accordance with the quantum field theory as well as without recourse 
to any flavor neutrino state. After a small modification, it will be seen that we can take 
into account the observation of $l^+_\sigma$ as schematically shown in
FIG.\ref{fig3}.
%which mimics someaccelerator experiments more closely than Fig.II.
\begin{figure}[h]
 \begin{center}
  \scalebox{0.9}[0.9]{%WinTpicVersion3.08
\unitlength 0.1in
\begin{picture}( 70.3000, 18.3000)(  8.0500,-26.2000)
% DOT 1 0 3 0
% 3 1620 1990 1620 1990 1620 1990
% 
\special{pn 13}%
\special{sh 1}%
\special{ar 1620 1990 10 10 0  6.28318530717959E+0000}%
\special{sh 1}%
\special{ar 1620 1990 10 10 0  6.28318530717959E+0000}%
\special{sh 1}%
\special{ar 1620 1990 10 10 0  6.28318530717959E+0000}%
% LINE 1 1 3 1
% 2 3020 1190 2550 1460
% 
\special{pn 13}%
\special{pa 3020 1190}%
\special{pa 2550 1460}%
\special{da 0.070}%
% STR 2 0 3 2
% 3 1980 2100 1980 2200 5 0
% $\nu_\sigma$
\put(18.5,-21){\makebox(0,0){$\nu_\sigma$}}%
% STR 2 0 3 3
% 3 2070 1400 2070 1500 5 0
% $l^+_\sigma$
\put(25,-16.3){\makebox(0,0){$l^+_\sigma$}}%
% STR 2 0 3 4
% 3 1120 1790 1120 1890 5 0
% $\pi^+$
\put(14,-18.9000){\makebox(0,0){$\pi^+$}}%
% VECTOR 1 0 3 5
% 2 1680 1990 2294 2074
% 
\special{pn 13}%
\special{pa 1680 1990}%
\special{pa 2294 2074}%
\special{fp}%
\special{sh 1}%
\special{pa 2294 2074}%
\special{pa 2232 2046}%
\special{pa 2242 2068}%
\special{pa 2226 2086}%
\special{pa 2294 2074}%
\special{fp}%
% VECTOR 1 0 3 6
% 2 1650 1960 2550 1446
% 
\special{pn 13}%
\special{pa 1650 1960}%
\special{pa 2550 1446}%
\special{fp}%
\special{sh 1}%
\special{pa 2550 1446}%
\special{pa 2482 1462}%
\special{pa 2504 1472}%
\special{pa 2502 1496}%
\special{pa 2550 1446}%
\special{fp}%
% VECTOR 1 0 3 7
% 2 1120 1990 1560 1990
% 
\special{pn 13}%
\special{pa 1120 1990}%
\special{pa 1560 1990}%
\special{fp}%
\special{sh 1}%
\special{pa 1560 1990}%
\special{pa 1494 1970}%
\special{pa 1508 1990}%
\special{pa 1494 2010}%
\special{pa 1560 1990}%
\special{fp}%
% LINE 1 0 3 8
% 4 2810 1270 2810 2070 1810 1270 2810 1270
% 
\special{pn 13}%
\special{pa 2810 1270}%
\special{pa 2810 2070}%
\special{fp}%
\special{pa 1810 1270}%
\special{pa 2810 1270}%
\special{fp}%
% LINE 1 0 3 9
% 2 2010 2410 2810 2070
% 
\special{pn 13}%
\special{pa 2010 2410}%
\special{pa 2810 2070}%
\special{fp}%
% LINE 1 0 3 10
% 2 2010 1610 2810 1270
% 
\special{pn 13}%
\special{pa 2010 1610}%
\special{pa 2810 1270}%
\special{fp}%
% LINE 1 0 3 11
% 2 1010 1610 1810 1270
% 
\special{pn 13}%
\special{pa 1010 1610}%
\special{pa 1810 1270}%
\special{fp}%
% BOX 1 0 3 12
% 2 1000 1610 2000 2410
% 
\special{pn 13}%
\special{pa 1000 1610}%
\special{pa 2000 1610}%
\special{pa 2000 2410}%
\special{pa 1000 2410}%
\special{pa 1000 1610}%
\special{fp}%
% ELLIPSE 2 0 3 13
% 4 3660 1430 3740 1580 3750 1550 3770 1560
% 
\special{pn 8}%
\special{ar 3660 1430 80 150  0.5602320 0.6181442}%
% LINE 2 0 3 14
% 4 2522 2165 6546 2392 2555 1796 6540 1731
% 
\special{pn 8}%
\special{pa 2522 2166}%
\special{pa 6546 2392}%
\special{fp}%
\special{pa 2556 1796}%
\special{pa 6540 1732}%
\special{fp}%
% SPLINE 2 0 3 15
% 42 6688 2076 6683 2127 6675 2175 6663 2221 6648 2263 6630 2300 6610 2332 6588 2357 6564 2374 6539 2386 6515 2388 6490 2383 6467 2369 6444 2349 6424 2321 6407 2288 6392 2249 6381 2206 6372 2158 6368 2109 6368 2058 6372 2008 6379 1958 6389 1912 6404 1868 6421 1830 6441 1798 6463 1770 6486 1751 6511 1739 6535 1734 6560 1738 6583 1749 6607 1768 6627 1793 6646 1825 6661 1863 6673 1905 6682 1951 6687 2001 6688 2051 6688 2051
% 
\special{pn 8}%
\special{pa 6688 2076}%
\special{pa 6686 2108}%
\special{pa 6682 2140}%
\special{pa 6676 2172}%
\special{pa 6668 2202}%
\special{pa 6660 2234}%
\special{pa 6648 2264}%
\special{pa 6634 2292}%
\special{pa 6618 2320}%
\special{pa 6600 2346}%
\special{pa 6576 2368}%
\special{pa 6548 2384}%
\special{pa 6516 2388}%
\special{pa 6486 2382}%
\special{pa 6460 2364}%
\special{pa 6438 2340}%
\special{pa 6420 2314}%
\special{pa 6406 2284}%
\special{pa 6394 2254}%
\special{pa 6386 2224}%
\special{pa 6378 2192}%
\special{pa 6372 2162}%
\special{pa 6370 2130}%
\special{pa 6368 2098}%
\special{pa 6368 2066}%
\special{pa 6370 2034}%
\special{pa 6374 2002}%
\special{pa 6378 1970}%
\special{pa 6384 1938}%
\special{pa 6390 1908}%
\special{pa 6402 1878}%
\special{pa 6414 1848}%
\special{pa 6428 1820}%
\special{pa 6446 1792}%
\special{pa 6466 1768}%
\special{pa 6492 1748}%
\special{pa 6522 1736}%
\special{pa 6554 1736}%
\special{pa 6582 1748}%
\special{pa 6608 1768}%
\special{pa 6628 1794}%
\special{pa 6644 1822}%
\special{pa 6656 1850}%
\special{pa 6666 1880}%
\special{pa 6674 1912}%
\special{pa 6682 1944}%
\special{pa 6686 1974}%
\special{pa 6688 2006}%
\special{pa 6688 2038}%
\special{pa 6688 2052}%
\special{sp}%
% SPLINE 2 0 3 16
% 41 2588 1993 2585 2021 2580 2047 2573 2073 2564 2096 2552 2116 2540 2134 2526 2147 2511 2157 2496 2162 2481 2164 2465 2161 2450 2154 2436 2142 2424 2127 2413 2108 2403 2087 2396 2063 2391 2036 2388 2009 2388 1981 2390 1954 2395 1927 2401 1901 2410 1877 2421 1856 2433 1838 2447 1823 2461 1813 2477 1806 2492 1804 2508 1806 2522 1812 2536 1823 2550 1837 2561 1855 2571 1876 2578 1899 2584 1924 2587 1952 2588 1980
% 
\special{pn 8}%
\special{pa 2588 1994}%
\special{pa 2584 2026}%
\special{pa 2578 2056}%
\special{pa 2568 2088}%
\special{pa 2554 2116}%
\special{pa 2534 2140}%
\special{pa 2508 2158}%
\special{pa 2476 2164}%
\special{pa 2448 2152}%
\special{pa 2426 2128}%
\special{pa 2410 2102}%
\special{pa 2398 2072}%
\special{pa 2392 2040}%
\special{pa 2388 2008}%
\special{pa 2388 1976}%
\special{pa 2392 1944}%
\special{pa 2398 1912}%
\special{pa 2408 1882}%
\special{pa 2422 1854}%
\special{pa 2442 1828}%
\special{pa 2468 1810}%
\special{pa 2500 1804}%
\special{pa 2528 1816}%
\special{pa 2552 1840}%
\special{pa 2568 1866}%
\special{pa 2578 1898}%
\special{pa 2586 1928}%
\special{pa 2588 1960}%
\special{pa 2588 1980}%
\special{sp}%
% POLYGON 1 0 3 17
% 5 6217 1410 6225 2620 7586 2612 7578 1402 6217 1410
% 
\special{pn 13}%
\special{pa 6218 1410}%
\special{pa 6226 2620}%
\special{pa 7586 2612}%
\special{pa 7578 1402}%
\special{pa 6218 1410}%
\special{fp}%
% STR 2 0 3 18
% 3 7070 1760 7070 1860 5 0
% $j_{(\rho)}^b(x)$
\put(70.7000,-18.6000){\makebox(0,0){$j_{(\rho)}^b(x)$}}%
% BOX 1 0 3 19
% 2 3050 1000 3190 1280
% 
\special{pn 13}%
\special{pa 3050 1000}%
\special{pa 3190 1000}%
\special{pa 3190 1280}%
\special{pa 3050 1280}%
\special{pa 3050 1000}%
\special{fp}%
% LINE 1 0 3 20
% 2 3050 990 3300 790
% 
\special{pn 13}%
\special{pa 3050 990}%
\special{pa 3300 790}%
\special{fp}%
% LINE 1 0 3 21
% 2 3190 1270 3440 1070
% 
\special{pn 13}%
\special{pa 3190 1270}%
\special{pa 3440 1070}%
\special{fp}%
% LINE 1 0 3 22
% 2 3190 1000 3440 800
% 
\special{pn 13}%
\special{pa 3190 1000}%
\special{pa 3440 800}%
\special{fp}%
% LINE 1 0 3 23
% 4 3440 1090 3440 810 3300 790 3420 800
% 
\special{pn 13}%
\special{pa 3440 1090}%
\special{pa 3440 810}%
\special{fp}%
\special{pa 3300 790}%
\special{pa 3420 800}%
\special{fp}%
\end{picture}%
}
 \end{center}
\caption{}
\label{fig3}
\end{figure}

The relevant weak interaction in (\ref{1.14}) is written as
\begin{align}
\mathcal{L}_{\text{int}}(x)=\sum_\rho \left[ \bar{\nu}_\rho(x)J_\rho(x)+\bar{J}_\rho(x)\nu_\rho(x) \right]; \label{1.21}\\
J_\rho(x)=\delta/\delta \bar{\nu}(x)\cdot\mathcal{L}_{\text{int}} 
\stackrel{\text{eff}}{=} if_{\pi\rho}v^a l_\rho(x)\partial_a\phi(x),\label{1.22}\\
\bar{J}_\rho(x)=\delta \mathcal{L}_{\text{int}}/\delta \nu(x) 
\stackrel{\text{eff}}{=} if^\ast_{\pi\rho}\bar{l}_\rho(x)v^a\partial_a\phi^\dagger(x),\label{1.23}
\end{align}
with $v^a=\gamma^a(1+\gamma_5)$. (We use the Kramers representation \cite{r16} of $\gamma$-matrices 
as explained in Refs.\cite{r6} and \cite{r7}) Following the natural definition of "currents" \cite{r17}, we consider 
the infinitesimal phase transformation
\begin{align}
\nu_F(x) \rightarrow \nu'_F(x)=e^{i\varphi(x)}\nu_F(x),\label{1.24}
\intertext{with}
\varphi(x)=
\begin{pmatrix}
\varphi_e(x) & & \\
& \varphi_\mu(x) & \\
& & \ddots
\end{pmatrix}, \nn
\end{align}
and obtain
\begin{align}
\mathcal{L}(x) &\rightarrow \mathcal{L}'(x)=\mathcal{L}(x)+\delta\mathcal{L}(x), \nn \\
\delta\mathcal{L}(x) &= -i\bar{\nu}_F(x)\left( \gamma^a\partial_a \varphi(x) 
+[M,\varphi(x)]\right)\nu_F(x)+\delta\mathcal{L}_{\text{int}}. \label{1.25}
\end{align}
By remembering $\mathcal{L}_{\text{int}}$ has no derivative of $\nu_F(x)$-field, we 
obtain the flavor neutrino current as
\begin{align}
j^a_{(\rho)}(x) &\equiv \del{\delta\mathcal{L}(x)}{(\partial_a\varphi_\rho(x))} \nn \\
&=-i\bar{\nu}_\rho(x) \gamma^a \nu_\rho(x) \nn \\
&=-i\sum_{j,l}{Z^{1/2}_{\rho j}}^\ast Z^{1/2}_{\rho l}\bar{\nu}_j(x)\gamma^a\nu_l(x); \label{1.26}
\end{align}
its 4-divergence is given by
\begin{align}
\partial_a j^a_{(\rho)}(x)&=\del{\delta\mathcal{L}(x)}{(\varphi_\rho(x))} \nn \\
&=-i\left[(\bar{\nu}_F(x)M)_\rho \nu_\rho(x)-\bar{\nu}_\rho(x) (M\nu_F(x))_\rho\right]
+\del{\delta \mathcal{L}_{\text{int}}(x)}{\varphi_\rho(x)}. \label{1.27}
\end{align}
The right-hand is called the " diffusion term " \cite{r13,r14}. In the interaction representation,
in which the $\nu_M(x)$-field satisfies the free equation
\begin{align}
(\sla{\partial}+M_{\text{diag}})\nu_M(x)=0, \label{1.28}
\end{align}
the total flavor current of neutrinos is seen to be conserved;
\begin{align}
\sum_\rho \partial_a j^a_{(\rho)}=0. \label{1.29}
\end{align}

The remaining part of the present paper is organized as follows. In Sec.II. we
perform some model calculations to see the relation of the current-operator
expectation values (1.19) to corresponding transition amplitudes. In Sec.III, we
investigate the structures of the expectation values for $j^a_{(\rho)}(x)$ with 
respect to the wave-packet state, decaying into $\nu_\sigma$ and $\bar{l}_\sigma$.
In Sec.IV, we discuss the structures of the same kind of the expectation values in 
the boson model for simplicity, in which the observation of a charged boson in 
decay products  is taken into consideration, and in Sec.V summarize results and future tasks.
The detailed analyses including numerical calculations are to be left to the subsequent paper \cite{r18}.

%% section 2 %%
\section{RELATION OF CURRENT EXPECTATION VALUE TO TRANSITION MATRIX ELEMENT}
\subsection{The case with no mixing}
We perform the simplest model calculation with the aim of seeing the relation of a 
current expectation value to the absolute square of the relevant transition matrix
element. First we examine the case of no mixing interaction.

Let's consider the simple situation where (pseud-)scalar particles ${A, B, C }$ have 
the weak interaction
\begin{align}
\mathcal{L}_{\text{int}}&=f_{ABC} \phi^\dagger_C(x) \phi_B(x) \phi_A(x)+H.c. \nn \\
&=-\mathcal{H}_{\text{int}}(x). \label{2.1}
\end{align}
An $A$-particle described by the complex field $\phi_A(x)$ decays into $\bar{B}$ and $C$
particles. We write the current of the $C$-field as
\begin{align}
j^{(C)}_b(x)&=i\left( \phi^\dagger_C(x) \partial_b \phi_C(x)
-\partial_b \phi^\dagger_C(x)\phi_C(x)\right) \nn \\
&\equiv i\phi^\dagger_C \olra{\partial}_b \phi_C(x). \label{2.2}
\end{align}

We execute, in the interaction representation, perturbative calculations in
the lowest order of $H_{\text{int}}$ by employing the plane wave expansion
\begin{align}
\phi_f(x)=\int\!\!d^3p\frac{1}{(2\pi)^{3/2} 2E_f(\vec{p})}
\left[ \alpha_f(\vec{p}) e^{ipx}+\beta^\dagger_f(\vec{p}) e^{-ipx} \right]
,\quad f=A,B,C; \label{2.3}
\end{align}
here, $px=\vec{p}\cdot\vec{x}-E_f(\vec{p})x^0$, $E_f(\vec{p})=\sqrt{\vec{p}^2+m^2}$;
the commutation relations among the expansion-coefficient operators are
\begin{align}
[\alpha_f(\vec{p}),\alpha^\dagger_g(\vec{q})]
=[\beta_f(\vec{p}),\beta^\dagger_g(\vec{q})]=2 E_f(\vec{p})\delta_{fg}\delta(\vec{p}-\vec{q}),
\quad \text{others}=0 ; \label{2.4}
\end{align}
the vacuum $\ket{0}$ is defined by
\begin{align}
\alpha_g(\vec{p})\ket{0}=\beta_g(\vec{p})\ket{0}=0 \qquad \text{for} \quad g=A,B,C. \label{2.5}
\end{align}
the number operator of $\phi_C$-field is given by
\begin{align}
N_C(x^0)&=i : \int\!\!d^3x j_4^{(C)}(x): \nn \\
&=\int\!\!d^3k\frac{1}{2E_C(\vec{k})} [\alpha^\dagger_C(\vec{k}) \alpha_C(\vec{k})-\beta^\dagger_C(\vec{k})\beta_C(\vec{k})]. \label{2.6}
\end{align}
(The symbol  : : means the normal ordering.) 

In the following we examine the forms 
of expectation values of the current and the number operator with respect to one 
$A$-particle state
\begin{align}
\ket{A(p;x^0)}=S(x^0,x^0_I)\alpha^\dagger_A(\vec{p})\ket{0}. \label{2.7}
\end{align}

We define the expectation value of the current with respect to the state \eqref{2.7} as  
\begin{align}
\mathcal{E}(A(x^0_I ; \text{C-current}(x))_b &\equiv 
\bra{0}\alpha_A(\vec{p}) S(x^0,x^0_I)^{-1} :j_b^{(c)}(x) : S(x^0,x^0_I)\alpha^\dagger_A(\vec{p})\ket{0}^{\text{con}}, \label{2.8}
\intertext{and also that of the number operator as}
\left<N_C(x^0)\right>_{A(p,x^0_I)} &\equiv i \int\!\!d^3x \mathcal{E}(A(x^0_I ; \text{C-current}(x))_4. \label{2.9}
\end{align}
We take out only the connected part in the expectation values, as designated in
R.H.S. of (\ref{2.8}). In the lowest order ( i.e. the second order ) of the weak interaction,
we have
\begin{align}
\left<N_C(x^0)\right>_{A(p,x^0_I)} =\bra{0}\alpha_A(\vec{p}) \int^{x^0}_{x^0_I}\!d^4z H_{\text{int}}(z)
N_C(x^0) \int^{x^0}_{x^0_I}\!d^4y H_{\text{int}}(y) \alpha^\dagger_A(\vec{p})\ket{0}, \label{2.10}
\end{align}
which consists of two parts ;
\begin{align}
\left<N_C(x^0)\right>_{A(p,x^0_I)}&=\left<n_C(x^0)\right>_{A(p,x^0_I)}+\left<\bar{n}_C(x^0)\right>_{A(p,x^0_I)}\label{2.11}, 
\intertext{where}
n_C &\equiv \int\!\!d^3k\frac{1}{2E_C(\vec{k})}\alpha^\dagger_C(\vec{k})\alpha_C(\vec{k}),\qquad
\bar{n}_C \equiv -\int\!\!d^3k\frac{1}{2E_C(\vec{k})}\beta^\dagger_C(\vec{k})\beta_C(\vec{k}). \label{2.12}
\end{align}

\begin{figure}[h]
 \begin{center}
  \begin{tabular}{cc}
   \scalebox{0.6}[0.6]{%WinTpicVersion3.08
\unitlength 0.1in
\begin{picture}(60,15)(-1,-14)
% LINE 1 0 3 0
% 2 10 510 1400 510
% 
\special{pn 13}%
\special{pa 10 510}%
\special{pa 1400 510}%
\special{fp}%
% CIRCLE 1 0 0 1
% 4 2600 910 2740 970 2740 970 2900 1040
% 
\special{pn 13}%
\special{ar 2600 910 152 152  0.4089078 6.2831853}%
\special{ar 2600 910 152 152  0.0000000 0.4048918}%
% LINE 1 0 3 2
% 2 3690 510 5080 510
% 
\special{pn 13}%
\special{pa 3690 510}%
\special{pa 5080 510}%
\special{fp}%
% ELLIPSE 1 0 3 3
% 4 2550 500 3690 920 970 540 3850 540
% 
\special{pn 13}%
\special{ar 2550 500 1140 420  0.0836505 3.0727144}%
% ELLIPSE 1 0 3 4
% 4 2550 520 3690 100 3850 480 970 480
% 
\special{pn 13}%
\special{ar 2550 520 1140 420  3.2104709 6.1995348}%
% VECTOR 1 0 3 5
% 2 4510 500 4500 500
% 
\special{pn 13}%
\special{pa 4510 500}%
\special{pa 4500 500}%
\special{fp}%
\special{sh 1}%
\special{pa 4500 500}%
\special{pa 4568 520}%
\special{pa 4554 500}%
\special{pa 4568 480}%
\special{pa 4500 500}%
\special{fp}%
% VECTOR 1 0 3 6
% 2 3450 760 3443 768
% 
\special{pn 13}%
\special{pa 3450 760}%
\special{pa 3444 768}%
\special{fp}%
\special{sh 1}%
\special{pa 3444 768}%
\special{pa 3502 732}%
\special{pa 3478 728}%
\special{pa 3472 706}%
\special{pa 3444 768}%
\special{fp}%
% VECTOR 1 0 3 7
% 2 3390 230 3384 222
% 
\special{pn 13}%
\special{pa 3390 230}%
\special{pa 3384 222}%
\special{fp}%
\special{sh 1}%
\special{pa 3384 222}%
\special{pa 3408 288}%
\special{pa 3416 266}%
\special{pa 3440 264}%
\special{pa 3384 222}%
\special{fp}%
% STR 2 0 3 8
% 3 2600 1120 2600 1220 5 0
% $n_C$
\put(26.0000,-12.2000){\makebox(0,0){$n_C$}}%
% STR 2 0 3 9
% 3 3570 810 3570 910 5 0
% $C$
\put(35.7000,-9.1000){\makebox(0,0){$C$}}%
% STR 2 0 3 10
% 3 3480 0 3480 100 5 0
% $\bar{B}$
\put(34.8000,-1.0000){\makebox(0,0){$\bar{B}$}}%
% STR 2 0 3 11
% 3 4650 210 4650 310 5 0
% $A$
\put(46.5000,-3.1000){\makebox(0,0){$A$}}%
% STR 2 0 3 12
% 3 460 210 460 310 5 0
% $A$
\put(6,-3.1000){\makebox(0,0){$A$}}%
\put(1,-3){\makebox(0,0){(a)}}%
% FUNC 1 0 3 0
% 9 4630 3130 5550 3540 4820 2940 5190 3280 4940 3100 4630 3130 5550 3540 0 3 0 0
% sin(x)
\special{pn 13}%
\end{picture}%
} &
   \scalebox{0.65}[0.65]{%WinTpicVersion3.08
\unitlength 0.1in
\begin{picture}(60,15)( -5,-17)
% STR 2 0 3 0
% 3 2620 760 2620 860 5 0
% $A$
\put(22,-8.6000){\makebox(0,0){$A$}}%
\put(-1.5,-7.6){\makebox(0,0){(b)}}
% STR 2 0 3 1
% 3 1870 830 1870 930 5 0
% $B$
\put(15,-10){\makebox(0,0){$B$}}%
% LINE 1 0 3 2
% 2 110 1080 2380 1080
% 
\special{pn 13}%
\special{pa 110 1080}%
\special{pa 2380 1080}%
\special{fp}%
% VECTOR 1 0 3 3
% 2 1710 1070 1694 1070
% 
\special{pn 13}%
\special{pa 1710 1070}%
\special{pa 1694 1070}%
\special{fp}%
\special{sh 1}%
\special{pa 1694 1070}%
\special{pa 1762 1090}%
\special{pa 1748 1070}%
\special{pa 1762 1050}%
\special{pa 1694 1070}%
\special{fp}%
% STR 2 0 3 4
% 3 2490 0 2490 100 5 0
% $A$
\put(25,-1.8){\makebox(0,0){$A$}}%
% STR 2 0 3 5
% 3 2260 1380 2260 1480 5 0
% $\bar{C}$
\put(22.6000,-14.8000){\makebox(0,0){$\bar{C}$}}%
% STR 2 0 3 6
% 3 1290 1690 1290 1790 5 0
% $\bar{n}_C$
\put(12.9000,-17.9000){\makebox(0,0){$\bar{n}_C$}}%
% VECTOR 1 0 3 7
% 2 2140 1330 2133 1338
% 
\special{pn 13}%
\special{pa 2140 1330}%
\special{pa 2134 1338}%
\special{fp}%
\special{sh 1}%
\special{pa 2134 1338}%
\special{pa 2192 1302}%
\special{pa 2168 1298}%
\special{pa 2162 1276}%
\special{pa 2134 1338}%
\special{fp}%
% ELLIPSE 1 0 3 8
% 4 1240 1070 2380 1490 -340 1110 2540 1110
% 
\special{pn 13}%
\special{ar 1240 1070 1140 420  0.0836505 3.0727144}%
% CIRCLE 1 0 0 9
% 4 1290 1480 1430 1540 1430 1540 1590 1610
% 
\special{pn 13}%
\special{ar 1290 1480 152 152  0.4089078 6.2831853}%
\special{ar 1290 1480 152 152  0.0000000 0.4048918}%
% LINE 1 0 3 10
% 2 100 1070 3153 57
% 
\special{pn 13}%
\special{pa 100 1070}%
\special{pa 3154 58}%
\special{fp}%
% LINE 1 0 3 11
% 2 2400 1060 1460 680
% 
\special{pn 13}%
\special{pa 2400 1060}%
\special{pa 1460 680}%
\special{fp}%
% LINE 1 0 3 12
% 2 30 90 1280 600
% 
\special{pn 13}%
\special{pa 30 90}%
\special{pa 1280 600}%
\special{fp}%
% VECTOR 1 0 3 13
% 4 2190 970 1816 820 1998 885 1998 885
% 
\special{pn 13}%
\special{pa 2190 970}%
\special{pa 1816 820}%
\special{fp}%
\special{sh 1}%
\special{pa 1816 820}%
\special{pa 1870 864}%
\special{pa 1866 840}%
\special{pa 1886 826}%
\special{pa 1816 820}%
\special{fp}%
\special{pa 1998 886}%
\special{pa 1998 886}%
\special{fp}%
% FUNC 1 0 3 0
% 9 4630 3130 5550 3540 4820 2940 5190 3280 4940 3100 4630 3130 5550 3540 0 3 0 0
% sin(x)
\special{pn 13}%
\end{picture}%
} 
  \end{tabular}
 \end{center}
\caption{Diagrams representing the expectation values of $n_c(x^0)$ and $\bar{n}_c(x^0)$.}
\label{fig4}
\end{figure}
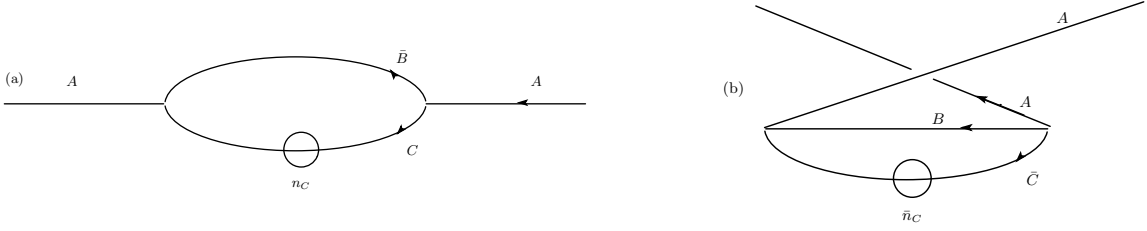

The first part, represented as FIG.\ref{fig4} (a), is given by
\begin{align}
\left<n_C(x^0)\right>_{A(p,x^0_I)}&=\int^{x^0}_{x^0_I}\!\!d^4z\int^{x^0}_{x^0_I}\!\!d^4y
\frac{|f_{ABC}|^2}{(2\pi)^3}e^{-ipz+ipy}\bra{0}\phi^\dagger_B(z)\phi_B(y)\ket{0}
\bra{0}\phi_C(z)n_C\phi^\dagger_C(y)\ket{0} \nn \\
&=\int^t_0\!\int^t_0dz^0dy^0 V|f_{ABC}|^2 \int\!\!d^3k 
\frac{ e^{i(E_A(\vec{p}) - E_B(\vec{q}) - E_C(\vec{k}) )(z^0-y^0) }}{(2\pi)^6 2E_B(\vec{q}) 2E_C(\vec{k})}\Biggr|_{\vec{q}=\vec{p}-\vec{k}},
\label{2.13}
\end{align}
here, $t \equiv x^0-x^0_I$, $V = \int\!\!d^3x 1$.
         Similarly, the second part in R.H.S. of (\ref{2.11}), represented as FIG.\ref{fig4}
(b), is given by
\begin{align}
\left< \bar{n}_C(x^0)\right>_{A(p,x^0_I)}&=\int^{x^0}_{x^0_I}\!\!d^4z\int^{x^0}_{x^0_I}\!\!d^4y
|f_{ABC}|^2\bra{0}\phi_B(z)\phi_B^\dagger(y)\ket{0} \nn \\
& \times \bra{0} \phi_C^\dagger(z) \bar{n}_C \phi_C(y) \ket{0}
\bra{0} \alpha_A(\vec{p}) \phi_A(z) \phi_A^\dagger(y) \alpha_A^\dagger(\vec{p}) \ket{0}^{\text{con}} \nn \\
&=-\int^t_0 \int^t_0\!dz^0 dy^0 V|f_{ABC}|^2 \int\!\!d^3k
\frac{ e^{-i(E_A(\vec{p}) + E_B(\vec{q}) + E_C(\vec{k}) )(z^0-y^0) }}{(2\pi)^6 2E_B(\vec{q}) 2E_C(\vec{k})}\Biggr|_{\vec{q}=-\vec{p}-\vec{k}}
\label{2.14}.
\end{align}
By adding the damping factor $e^{-\epsilon |z^0|-\epsilon |y^0|}$, $\epsilon >0$, to the
integrands of (\ref{2.13}) and (\ref{2.14}), we obtain
\begin{align}
\left< n_C(x^0) \right>_{A(p,x^0_I)}&=V|f_{ABC}|^2 \int\!\!d^3k \frac{1}{(2\pi)^6 2E_B(\vec{q}) 2E_C(\vec{k})} \nn \\
&\times\frac{ e^{i(E_A(\vec{p}) - E_B(\vec{q}) - E_C(\vec{k}) +i\epsilon)t }-1}{E_A(\vec{p}) - E_B(\vec{q}) - E_C(\vec{k}) +i\epsilon}\cdot
\frac{ e^{-i(E_A(\vec{p}) - E_B(\vec{q}) - E_C(\vec{k}) -i\epsilon)t }-1}{E_A(\vec{p}) - E_B(\vec{q}) - E_C(\vec{k}) -i\epsilon}\Biggr|_{\vec{q}=\vec{p}-\vec{k}}, \label{2.15}\\
\left< \bar{n}_C(x^0) \right>_{A(p,x^0_I)}&=-V|f_{ABC}|^2 \int\!\!d^3k \frac{1}{(2\pi)^6 2E_B(\vec{q}) 2E_C(\vec{k})} \nn \\
&\times\frac{ e^{-i(E_A(\vec{p})+E_B(\vec{q})+E_C(\vec{k})-i\epsilon)t }-1}{E_A(\vec{p})+E_B(\vec{q})+E_C(\vec{k})-i\epsilon}\cdot
\frac{ e^{i(E_A(\vec{p})+E_B(\vec{q})+E_C(\vec{k})+i\epsilon)t
 }-1}{E_A(\vec{p})+E_B(\vec{q})+E_C(\vec{k})+i\epsilon}\Biggr|_{\vec{q}=-\vec{p}-\vec{k}}. \label{2.16}
\end{align}
Thus we see that due to the higher configuration in the intermediate
state of FIG.\ref{fig4} (b) in comparison with FIG.\ref{fig4} (a), the main contribution
to $\left< N_C(x^0) \right>_{A(p,x^0_I)}$ comes
from $\left< n_C(x^0) \right>_{A(p,x^0_I)}$;  i.e.
\begin{align}
\left< N_C(x^0) \right>_{A(p,x^0_I)} \simeq \left< n_C(x^0)
 \right>_{A(p,x^0_I)}. \label{2.17}
\end{align}

On the other hand, the transition amplitude from the initial A state at $x^0_I$ to
$\bar{B}(q)$ and $C(k)$ at $x^0$ is , in the first order of the weak interaction,
given by
\begin{align}
M(A(p,x^0_I)&\rightarrow \bar{B}(q)+C; x^0) \nn \\
&=\bra{0} \beta_B(\vec{q}) \alpha_C(\vec{k}) 
i\!\int^{x^0}_{x^0_I}\!\!d^4z f_{ABC} \phi_C^\dagger(z)\phi_B(z)\phi_A(z)
\alpha_A^\dagger(\vec{p})\ket{0} \nn \\
&=i f_{ABC} \int^{x^0}_{x^0_I}\!d^4z \frac{e^{iz(p-q-k)}}{(2\pi)^{9/2}}, \label{2.18}
\intertext{from which we obtain}
\int\frac{d^3q}{2E_B(\vec{q})}\int\frac{d^3k}{2E_C(\vec{k})}|M|^2 &=|f_{ABC}|^2 V
 \int\!\!d^3q \int\!\!d^3k \frac{\delta(\vec{p}-\vec{q}-\vec{k})}{(2\pi)^6 2E_B(\vec{q}) 2E_C(\vec{k})} \nn \\
&\times \int^{x^0_I}_{x^0}\!\int^{x^0_I}_{x^0}\!\!dz^0dy^0 e^{i(z^0-y^0)(E_A(\vec{p})-E_B(\vec{q})-E_C(\vec{k}))} \nn \\
&=\left< n_C(x^0) \right>_{A(p,x^0_I)} \simeq
 \ev{N_C(x^0)}_{A(p,x^0_I)}. \label{2.19}
\end{align}
Diagramatically we may express the above equality as FIG.\ref{fig5}.
\begin{figure}[h]
 \begin{center}
  \scalebox{0.5}[0.5]{%WinTpicVersion3.08
\unitlength 0.1in
\begin{picture}(20,15)(30,-20)
% LINE 1 0 3 0
% 2 2060 1400 3450 1400
% 
\special{pn 13}%
\special{pa 2060 1400}%
\special{pa 3450 1400}%
\special{fp}%
% STR 2 0 3 0
% 3 3520 2700 3520 2800 5 0
%\scalebox{3.5}[3.5]{$\int^{x^0}_{x^0_I}\int^{x^0}_{x^0_I}d^4zd^4y$}
\put(6,-14){\makebox(0,0){\scalebox{3.5}[3.5]{$\int^{x^0}_{x^0_I}\int^{x^0}_{x^0_I}d^4zd^4y$}}}%
% STR 2 0 3 0
% 3 2670 1120 2670 1220 5 0
% {\Large $A(\vec{p},x^0)$}
\put(26.7000,-12.2000){\makebox(0,0){{\Large$A(\vec{p},x^0)$}}}%
% STR 2 0 3 0
% 3 6700 1100 6700 1200 5 0
% {\Large$A(\vec{p},x^0)$}
\put(67.0000,-12.0000){\makebox(0,0){{\Large$A(\vec{p},x^0)$}}}%
%% STR 2 0 3 0
%% 3 5530 890 5530 990 5 0
%% {\Large$\bar{B}$}
\put(55.3000,-9.9000){\makebox(0,0){{\Large$\bar{B}$}}}%
% STR 2 0 3 0
% 3 5620 1700 5620 1800 5 0
% {\Large$C$}
\put(56.2000,-18.0000){\makebox(0,0){{\Large$C$}}}%
% STR 2 0 3 0
% 3 4650 2010 4650 2110 5 0
% {\Large$N_C(x^0)$}
\put(46.5000,-21.1000){\makebox(0,0){{\Large$N_C(x^0)$}}}%
% VECTOR 1 0 3 0
% 2 5440 1120 5434 1112
% 
\special{pn 13}%
\special{pa 5440 1120}%
\special{pa 5434 1112}%
\special{fp}%
\special{sh 1}%
\special{pa 5434 1112}%
\special{pa 5458 1178}%
\special{pa 5466 1156}%
\special{pa 5490 1154}%
\special{pa 5434 1112}%
\special{fp}%
% VECTOR 1 0 3 0
% 2 5500 1650 5493 1658
% 
\special{pn 13}%
\special{pa 5500 1650}%
\special{pa 5494 1658}%
\special{fp}%
\special{sh 1}%
\special{pa 5494 1658}%
\special{pa 5552 1622}%
\special{pa 5528 1618}%
\special{pa 5522 1596}%
\special{pa 5494 1658}%
\special{fp}%
% VECTOR 1 0 3 0
% 2 6560 1390 6550 1390
% 
\special{pn 13}%
\special{pa 6560 1390}%
\special{pa 6550 1390}%
\special{fp}%
\special{sh 1}%
\special{pa 6550 1390}%
\special{pa 6618 1410}%
\special{pa 6604 1390}%
\special{pa 6618 1370}%
\special{pa 6550 1390}%
\special{fp}%
% ELLIPSE 1 0 3 0
% 4 4600 1410 5740 990 5900 1370 3020 1370
% 
\special{pn 13}%
\special{ar 4600 1410 1140 420  3.2104709 6.1995348}%
% ELLIPSE 1 0 3 0
% 4 4600 1390 5740 1810 3020 1430 5900 1430
% 
\special{pn 13}%
\special{ar 4600 1390 1140 420  0.0836505 3.0727144}%
% LINE 1 0 3 0
% 2 5740 1400 7130 1400
% 
\special{pn 13}%
\special{pa 5740 1400}%
\special{pa 7130 1400}%
\special{fp}%
% CIRCLE 1 0 0 0
% 4 4650 1800 4790 1860 4790 1860 4950 1930
% 
\special{pn 13}%
\special{ar 4650 1800 152 152  0.4089078 6.2831853}%
\special{ar 4650 1800 152 152  0.0000000 0.4048918}%
% FUNC 1 0 3 0
% 9 4630 3130 5550 3540 4820 2940 5190 3280 4940 3100 4630 3130 5550 3540 0 3 0 0
% sin(x)
\special{pn 13}%
\end{picture}%
}\\
  \scalebox{0.5}[0.5]{%WinTpicVersion3.08
\unitlength 0.1in
\begin{picture}( 18.0000, 11.9000)(65,-16)
% VECTOR 1 0 3 0
% 6 8210 1420 7600 820 8200 1410 7600 2010 9400 1410 8800 1410
% 
\special{pn 13}%
\special{pa 8210 1420}%
\special{pa 7600 820}%
\special{fp}%
\special{sh 1}%
\special{pa 7600 820}%
\special{pa 7634 882}%
\special{pa 7638 858}%
\special{pa 7662 852}%
\special{pa 7600 820}%
\special{fp}%
\special{pa 8200 1410}%
\special{pa 7600 2010}%
\special{fp}%
\special{sh 1}%
\special{pa 7600 2010}%
\special{pa 7662 1978}%
\special{pa 7638 1972}%
\special{pa 7634 1950}%
\special{pa 7600 2010}%
\special{fp}%
\special{pa 9400 1410}%
\special{pa 8800 1410}%
\special{fp}%
\special{sh 1}%
\special{pa 8800 1410}%
\special{pa 8868 1430}%
\special{pa 8854 1410}%
\special{pa 8868 1390}%
\special{pa 8800 1410}%
\special{fp}%
% LINE 1 0 3 1
% 2 8860 1410 8200 1410
% 
\special{pn 13}%
\special{pa 8860 1410}%
\special{pa 8200 1410}%
\special{fp}%
% STR 2 0 3 2
% 3 6785 1310 6785 1410 5 0
% \scalebox{3.5}[3.5]{$\int^{x^0}_{x^0_I}\!\!d^4z$}
\put(67.8500,-14.1000){\makebox(0,0){\scalebox{3.5}[3.5]{$\int^{x^0}_{x^0_I}\!\!d^4z$}}}%
% STR 2 0 3 3
% 3 4050 1310 4050 1410 5 0
% \scalebox{3.5}[3.5]{$\simeq\int\!\!\int\!\!\frac{d^3qd^3k}{2E_B(\vec{q})\ 2E_C(\vec{k})}$}
\put(40.5000,-14.1000){\makebox(0,0){\scalebox{3.5}[3.5]{$\simeq\int\!\!\int\!\!\frac{d^3qd^3k}{2E_B(\vec{q})\ 2E_C(\vec{k})}$}}}%
% STR 2 0 3 4
% 3 8020 820 8020 920 5 0
% {\Large$\bar{B}(q)$}
\put(80.2000,-9.2000){\makebox(0,0){{\Large$\bar{B}(q)$}}}%
% STR 2 0 3 5
% 3 8090 1720 8090 1820 5 0
% {\Large$C(k)$}
\put(80.9000,-18.2000){\makebox(0,0){{\Large$C(k)$}}}%
% LINE 1 0 3 6
% 2 5980 800 5980 2090
% 
\special{pn 13}%
\special{pa 5980 800}%
\special{pa 5980 2090}%
\special{fp}%
% LINE 1 0 3 7
% 2 9550 800 9550 2090
% 
\special{pn 13}%
\special{pa 9550 800}%
\special{pa 9550 2090}%
\special{fp}%
% STR 2 0 3 8
% 3 8320 1160 8320 1260 5 0
% {\Large$(z)$}
\put(83.2000,-12.6000){\makebox(0,0){{\Large$(z)$}}}%
% STR 2 0 3 9
% 3 9130 1160 9130 1260 5 0
% {\Large$A(p)$}
\put(91.3000,-12.6000){\makebox(0,0){{\Large$A(p)$}}}%
% STR 2 0 3 10
% 3 9670 650 9670 750 5 0
% {\Large$2$}
\put(96.7000,-7.5000){\makebox(0,0){{\Large$2$}}}%
\end{picture}%
}
 \end{center}
\caption{Relation of the expectation value of number operator to the
 transition probability.}
\label{fig5}
\end{figure}

A derivation which is somewhat more general than that explained above is
given when we assume the $\bar{n}_C$-contribution to the expectation values (\ref{2.8})
is smaller than $n_C$-contribution, because the former arises from higher configurations
; i.e.
\begin{align}
\ev{N_C(x^0)}_{A(p,x^0_I)} &= \bra{0} \alpha_A(\vec{p}) S^{-1}(x^0,x^0_I) [n_C+\bar{n}_C]
S(x^0,x^0_I) \alpha_A^\dagger(\vec{p}) \ket{0}^{\text{con}}, \nn \\
&\simeq \int \frac{d^3k}{2E_C(\vec{k})} \bra{0} \alpha_A(\vec{p})S^{-1}(x^0,x^0_I) \alpha_C^\dagger(\vec{k}) \ket{0}^{\text{con}}
\bra{0} \alpha_C(\vec{k})S(x^0,x^0_I) \alpha_A^\dagger(\vec{p}) \ket{0}^{\text{con}} \nn \\
&=\int \frac{d^3k}{2E_C(\vec{k})} \sum_{\text{phase vol}}|\bra{C(k);X}
 S(x^0,x^0_I) \ket{\alpha_A(\vec{p})}^{\text{con}}|^2. \label{2.20}
\end{align}

A similar consideration applies also to the weak interaction process, in which
a source and a target are included and the particle C scattered from the target is
observed, as shown in FIG.\ref{fig6} :
\begin{figure}[h]
 \begin{center}
  \scalebox{0.9}[0.9]{%WinTpicVersion3.08
\unitlength 0.1in
\begin{picture}(10,15)(41, -18)
% LINE 1 0 3 0
% 2 2160 940 2595 940
% 
\special{pn 13}%
\special{pa 2160 940}%
\special{pa 2596 940}%
\special{fp}%
% STR 2 0 3 1
% 3 7000 466 7000 566 5 0
% \large$2$
\put(70.0000,-5.6600){\makebox(0,0){\large$2$}}%
% LINE 1 0 3 2
% 2 6900 576 6900 1766
% 
\special{pn 13}%
\special{pa 6900 576}%
\special{pa 6900 1766}%
\special{fp}%
% LINE 1 0 3 3
% 2 5520 576 5520 1766
% 
\special{pn 13}%
\special{pa 5520 576}%
\special{pa 5520 1766}%
\special{fp}%
% STR 2 0 3 4
% 3 4625 1201 4625 1301 5 0
% \scalebox{2.5}[2.5]{$\simeq$}
\put(46.2500,-13.0100){\makebox(0,0){\scalebox{2.5}[2.5]{$\simeq$}}}%
% LINE 1 0 3 5
% 2 6255 1620 5775 1620
% 
\special{pn 13}%
\special{pa 6256 1620}%
\special{pa 5776 1620}%
\special{fp}%
% STR 2 0 3 6
% 3 5780 1741 5780 1841 5 0
% \large$(x^0)$
\put(57.8000,-18.4100){\makebox(0,0){\large$(x^0)$}}%
% STR 2 0 3 7
% 3 6135 1681 6135 1781 5 0
% \large$C$
\put(61.3500,-17.8100){\makebox(0,0){\large$C$}}%
% STR 2 0 3 8
% 3 6435 1211 6435 1311 5 0
% \large$I$
\put(64.3500,-13.1100){\makebox(0,0){\large$I$}}%
% STR 2 0 3 9
% 3 6065 611 6065 711 5 0
% \large$X$
\put(60.6500,-7.1100){\makebox(0,0){\large$X$}}%
% STR 2 0 3 10
% 3 6595 1671 6595 1771 5 0
% \large$B$
\put(65.9500,-17.7100){\makebox(0,0){\large$B$}}%
% STR 2 0 3 11
% 3 6585 721 6585 821 5 0
% \large$A$
\put(65.8500,-8.2100){\makebox(0,0){\large$A$}}%
% STR 2 0 3 12
% 3 5155 1442 5155 1542 5 0
% \scalebox{1}[1]{phase vol}
\put(51.5500,-15.4200){\makebox(0,0){\scalebox{1}[1]{phase vol}}}%
% STR 2 0 3 13
% 3 5135 1202 5135 1302 5 0
% \scalebox{2.5}[2.5]{$\sum$}
\put(51.3500,-13.0200){\makebox(0,0){\scalebox{2.5}[2.5]{$\sum$}}}%
% VECTOR 1 0 3 14
% 2 6195 1624 6050 1624
% 
\special{pn 13}%
\special{pa 6196 1624}%
\special{pa 6050 1624}%
\special{fp}%
\special{sh 1}%
\special{pa 6050 1624}%
\special{pa 6118 1644}%
\special{pa 6104 1624}%
\special{pa 6118 1604}%
\special{pa 6050 1624}%
\special{fp}%
% VECTOR 1 0 3 15
% 2 6775 1624 6630 1624
% 
\special{pn 13}%
\special{pa 6776 1624}%
\special{pa 6630 1624}%
\special{fp}%
\special{sh 1}%
\special{pa 6630 1624}%
\special{pa 6698 1644}%
\special{pa 6684 1624}%
\special{pa 6698 1604}%
\special{pa 6630 1624}%
\special{fp}%
% DOT 1 0 3 16
% 2 5753 1624 5753 1624
% 
\special{pn 13}%
\special{sh 1}%
\special{ar 5754 1624 10 10 0  6.28318530717959E+0000}%
\special{sh 1}%
\special{ar 5754 1624 10 10 0  6.28318530717959E+0000}%
% LINE 1 1 3 17
% 2 6318 950 6253 1624
% 
\special{pn 13}%
\special{pa 6318 950}%
\special{pa 6254 1624}%
\special{da 0.070}%
% DOT 1 0 3 18
% 2 6275 1624 6275 1624
% 
\special{pn 13}%
\special{sh 1}%
\special{ar 6276 1624 10 10 0  6.28318530717959E+0000}%
\special{sh 1}%
\special{ar 6276 1624 10 10 0  6.28318530717959E+0000}%
% LINE 1 0 3 19
% 2 6275 1624 6775 1624
% 
\special{pn 13}%
\special{pa 6276 1624}%
\special{pa 6776 1624}%
\special{fp}%
% VECTOR 1 0 3 20
% 2 6304 950 5888 774
% 
\special{pn 13}%
\special{pa 6304 950}%
\special{pa 5888 774}%
\special{fp}%
\special{sh 1}%
\special{pa 5888 774}%
\special{pa 5942 818}%
\special{pa 5938 796}%
\special{pa 5958 782}%
\special{pa 5888 774}%
\special{fp}%
% LINE 1 0 3 21
% 2 6014 825 5755 711
% 
\special{pn 13}%
\special{pa 6014 826}%
\special{pa 5756 712}%
\special{fp}%
% LINE 1 0 3 22
% 2 6340 950 6775 950
% 
\special{pn 13}%
\special{pa 6340 950}%
\special{pa 6776 950}%
\special{fp}%
% DOT 1 0 3 23
% 2 6318 950 6318 950
% 
\special{pn 13}%
\special{sh 1}%
\special{ar 6318 950 10 10 0  6.28318530717959E+0000}%
\special{sh 1}%
\special{ar 6318 950 10 10 0  6.28318530717959E+0000}%
% VECTOR 1 0 3 24
% 2 6635 951 6490 951
% 
\special{pn 13}%
\special{pa 6636 952}%
\special{pa 6490 952}%
\special{fp}%
\special{sh 1}%
\special{pa 6490 952}%
\special{pa 6558 972}%
\special{pa 6544 952}%
\special{pa 6558 932}%
\special{pa 6490 952}%
\special{fp}%
% STR 2 0 3 25
% 3 2780 1671 2780 1771 5 0
% \large$C$
\put(27.8000,-17.7100){\makebox(0,0){\large$C$}}%
% STR 2 0 3 26
% 3 2250 1661 2250 1761 5 0
% \large$B$
\put(22.5000,-17.6100){\makebox(0,0){\large$B$}}%
% STR 2 0 3 27
% 3 2450 1201 2450 1301 5 0
% \large$I$
\put(24.5000,-13.0100){\makebox(0,0){\large$I$}}%
% STR 2 0 3 28
% 3 2870 601 2870 701 5 0
% \large$X$
\put(28.7000,-7.0100){\makebox(0,0){\large$X$}}%
% STR 2 0 3 29
% 3 2310 711 2310 811 5 0
% \large$A$
\put(23.1000,-8.1100){\makebox(0,0){\large$A$}}%
% STR 2 0 3 30
% 3 3170 1731 3170 1831 5 0
% \large$N_C(x^0)$
\put(31.7000,-18.3100){\makebox(0,0){\large$N_C(x^0)$}}%
% STR 2 0 3 31
% 3 3550 1671 3550 1771 5 0
% \large$C$
\put(35.5000,-17.7100){\makebox(0,0){\large$C$}}%
% STR 2 0 3 32
% 3 3850 1201 3850 1301 5 0
% \large$I$
\put(38.5000,-13.0100){\makebox(0,0){\large$I$}}%
% STR 2 0 3 33
% 3 3480 601 3480 701 5 0
% \large$X$
\put(34.8000,-7.0100){\makebox(0,0){\large$X$}}%
% STR 2 0 3 34
% 3 4010 1661 4010 1761 5 0
% \large$B$
\put(40.1000,-17.6100){\makebox(0,0){\large$B$}}%
% STR 2 0 3 35
% 3 4000 711 4000 811 5 0
% \large$A$
\put(40.0000,-8.1100){\makebox(0,0){\large$A$}}%
% STR 2 0 3 36
% 3 1800 1431 1800 1531 5 0
% \scalebox{1}[1]{phase vol}
\put(18.0000,-15.3100){\makebox(0,0){\scalebox{1}[1]{phase vol}}}%
% STR 2 0 3 37
% 3 1780 1191 1780 1291 5 0
% \scalebox{2.5}[2.5]{$\sum$}
\put(17.8000,-12.9100){\makebox(0,0){\scalebox{2.5}[2.5]{$\sum$}}}%
% VECTOR 1 0 3 38
% 2 3610 1614 3465 1614
% 
\special{pn 13}%
\special{pa 3610 1614}%
\special{pa 3466 1614}%
\special{fp}%
\special{sh 1}%
\special{pa 3466 1614}%
\special{pa 3532 1634}%
\special{pa 3518 1614}%
\special{pa 3532 1594}%
\special{pa 3466 1614}%
\special{fp}%
% VECTOR 1 0 3 39
% 2 4190 1614 4045 1614
% 
\special{pn 13}%
\special{pa 4190 1614}%
\special{pa 4046 1614}%
\special{fp}%
\special{sh 1}%
\special{pa 4046 1614}%
\special{pa 4112 1634}%
\special{pa 4098 1614}%
\special{pa 4112 1594}%
\special{pa 4046 1614}%
\special{fp}%
% DOT 1 0 3 40
% 2 3168 1614 3168 1614
% 
\special{pn 13}%
\special{sh 1}%
\special{ar 3168 1614 10 10 0  6.28318530717959E+0000}%
\special{sh 1}%
\special{ar 3168 1614 10 10 0  6.28318530717959E+0000}%
% LINE 1 0 3 41
% 2 3661 1614 2696 1614
% 
\special{pn 13}%
\special{pa 3662 1614}%
\special{pa 2696 1614}%
\special{fp}%
% LINE 1 1 3 42
% 2 3733 940 3668 1614
% 
\special{pn 13}%
\special{pa 3734 940}%
\special{pa 3668 1614}%
\special{da 0.070}%
% LINE 1 1 3 43
% 2 2675 1614 2595 940
% 
\special{pn 13}%
\special{pa 2676 1614}%
\special{pa 2596 940}%
\special{da 0.070}%
% DOT 1 0 3 44
% 2 2675 1614 2675 1614
% 
\special{pn 13}%
\special{sh 1}%
\special{ar 2676 1614 10 10 0  6.28318530717959E+0000}%
\special{sh 1}%
\special{ar 2676 1614 10 10 0  6.28318530717959E+0000}%
% DOT 1 0 3 45
% 2 3690 1614 3690 1614
% 
\special{pn 13}%
\special{sh 1}%
\special{ar 3690 1614 10 10 0  6.28318530717959E+0000}%
\special{sh 1}%
\special{ar 3690 1614 10 10 0  6.28318530717959E+0000}%
% LINE 1 0 3 46
% 2 3690 1614 4190 1614
% 
\special{pn 13}%
\special{pa 3690 1614}%
\special{pa 4190 1614}%
\special{fp}%
% LINE 1 0 3 47
% 2 2160 1614 2660 1614
% 
\special{pn 13}%
\special{pa 2160 1614}%
\special{pa 2660 1614}%
\special{fp}%
% DOT 1 0 3 48
% 2 2595 940 2595 940
% 
\special{pn 13}%
\special{sh 1}%
\special{ar 2596 940 10 10 0  6.28318530717959E+0000}%
\special{sh 1}%
\special{ar 2596 940 10 10 0  6.28318530717959E+0000}%
% VECTOR 1 0 3 49
% 2 3719 940 3303 764
% 
\special{pn 13}%
\special{pa 3720 940}%
\special{pa 3304 764}%
\special{fp}%
\special{sh 1}%
\special{pa 3304 764}%
\special{pa 3358 808}%
\special{pa 3352 786}%
\special{pa 3372 772}%
\special{pa 3304 764}%
\special{fp}%
% LINE 1 0 3 50
% 2 3429 815 3170 701
% 
\special{pn 13}%
\special{pa 3430 816}%
\special{pa 3170 702}%
\special{fp}%
% LINE 1 0 3 51
% 2 3755 940 4190 940
% 
\special{pn 13}%
\special{pa 3756 940}%
\special{pa 4190 940}%
\special{fp}%
% DOT 1 0 3 52
% 2 3733 940 3733 940
% 
\special{pn 13}%
\special{sh 1}%
\special{ar 3734 940 10 10 0  6.28318530717959E+0000}%
\special{sh 1}%
\special{ar 3734 940 10 10 0  6.28318530717959E+0000}%
% VECTOR 1 0 3 53
% 2 4050 941 3905 941
% 
\special{pn 13}%
\special{pa 4050 942}%
\special{pa 3906 942}%
\special{fp}%
\special{sh 1}%
\special{pa 3906 942}%
\special{pa 3972 962}%
\special{pa 3958 942}%
\special{pa 3972 922}%
\special{pa 3906 942}%
\special{fp}%
% LINE 1 0 3 54
% 2 2595 940 3153 712
% 
\special{pn 13}%
\special{pa 2596 940}%
\special{pa 3154 712}%
\special{fp}%
\end{picture}%
}
 \end{center}
\caption{Relation of the expectation value of number operator to the
 transition probability. Both include production and detection processes. }
\label{fig6}
\end{figure}

In the case where X is one D-particle state,  RHS of FIG.\ref{fig6}  corresponds to
the process examined by Yabuki and Ishikawa \cite{r11} when the mixing exists in the
intermediate state.

\subsection{The case with mixing}
Along the aim of the present paper, it is important for us to consider the case
where there exists a "flavor" degree of freedom with mixing among (pseudo-)scalar
fields $\phi_{C\rho}(x)$, $\rho= 1,2,\cdots$. The relevant Lagrangian is
\begin{align}
\mathcal{L}(x)&=\mathcal{L}_0(\text{$\phi_A$- and $\phi_B$-fields})
-(\partial^b \phi^\dagger_C(x)\cdot \partial_b\phi_C(x)-\phi^\dagger_C(x)M^2\phi_C(x)) \nn \\
& +\sum_\rho \left\{ f_{ABC\rho} \phi_{C\rho}^\dagger(x) \phi_B(x) \phi_A(x)+H.c. \right\}, \label{2.21}\\
\phi_C(x)&=
 \begin{pmatrix}
  \phi_e(x) \\
  \phi_\mu(x) \\
  \vdots
 \end{pmatrix}, \qquad
M^2=
 \begin{pmatrix}
  m^2_{ee} & m^2_{e\mu} & \cdots \\
  m^2_{\mu e} & m^2_{\mu \mu} &\cdots \\
  \vdots & \vdots & \ddots
 \end{pmatrix}, \qquad
(M^2)^\dagger=M^2. \label{2.22}
\end{align}
Similarly to (\ref{1.11})-(\ref{1.13}), we have
\begin{align}
&{Z^{1/2}}^\dagger M^2 Z^{1/2}=M^2_{\text{diag}}=
 \begin{pmatrix}
  m^2_1 & & \\
  & m^2_2 & \\
  & & \ddots 
 \end{pmatrix}, \label{2.23} \\
&\phi_C(x)=Z^{1/2}\phi_C(x)^{(M)}, \qquad
\phi_C^{(M)}(x)=
 \begin{pmatrix}
  \phi_{C1}(x) \\
  \phi_{C2}(x) \\
  \vdots
 \end{pmatrix}, \qquad
{Z^{1/2}}^\dagger Z^{1/2} =I. \label{2.24}
\end{align}
We assume $m^2_j >0$ for all $j=1,2,\cdots$, and now consider the interaction representation, 
where $\phi_g$-fields, $g = A, B, C_j$'s , satisfy the
free equations of motion with masses $m_g$'s, and Fock space is constructed on
the vacuum $\ket{0}$,  defined in the same way as (\ref{2.5}).

We examine the structures of the expectation value of $N_{C\rho}(x^0)$ defined by
\begin{align}
N_{C\rho}(x^0) \equiv i:\int\!\!d^3x j_4^{(C\rho)}(x): \label{2.25}
\intertext{with}
j_b^{(C\rho)}(x) = i\phi^\dagger_{C\rho}(x)\olra{\partial}_b \phi_{C\rho}(x),\
 \quad \rho=e,\mu,\cdots ; \label{2.26}
\end{align}
here, needless to explain, the symbol $: \ :$  represents the normal ordering
with respect to \\ $\{\alpha_{Cj}(\vec{k}),\beta_{Cj}(\vec{k}),\ j=1,2,\cdots \}$ and their Hermitean
conjugates. Concretely we have
\begin{align}
N_{C\rho}(x^0)&
%i^2\sum_{j,l} {Z^{1/2}_{\rho j}}^\ast Z^{1/2}_{\rho l} \int\!\!d^3x\int\!\!dk^3\int\!\!d^3q
% \frac{1}{(2\pi)^3 2E_{Cj}(\vec{k})2E_{Cl}(\vec{q})} \nn\\
%&\times :[ (\alpha_{Cj}^\dagger(\vec{k})e^{-ikx}+\beta_{Cj}(\vec{k})e^{ikx})E_{Cl}(\vec{q})
% (-\alpha_{Cl}(\vec{q})e^{iqx}+\beta_{Cj}^\dagger(\vec{q})e^{-iqx}) \nn \\
%&\ -E_{Cl}(\vec{k})(\alpha_{Cj}^\dagger(\vec{k})e^{-ikx}-\beta^\dagger_{Cj}(\vec{k})e^{ikx})
% (\alpha_{Cl}(\vec{q})e^{iqx}+\beta_{Cl}^\dagger(\vec{q})e^{-iqx})]: \nn \\
%&
=\sum_{j,l}{Z^{1/2}_{\rho j}}^\ast Z^{1/2}_{\rho l}\int d^3k \frac{1}{4}\left[\left(\frac{1}{E_{Cj}(\vec{k})}
 +\frac{1}{E_{Cl}(\vec{k})}\right) \right. \nn \\
&\times \left[\alpha^\dagger_{Cj}(\vec{k})\alpha_{Cl}(\vec{k})e^{i(E_{Cj}(\vec{k})-E_{Cl}(\vec{k}))x^0}
 -\beta_{Cl}^\dagger(\vec{k})\beta_{Cj}(\vec{k})e^{i(E_{Cl}(\vec{k})-E_{Cj}(\vec{k}))x^0} \right] \nn \\
&+\left(-\frac{1}{E_{Cj}(\vec{k})}+\frac{1}{E_{Cl}(\vec{k})}\right) 
 \alpha^\dagger_{Cj}(\vec{k})\beta^\dagger_{Cl}(-\vec{k})e^{i(E_{Cj}(\vec{k})+E_{Cl}(\vec{k}))x^0} \nn \\
&\left. +\left(\frac{1}{E_{Cj}(\vec{k})}-\frac{1}{E_{Cl}(\vec{k})}\right) 
 \beta_{Cj}(-\vec{k})\alpha_{Cl}(\vec{k})e^{-i(E_{Cj}(\vec{k})+E_{Cl}(\vec{k}))x^0}\right]; \label{2.27}
\end{align}
here, $\beta_{Cj}(\vec{q})$ with $\{\vec{q}=-\vec{k}, q_0=E_{Cj}(\vec{k})\}$ is written
simply as $\beta_{Cj}(-\vec{k})$.

\begin{figure}[h]
 \begin{center}
  \scalebox{0.9}[0.9]{%WinTpicVersion3.08
\unitlength 0.1in
\begin{picture}( 77.8500, 14.2000)(1,-19.5000)
% ELLIPSE 1 0 3 0
% 4 2610 1215 3430 1655 3430 1655 3660 1785
% 
\special{pn 13}%
\special{ar 2610 1216 820 440  0.7910799 6.2831853}%
\special{ar 2610 1216 820 440  0.0000000 0.7853982}%
% VECTOR 1 0 3 1
% 2 6850 830 6650 880
% 
\special{pn 13}%
\special{pa 6850 830}%
\special{pa 6650 880}%
\special{fp}%
\special{sh 1}%
\special{pa 6650 880}%
\special{pa 6720 884}%
\special{pa 6702 868}%
\special{pa 6710 844}%
\special{pa 6650 880}%
\special{fp}%
% STR 2 0 3 2
% 3 5510 1555 5510 1655 5 0
% \large$\bar{C}_l$
\put(55.1000,-16.5500){\makebox(0,0){\large$\bar{C}_l$}}%
% STR 2 0 3 3
% 3 6930 1535 6930 1635 5 0
% \large$\bar{C}_j$
\put(69.3000,-16.3500){\makebox(0,0){\large$\bar{C}_j$}}%
% STR 2 0 3 4
% 3 6660 1225 6660 1325 5 0
% \large$B$
\put(66.6000,-13.2500){\makebox(0,0){\large$B$}}%
% STR 2 0 3 5
% 3 6780 625 6780 725 5 0
% \large$A$
\put(67.8000,-7.2500){\makebox(0,0){\large$A$}}%
% STR 2 0 3 6
% 3 5030 1115 5030 1215 5 0
% {\large (2)}
\put(50.3000,-12.1500){\makebox(0,0){{\large (2)}}}%
% VECTOR 1 0 3 7
% 2 6970 1375 6960 1385
% 
\special{pn 13}%
\special{pa 6970 1376}%
\special{pa 6960 1386}%
\special{fp}%
\special{sh 1}%
\special{pa 6960 1386}%
\special{pa 7022 1352}%
\special{pa 6998 1348}%
\special{pa 6994 1324}%
\special{pa 6960 1386}%
\special{fp}%
% DOT 1 0 3 8
% 2 6210 1655 6210 1665
% 
\special{pn 13}%
\special{sh 1}%
\special{ar 6210 1656 10 10 0  6.28318530717959E+0000}%
\special{sh 1}%
\special{ar 6210 1666 10 10 0  6.28318530717959E+0000}%
% ELLIPSE 1 0 3 9
% 4 6220 1205 7040 1665 5310 1205 7240 1205
% 
\special{pn 13}%
\special{ar 6220 1206 820 460  6.2831853 6.2831853}%
\special{ar 6220 1206 820 460  0.0000000 3.1415927}%
% LINE 1 0 3 10
% 2 7040 1195 5390 1195
% 
\special{pn 13}%
\special{pa 7040 1196}%
\special{pa 5390 1196}%
\special{fp}%
% VECTOR 1 0 3 11
% 2 6680 1195 6480 1195
% 
\special{pn 13}%
\special{pa 6680 1196}%
\special{pa 6480 1196}%
\special{fp}%
\special{sh 1}%
\special{pa 6480 1196}%
\special{pa 6548 1216}%
\special{pa 6534 1196}%
\special{pa 6548 1176}%
\special{pa 6480 1196}%
\special{fp}%
% LINE 1 0 3 12
% 2 5400 1185 7000 785
% 
\special{pn 13}%
\special{pa 5400 1186}%
\special{pa 7000 786}%
\special{fp}%
% LINE 1 0 3 13
% 2 7050 1175 6320 1005
% 
\special{pn 13}%
\special{pa 7050 1176}%
\special{pa 6320 1006}%
\special{fp}%
% LINE 1 0 3 14
% 2 6120 955 5400 785
% 
\special{pn 13}%
\special{pa 6120 956}%
\special{pa 5400 786}%
\special{fp}%
% STR 2 0 3 15
% 3 5580 625 5580 725 5 0
% \large$A$
\put(55.8000,-7.2500){\makebox(0,0){\large$A$}}%
% STR 2 0 3 16
% 3 6210 1765 6210 1865 5 0
% \large$(\bar{C}_l^\dagger \bar{C}_j)$
\put(62.1000,-18.6500){\makebox(0,0){\large$(\bar{C}_l^\dagger \bar{C}_j)$}}%
% STR 2 0 3 17
% 3 1280 985 1280 1085 5 0
% \large$A$
\put(12.8000,-10.8500){\makebox(0,0){\large$A$}}%
% STR 2 0 3 18
% 3 2610 1765 2610 1865 5 0
% \large$(C_j^\dagger C_l)$
\put(26.1000,-18.6500){\makebox(0,0){\large$(C_j^\dagger C_l)$}}%
% STR 2 0 3 19
% 3 1910 1555 1910 1655 5 0
% \large$C_j$
\put(19.1000,-16.5500){\makebox(0,0){\large$C_j$}}%
% STR 2 0 3 20
% 3 3330 1535 3330 1635 5 0
% \large$C_l$
\put(33.3000,-16.3500){\makebox(0,0){\large$C_l$}}%
% STR 2 0 3 21
% 3 2610 515 2610 615 5 0
% \large$\bar{B}$
\put(26.1000,-6.1500){\makebox(0,0){\large$\bar{B}$}}%
% STR 2 0 3 22
% 3 3910 985 3910 1085 5 0
% \large$A$
\put(39.1000,-10.8500){\makebox(0,0){\large$A$}}%
% STR 2 0 3 23
% 3 630 1115 630 1215 5 0
% {\large (1)}
\put(6.3000,-12.1500){\makebox(0,0){{\large (1)}}}%
% VECTOR 1 0 3 24
% 2 3310 1445 3300 1455
% 
\special{pn 13}%
\special{pa 3310 1446}%
\special{pa 3300 1456}%
\special{fp}%
\special{sh 1}%
\special{pa 3300 1456}%
\special{pa 3362 1422}%
\special{pa 3338 1418}%
\special{pa 3334 1394}%
\special{pa 3300 1456}%
\special{fp}%
% VECTOR 1 0 3 25
% 4 3340 1015 3330 1005 3350 1015 3350 1015
% 
\special{pn 13}%
\special{pa 3340 1016}%
\special{pa 3330 1006}%
\special{fp}%
\special{sh 1}%
\special{pa 3330 1006}%
\special{pa 3364 1066}%
\special{pa 3368 1044}%
\special{pa 3392 1038}%
\special{pa 3330 1006}%
\special{fp}%
\special{pa 3350 1016}%
\special{pa 3350 1016}%
\special{fp}%
% DOT 1 0 3 26
% 2 2610 1655 2610 1665
% 
\special{pn 13}%
\special{sh 1}%
\special{ar 2610 1656 10 10 0  6.28318530717959E+0000}%
\special{sh 1}%
\special{ar 2610 1666 10 10 0  6.28318530717959E+0000}%
% VECTOR 1 0 3 27
% 2 4010 1215 3810 1215
% 
\special{pn 13}%
\special{pa 4010 1216}%
\special{pa 3810 1216}%
\special{fp}%
\special{sh 1}%
\special{pa 3810 1216}%
\special{pa 3878 1236}%
\special{pa 3864 1216}%
\special{pa 3878 1196}%
\special{pa 3810 1216}%
\special{fp}%
% LINE 1 0 3 28
% 2 3410 1215 4190 1215
% 
\special{pn 13}%
\special{pa 3410 1216}%
\special{pa 4190 1216}%
\special{fp}%
% LINE 1 0 3 29
% 2 1010 1215 1790 1215
% 
\special{pn 13}%
\special{pa 1010 1216}%
\special{pa 1790 1216}%
\special{fp}%
% VECTOR 1 0 3 30
% 2 5790 880 5600 830
% 
\special{pn 13}%
\special{pa 5790 880}%
\special{pa 5600 830}%
\special{fp}%
\special{sh 1}%
\special{pa 5600 830}%
\special{pa 5660 866}%
\special{pa 5652 844}%
\special{pa 5670 828}%
\special{pa 5600 830}%
\special{fp}%
\end{picture}%
}
 \end{center}
\caption{Diagram representing \eqref{2.35} and \eqref{2.36}, respectively}
\label{fig7}
\end{figure}
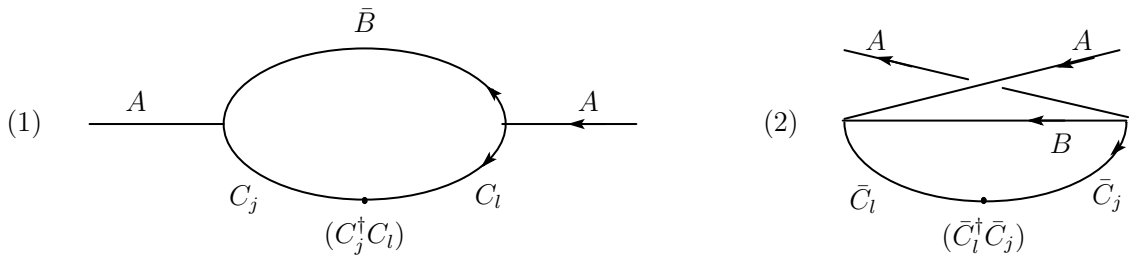

Similarly to (\ref{2.8}), we write the expectation value of $j_b^{(C\rho)}(x)$ as
\begin{align}
\mathcal{E}(A(x^0_I); C_\rho\text{-cur}(x))_b \equiv \bra{0}\alpha_A(\vec{p}) S(x^0,x^0_I)^{-1}
 :j_b^{(C\rho)}(x):S(x^0,x^0_I)
 \alpha^\dagger_A(\vec{p})\ket{0}^{\text{con}}, \label{2.28}
\end{align}
which leads to
\begin{align}
\langle N_{C\rho}(x^0) \rangle_{A(p,x^0_I)} \equiv i\int d^3x
 \mathcal{E}(A(x^0_I); C_\rho\text{-cur}(x))_4. \label{2.29}
\end{align}
As we see from (\ref{2.27}), $N_{C\rho}(x^0)$ consists of 4 terms, which we
write ${N_{C\rho}(x^0)}^{(n)} , n = 1, 2, 3, 4,$ in order;
\begin{align}
{N_{C\rho}(x^0)}^{(1)}&=\sum_{j,l} {Z^{1/2}_{\rho j}}^\ast Z^{1/2}_{\rho l} \int\!\!d^3k
 \frac{1}{4}\left[\frac{1}{2E_{Cj}(\vec{k})}+\frac{1}{2E_{Cl}(\vec{q})}\right] \nn \\
&\times \alpha^\dagger_{Cj}(\vec{k})\alpha_{Cl}(\vec{k})e^{i(E_{Cj}(\vec{k})-E_{Cl}(\vec{k}))x^0}, \label{2.30}\\
{N_{C\rho}(x^0)}^{(3)}&=\left[ {N_{C\rho}(x^0)}^{(4)}\right]^\dagger.\label{2.31}
\end{align}
In the lowest (i.e. the second ) order contribution with respect to the weak
interaction in (\ref{2.21}),  we have for $n = 1, 2$,
\begin{align}
\langle {N_{C\rho}(x^0)}^{(n)} \rangle_{A(p,x^0_I)}&=\bra{0}\alpha_A(\vec{p})
 \int^{x^0}_{x^0_I}\!d^4z H_{\text{int}}(z){N_{C\rho}(x^0)}^{(n)}
 \int^{x^0}_{x^0_I}\!d^4y H_{\text{int}}(y) 
 \alpha^\dagger_A(\vec{p})\ket{0}^{\text{con}}, \label{2.32} \\
\langle {N_{C\rho}(x^0)}^{(3+4)} \rangle_{A(p,x^0_I)}&=i^2\bra{0}\alpha_A(\vec{p})
 \int^{x^0}_{x^0_I}\!\!\int^{x^0}_{x^0_I}\!d^4zd^4y
 [H_{\text{int}}(z)H_{\text{int}}(y){N_{C\rho}(x^0)}^{(3)} \nn\\
&+{N_{C\rho}(x^0)}^{(4)}H_{\text{int}}(y)H_{\text{int}}(z)]
 \alpha^\dagger_A(\vec{p})\ket{0}^{\text{con}}. \label{2.33}
\end{align}
In the following, we assume only one flavor $\sigma$ is dominant in $H_{\text{int}}(x)$,
similarly to $\pi^+$ decay (i.e. $\pi^+ \rightarrow \mu^+ +\nu_\mu$) , so that we take
\begin{align}
H_{\text{int}}(x)=-f_{ABC\sigma}\phi^\dagger_{C\sigma}(x)\phi_B(x)\phi_A(x)+H.c.
 .\label{2.34}
\end{align}
Then we obtain, as the respective contributions from the diagrams FIG's.\ref{fig7}
(1) and (2),
\begin{align}
\langle {N_{C\rho}(x^0)}^{(1)} \rangle_{A(p,x^0_I)}&=V|f_{ABC\sigma}|^2\sum_{j,l} 
 {Z^{1/2}_{\rho j}}^\ast Z^{1/2}_{\rho l}Z^{1/2}_{\sigma j}{Z^{1/2}_{\sigma j}}^\ast
 \int^t_0\!\int^t_0 dz^0 dy^0 \int\frac{d^3k}{(2\pi)^6} \nn \\
&\times e^{i(E_{Cj}(\vec{k})-E_{Cl}(\vec{k}))t}
 \frac{1}{4}\left[ \frac{1}{E_{Cj}(\vec{k})}+\frac{1}{E_{Cl}(\vec{k})} \right]\frac{1}{2E_B(\vec{q})} \nn \\ 
&\times e^{i( E_A(\vec{p})-E_B(\vec{q})-E_{Cj}(\vec{k}))z^0} 
 e^{i(-E_A(\vec{p})+E_B(\vec{q})+E_{Cl}(\vec{k}))y^0} \bigg|_{\vec{q}=\vec{p}-\vec{k}}, \label{2.35} \\
\langle {N_{C\rho}(x^0)}^{(2)} \rangle_{A(p,x^0_I)}&=-V|f_{ABC\sigma}|^2\sum_{j,l} 
 {Z^{1/2}_{\rho j}}^\ast Z^{1/2}_{\rho l} Z^{1/2}_{\sigma j} {Z^{1/2}_{\sigma j}}^\ast
 \int^t_0\!\int^t_0 dz^0 dy^0 \int\frac{d^3k}{(2\pi)^6} \nn \\
&\times e^{i(E_{Cl}(\vec{k})-E_{Cj}(\vec{k}))t}
 \frac{1}{4}\left[ \frac{1}{E_{Cj}(\vec{k})}+\frac{1}{E_{Cl}(\vec{k})} \right]\frac{1}{2E_B(\vec{q})} \nn \\
&\times e^{-i( E_A(\vec{p})+E_B(\vec{q})+E_{Cj}(\vec{k}))z^0} 
 e^{i(E_A(\vec{p})+E_B(\vec{q})+E_{Cl}(\vec{k}))y^0} \bigg|_{\vec{q}=-\vec{p}-\vec{k}} ; \label{2.36}
\end{align}
here, $t = x^0-x^0_I$. 

Due to the same reason as explained in the preceding subsection A, 
$\langle {N_{C\rho}(x^0)}^{(1)} \rangle_{A(p,x^0_I)}$ contributes dominantly over
$\langle {N_{C\rho}(x^0)}^{(2)} \rangle_{A(p,x^0_I)}$.

As to $\langle {N_{C\rho}(x^0)}^{(3)} \rangle_{A(p,x^0_I)}$, the daigrams represented by
FIG.\ref{fig8} contribute to it, the form of which is concretely given as
\begin{align}
\langle {N_{C\rho}(x^0)}^{(3)} \rangle_{A(p,x^0_I)} &= V|f_{ABC\sigma}|^2
 \sum_{j,l}{Z^{1/2}_{\rho j}}^\ast Z^{1/2}_{\rho l}Z^{1/2}_{\sigma j}{Z^{1/2}_{\sigma j}}^\ast
 \int^t_0\!dz^0\!\int^{z^0}_0\!dy^0 \nn\\
&\times \int\frac{d^3k}{(2\pi)^6} \frac{1}{4}
 \left[\frac{1}{E_{Cj}(\vec{k})}-\frac{1}{E_{Cl}(\vec{k})}\right]
 \frac{1}{2E_B(\vec{q})}e^{i(E_{Cj}(\vec{k})+E_{Cl}(\vec{k}))t}\nn \\
&\times \left[ e^{i( E_A(\vec{p})-E_B(\vec{q})-E_{Cj}(\vec{k}))z^0}
 e^{i( E_A(\vec{p})+E_B(\vec{q})-E_{Cj}(\vec{k}))y^0}\bigg|_{\vec{q}=\vec{p}-\vec{k}} \right. \nn \\
&\ +e^{i(-E_A(\vec{p})-E_B(\vec{q})-E_{Cl}(\vec{k}))z^0}
  e^{i(E_A(\vec{p})+E_B(\vec{q})-E_{Cj}(\vec{k}))y^0}\bigg|_{\vec{q}=-\vec{p}+\vec{k}}
  \left. \right] \nn \\
&=\left[ \langle {N_{C\rho}(x^0)}^{(4)} \rangle_{A(p,x^0_I)} \right]^\ast . \label{2.37}
\end{align}
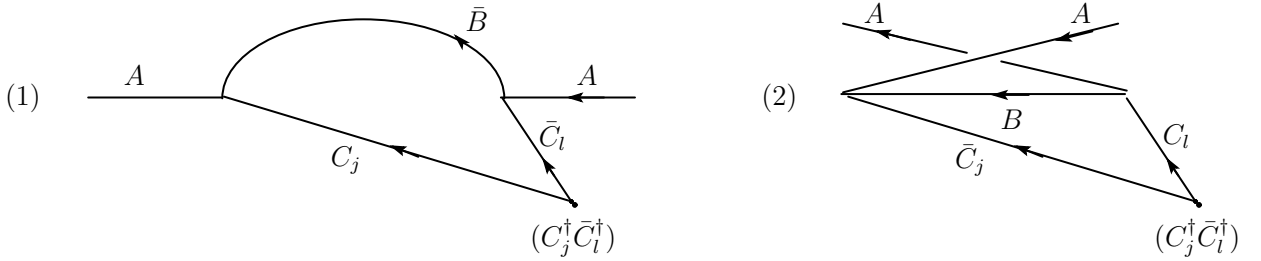
\begin{figure}[h]
 \begin{center}
  \scalebox{0.9}[0.9]{%WinTpicVersion3.08
\unitlength 0.1in
\begin{picture}( 91.2500, 14.7500)(1.5,-21.1500)
% VECTOR 1 0 3 0
% 2 5790 880 5600 830
% 
\special{pn 13}%
\special{pa 5790 880}%
\special{pa 5600 830}%
\special{fp}%
\special{sh 1}%
\special{pa 5600 830}%
\special{pa 5660 866}%
\special{pa 5652 844}%
\special{pa 5670 828}%
\special{pa 5600 830}%
\special{fp}%
% VECTOR 1 0 3 1
% 2 6560 1560 6410 1500
% 
\special{pn 13}%
\special{pa 6560 1560}%
\special{pa 6410 1500}%
\special{fp}%
\special{sh 1}%
\special{pa 6410 1500}%
\special{pa 6464 1544}%
\special{pa 6460 1520}%
\special{pa 6480 1506}%
\special{pa 6410 1500}%
\special{fp}%
% DOT 1 0 3 2
% 2 7470 1840 7470 1840
% 
\special{pn 13}%
\special{sh 1}%
\special{ar 7470 1840 10 10 0  6.28318530717959E+0000}%
\special{sh 1}%
\special{ar 7470 1840 10 10 0  6.28318530717959E+0000}%
% VECTOR 1 0 3 3
% 2 7350 1670 7290 1580
% 
\special{pn 13}%
\special{pa 7350 1670}%
\special{pa 7290 1580}%
\special{fp}%
\special{sh 1}%
\special{pa 7290 1580}%
\special{pa 7310 1648}%
\special{pa 7320 1624}%
\special{pa 7344 1624}%
\special{pa 7290 1580}%
\special{fp}%
% DOT 2 0 3 4
% 3 7450 1820 7450 1820 7450 1820
% 
\special{pn 8}%
\special{sh 1}%
\special{ar 7450 1820 10 10 0  6.28318530717959E+0000}%
\special{sh 1}%
\special{ar 7450 1820 10 10 0  6.28318530717959E+0000}%
\special{sh 1}%
\special{ar 7450 1820 10 10 0  6.28318530717959E+0000}%
% LINE 1 0 3 5
% 4 7050 1220 7450 1820 7450 1820 5430 1210
% 
\special{pn 13}%
\special{pa 7050 1220}%
\special{pa 7450 1820}%
\special{fp}%
\special{pa 7450 1820}%
\special{pa 5430 1210}%
\special{fp}%
% STR 2 0 3 6
% 3 7460 1930 7460 2030 5 0
% \large$(C_j^\dagger \bar{C}_l^\dagger)$
\put(74.6000,-20.3000){\makebox(0,0){\large$(C_j^\dagger \bar{C}_l^\dagger)$}}%
% STR 2 0 3 7
% 3 6140 1480 6140 1580 5 0
% \large$\bar{C}_j$
\put(61.4000,-15.8000){\makebox(0,0){\large$\bar{C}_j$}}%
% STR 2 0 3 8
% 3 7340 1340 7340 1440 5 0
% \large$C_l$
\put(73.4000,-14.4000){\makebox(0,0){\large$C_l$}}%
% VECTOR 1 0 3 9
% 2 2930 1560 2780 1500
% 
\special{pn 13}%
\special{pa 2930 1560}%
\special{pa 2780 1500}%
\special{fp}%
\special{sh 1}%
\special{pa 2780 1500}%
\special{pa 2834 1544}%
\special{pa 2830 1520}%
\special{pa 2850 1506}%
\special{pa 2780 1500}%
\special{fp}%
% DOT 1 0 3 10
% 2 3840 1840 3840 1840
% 
\special{pn 13}%
\special{sh 1}%
\special{ar 3840 1840 10 10 0  6.28318530717959E+0000}%
\special{sh 1}%
\special{ar 3840 1840 10 10 0  6.28318530717959E+0000}%
% VECTOR 1 0 3 11
% 2 3720 1670 3660 1580
% 
\special{pn 13}%
\special{pa 3720 1670}%
\special{pa 3660 1580}%
\special{fp}%
\special{sh 1}%
\special{pa 3660 1580}%
\special{pa 3680 1648}%
\special{pa 3690 1624}%
\special{pa 3714 1624}%
\special{pa 3660 1580}%
\special{fp}%
% DOT 2 0 3 12
% 3 3820 1820 3820 1820 3820 1820
% 
\special{pn 8}%
\special{sh 1}%
\special{ar 3820 1820 10 10 0  6.28318530717959E+0000}%
\special{sh 1}%
\special{ar 3820 1820 10 10 0  6.28318530717959E+0000}%
\special{sh 1}%
\special{ar 3820 1820 10 10 0  6.28318530717959E+0000}%
% LINE 1 0 3 13
% 4 3420 1220 3820 1820 3820 1820 1800 1210
% 
\special{pn 13}%
\special{pa 3420 1220}%
\special{pa 3820 1820}%
\special{fp}%
\special{pa 3820 1820}%
\special{pa 1800 1210}%
\special{fp}%
% VECTOR 1 0 3 14
% 2 3160 870 3150 860
% 
\special{pn 13}%
\special{pa 3160 870}%
\special{pa 3150 860}%
\special{fp}%
\special{sh 1}%
\special{pa 3150 860}%
\special{pa 3184 922}%
\special{pa 3188 898}%
\special{pa 3212 894}%
\special{pa 3150 860}%
\special{fp}%
% ELLIPSE 1 0 3 15
% 4 2610 1220 3430 760 3630 1220 1700 1220
% 
\special{pn 13}%
\special{ar 2610 1220 820 460  3.1415927 6.2831853}%
% VECTOR 1 0 3 16
% 2 6850 830 6650 880
% 
\special{pn 13}%
\special{pa 6850 830}%
\special{pa 6650 880}%
\special{fp}%
\special{sh 1}%
\special{pa 6650 880}%
\special{pa 6720 884}%
\special{pa 6702 868}%
\special{pa 6710 844}%
\special{pa 6650 880}%
\special{fp}%
% STR 2 0 3 17
% 3 6380 1240 6380 1340 5 0
% \large$B$
\put(63.8000,-13.4000){\makebox(0,0){\large$B$}}%
% STR 2 0 3 18
% 3 6780 625 6780 725 5 0
% \large$A$
\put(67.8000,-7.2500){\makebox(0,0){\large$A$}}%
% STR 2 0 3 19
% 3 5030 1115 5030 1215 5 0
% {\large (2)}
\put(50.3000,-12.1500){\makebox(0,0){{\large (2)}}}%
% LINE 1 0 3 20
% 2 7040 1195 5390 1195
% 
\special{pn 13}%
\special{pa 7040 1196}%
\special{pa 5390 1196}%
\special{fp}%
% VECTOR 1 0 3 21
% 2 6480 1200 6280 1200
% 
\special{pn 13}%
\special{pa 6480 1200}%
\special{pa 6280 1200}%
\special{fp}%
\special{sh 1}%
\special{pa 6280 1200}%
\special{pa 6348 1220}%
\special{pa 6334 1200}%
\special{pa 6348 1180}%
\special{pa 6280 1200}%
\special{fp}%
% LINE 1 0 3 22
% 2 5400 1185 7000 785
% 
\special{pn 13}%
\special{pa 5400 1186}%
\special{pa 7000 786}%
\special{fp}%
% LINE 1 0 3 23
% 2 7050 1175 6320 1005
% 
\special{pn 13}%
\special{pa 7050 1176}%
\special{pa 6320 1006}%
\special{fp}%
% LINE 1 0 3 24
% 2 6120 955 5400 785
% 
\special{pn 13}%
\special{pa 6120 956}%
\special{pa 5400 786}%
\special{fp}%
% STR 2 0 3 25
% 3 5580 625 5580 725 5 0
% \large$A$
\put(55.8000,-7.2500){\makebox(0,0){\large$A$}}%
% STR 2 0 3 26
% 3 1280 985 1280 1085 5 0
% \large$A$
\put(12.8000,-10.8500){\makebox(0,0){\large$A$}}%
% STR 2 0 3 27
% 3 3830 1930 3830 2030 5 0
% \large$(C_j^\dagger \bar{C}_l^\dagger)$
\put(38.3000,-20.3000){\makebox(0,0){\large$(C_j^\dagger \bar{C}_l^\dagger)$}}%
% STR 2 0 3 28
% 3 2510 1480 2510 1580 5 0
% \large$C_j$
\put(25.1000,-15.8000){\makebox(0,0){\large$C_j$}}%
% STR 2 0 3 29
% 3 3710 1340 3710 1440 5 0
% \large$\bar{C}_l$
\put(37.1000,-14.4000){\makebox(0,0){\large$\bar{C}_l$}}%
% STR 2 0 3 30
% 3 3270 660 3270 760 5 0
% \large$\bar{B}$
\put(32.7000,-7.6000){\makebox(0,0){\large$\bar{B}$}}%
% STR 2 0 3 31
% 3 3910 985 3910 1085 5 0
% \large$A$
\put(39.1000,-10.8500){\makebox(0,0){\large$A$}}%
% STR 2 0 3 32
% 3 630 1115 630 1215 5 0
% {\large (1)}
\put(6.3000,-12.1500){\makebox(0,0){{\large (1)}}}%
% VECTOR 1 0 3 33
% 2 4010 1215 3810 1215
% 
\special{pn 13}%
\special{pa 4010 1216}%
\special{pa 3810 1216}%
\special{fp}%
\special{sh 1}%
\special{pa 3810 1216}%
\special{pa 3878 1236}%
\special{pa 3864 1216}%
\special{pa 3878 1196}%
\special{pa 3810 1216}%
\special{fp}%
% LINE 1 0 3 34
% 2 3410 1215 4190 1215
% 
\special{pn 13}%
\special{pa 3410 1216}%
\special{pa 4190 1216}%
\special{fp}%
% LINE 1 0 3 35
% 2 1010 1215 1790 1215
% 
\special{pn 13}%
\special{pa 1010 1216}%
\special{pa 1790 1216}%
\special{fp}%
\end{picture}%
}
 \end{center}
\caption{Diagrams representing \eqref{2.37}}
\label{fig8}
\end{figure}

Due to the factor $1/E_{Cj}(\vec{k})-1/E_{Cl}(\vec{k})$ in the integrand,
$\langle {N_{C\rho}(x^0)}^{(3+4)} \rangle_{A(p,x^0_I)}$ has its magnitude much smaller 
than that of $\langle {N_{C\rho}(x^0)}^{(1)} \rangle_{A(p,x^0_I)}$.

From the model calculation explained above, though more detailed numerical analyses are 
necessary, we are possible to regard the magnitude of $\langle {N_{C\rho}(x^0)}^{(1)} \rangle_{A(p,x^0_I)}$
as dominant over other $\langle {N_{C\rho}(x^0)}^{(n)} \rangle_{A(p,x^0_I)}$'s ; i.e.
\begin{align}
\langle N_{C\rho}(x^0) \rangle_{A(p,x^0_I)} \simeq \langle {N_{C\rho}(x^0)}^{(1)} \rangle_{A(p,x^0_I)}. \label{2.38}
\end{align}

If we add by hand the term due to the lifetime of A-particle in the way as
\begin{align}
e^{iE_A(\vec{p})z^0} \rightarrow e^{(iE_A(\vec{p})-\frac{\Gamma_A(\vec{p})}{2})z^0}, \nn \\
e^{-iE_A(\vec{p})y^0} \rightarrow e^{(-iE_A(\vec{p})-\frac{\Gamma_A(\vec{p})}{2})y^0}. \label{2.39}
\end{align}
in (\ref{2.35}), we obtain
\begin{align}
\langle {N_{C\rho}(x^0)}^{(1)} \rangle_{A(p,x^0_I)}&=V|f_{ABC\sigma}|^2\sum_{j,l} 
 {Z^{1/2}_{\rho j}}^\ast Z^{1/2}_{\rho l}Z^{1/2}_{\sigma j}{Z^{1/2}_{\sigma j}}^\ast \nn \\
&\times \int\frac{d^3k}{(2\pi)^6} e^{i(E_{Cj}(\vec{k})-E_{Cl}(\vec{k}))t} \nn 
 \frac{1}{4}\left[ \frac{1}{E_{Cj}(\vec{k})}+\frac{1}{E_{Cl}(\vec{k})} \right]\frac{1}{2E_B(\vec{q})} \nn \\
&\frac{e^{i( E_A(\vec{p})-E_B(\vec{q})-E_{Cj}(\vec{k})+i\Gamma_A(\vec{p})/2)t}-1}{E_A(\vec{p})-E_B(\vec{q})-E_{Cj}(\vec{k})+i\Gamma_A(\vec{p})/2}\cdot
\frac{e^{i(-E_A(\vec{p})+E_B(\vec{q})+E_{Cl}(\vec{k})+i\Gamma_A(\vec{p})/2)t}-1}{E_A(\vec{p})-E_B(\vec{q})-E_{Cl}(\vec{k})-i\Gamma_A(\vec{p})/2} \bigg|_{\vec{q}=\vec{p}-\vec{k}}, \label{2.40}
\end{align}

Under the condition $\Gamma_A(\vec{p})t \gg 1$. we have
\begin{align}
\langle {N_{C\rho}(x^0)}^{(1)} \rangle_{A(p,x^0_I)} &\simeq V|f_{ABC\sigma}|^2\sum_{j,l}
 {Z^{1/2}_{\rho j}}^\ast Z^{1/2}_{\rho l}Z^{1/2}_{\sigma j}{Z^{1/2}_{\sigma j}}^\ast 
 \int\!\frac{d^3k}{(2\pi)^6}e^{i(E_{Cj}(\vec{k})-E_{Cl}(\vec{k}))t} G_{AB;jl}(\vec{p},\vec{q}=\vec{p}-\vec{k},k), \label{2.41} \\
G_{AB;jl}(\vec{p},\vec{q},k) &\equiv \frac{1}{4}\left[\frac{1}{E_{Cj}(\vec{k})}+\frac{1}{E_{Cl}(\vec{k})}\right]\frac{1}{2E_B(\vec{q})} 
 \frac{1}{\left[E_A(\vec{p})-E_B(\vec{q})-E_{Cj}(\vec{k})+i\Gamma(\vec{p})/2\right]} \nn \\
&\times \frac{1}{\left[E_A(\vec{p})-E_B(\vec{q})-E_{Cl}(\vec{k})-i\Gamma(\vec{p})/2\right]} \label{2.42}
\end{align}
E.q.\eqref{2.40} should be compared with E.q.\eqref{2.15}. It will be meaningful to examine whether $G_{AB;jl}$ depends weakly on $\{j,l\}$,
and then,  whether (\ref{2.41}) depends on $t$, characteristically different from the
standard formula (\ref{1.5}).

%% section 3 %%
\section{EXPECTATION VALUES OF FLAVOR-NEUTRINO CURRENTS WITH RESPECT TO ONE-PION WAVE-PACKET STATE}
\subsection{Starting Relations}
We examine the case where flavor neutrinos are produced through a boson decay  such as $\pi^+ \rightarrow \mu^+ +\nu_\mu$; 
for convenience, nuetrinos with only one flavor $\sigma$ are produced. 
As the initial $\pi^+$-state $\ket{\Psi(\ev{\vec{p}}; \vec{X}, x^0_I)}$, we adopt, in accordance with Giunti et al.\cite{r10},
\begin{align}
\ket{\Psi(\ev{\vec{p}}; \vec{X}, x^0_I)}&=\int\!\!d^3p A^\ast_\pi(\vec{p},\ev{\vec{p}}; \vec{X}, x^0_I)
 \frac{1}{\sqrt{2E_\pi(\vec{p})}} \alpha^\dagger_\pi(\vec{p})\ket{0}, \label{3.1}  \\
 A^\ast_\pi(\vec{p},\ev{\vec{p}}; \vec{X}, x^0_I) &=A^\ast_\pi(\vec{p}, \ev{\vec{p}}) e^{-i\vec{p}\cdot \vec{X}+iE_\pi(\vec{p})x^0_I} \nn \\
 &=\frac{1}{(\sqrt{2\pi}\sigma_\pi)^{3/2}} \exp\left[ -\frac{(\vec{p}-\ev{\vec{p}})^2}{4\sigma^2_\pi}-i\vec{p}\cdot \vec{X}+iE_\pi(\vec{p})x^0_I \right].
\label{3.2}
\end{align}
Here, the operator $\alpha_\pi(\vec{p})$ appears in the plane-wave expansion of the $\pi^+$-field, as 
already given by (\ref{2.3}). The state normarization is given by
\begin{align}
&\bracket{\Psi(\ev{\vec{p}}; \vec{X}, x^0_I)}{\Psi(\ev{\vec{p}}'; \vec{X}', x^0_I)} \nn \\
 &\ =\exp\left[ -\frac{\sigma^2_\pi}{2}(\vec{X}-\vec{X}')^2 
  -\frac{1}{8\sigma^2_\pi}(\ev{\vec{p}}-\ev{\vec{p}}')^2 
  -i\frac{(\ev{\vec{p}}+\ev{\vec{p}}') \cdot (\vec{X}-\vec{X}')}{2}\right] \nn \\
&\xrightarrow{\vec{X}=\vec{X}'}\exp \left[ -\frac{1}{8\sigma^2_\pi}(\ev{\vec{p}}-\ev{\vec{p}}')^2 \right]
 \xrightarrow{\ev{\vec{p}}=\ev{\vec{p}}'} 1. \label{3.3}
\end{align}

We have the following relations :
\begin{enumerate}
\item[(i)] 
\begin{align}
&\int\!\!d^3p A^\ast_\pi(\vec{p},\ev{\vec{p}}; \vec{X}, x^0_I) (-i\del{}{\vec{X}})A_\pi(\vec{p},\ev{\vec{p}}; \vec{X}, x^0_I) \nn \\
&\ =\int\!\!d^3p \ \vec{p} \ |A_\pi(\vec{p},\ev{\vec{p}})|^2 = \ev{\vec{p}}, \label{3.4}
\end{align}
\item[(ii)]
\begin{align}
&\int\!\!d^3p A^\ast_\pi(\vec{p},\ev{\vec{p}}; \vec{X}, x^0) (-i\del{}{x^0})A_\pi(\vec{p},\ev{\vec{p}}; \vec{X}, x^0) \nn \\
&\ =\int\!\!d^3p E_\pi(\vec{p}) |A_\pi(\vec{p},\ev{\vec{p}})|^2 \nn \\
&\ \simeq \int\!\!d^3p [E_\pi(\ev{\vec{p}})+\ev{\vec{v}_\pi}(\vec{p}-\ev{\vec{p}})] |A_\pi(\vec{p},\ev{\vec{p}})|^2
 = E_\pi(\ev{\vec{p}}), \label{3.5}
\intertext{where}
\ev{\vec{v}_\pi} &\equiv \left[\del{E_\pi}{\vec{p}}\right]_{\vec{p}=\ev{\vec{p}}}=\frac{\ev{\vec{p}}}{E_\pi(\ev{\vec{p}})} ; \label{3.6}
\end{align}
hereafter $E_\pi(\ev{\vec{p}})$ is written as $\ev{E_\pi}$ for simplicity. 
\item[(iii)] By employing the above relations (\ref{3.3}) - (\ref{3.5}) and the energy and momentum 
operators of $\pi^\pm$-field 
\begin{align}
 \begin{pmatrix}
  \hat{H} \\
  \hat{\vec{P}}
 \end{pmatrix}
=\int\!\frac{d^3p}{2E_\pi(\vec{p})}
 \begin{pmatrix}
  E_\pi(\vec{p}) \\
  \vec{p}
 \end{pmatrix}
[\alpha^\dagger_\pi\vec{p})\alpha_\pi(\vec{p})+\beta^\dagger_\pi(-\vec{p})\beta_\pi(-\vec{p})], \label{3.7}
\intertext{we easily obtain}
\bra{\Psi(\ev{\vec{p}}; \vec{X}, x^0_I)}
 \begin{pmatrix}
  \hat{H} \\
  \hat{\vec{P}}
 \end{pmatrix}
\ket{\Psi(\ev{\vec{p}}; \vec{X}, x^0_I)}
 \begin{pmatrix}
  \simeq \ev{E_\pi} \\
  = \ev{\vec{p}}
 \end{pmatrix}. \label{3.8}
\end{align}
\item[(iv)] The wave packet in the coordinate space is defined by 
\begin{align}
A^\ast_\pi(\vec{x},x^0_I ; \vec{X},x^0_I, \ev{\vec{p}}) 
 \equiv \int\!\frac{d^3p}{(2\pi)^{3/2}}A^\ast_\pi(\vec{p},\ev{\vec{p}} ; \vec{X},x^0_I)
 e^{i(\vec{p}\cdot\vec{x}-E_\pi(\vec{p})x^0)}. \label{3.9}
\end{align}
By using the approximate form of $E_\pi(\vec{p})$ in (\ref{3.5}), we obtain 
\begin{align}
&A^\ast_\pi(\vec{x},x^0_I ; \vec{X},x^0_I, \ev{\vec{p}}) \nn \\
&\simeq \frac{1}{(\sqrt{2\pi}\sigma_{\pi x})^{3/2}} 
 \exp\left[ i\ev{\vec{p}}\cdot(\vec{x}-\vec{X})-i\ev{E_\pi}(x^0-x^0_I)
  -\frac{[(\vec{x}-\vec{X})-\ev{\vec{v_\pi}}(x^0-x^0_I) ]^2}{4\sigma^2_{\pi x}} \right], \label{3.10}
\end{align}
where $\sigma_\pi \sigma_{\pi x}= 1/2$. The normarization is 
\begin{align}
&\int\!\!d^3x A^\ast_\pi(\vec{x},x^0_I ; \vec{X},x^0_I, \ev{\vec{p}}) A_\pi(\vec{x},x^0_I ; \vec{X}',x^0_I, \ev{\vec{p}}') \nn \\
&=\int\!\!d^3p A^\ast_\pi(\vec{p},\ev{\vec{p}}; \vec{X}, x^0_I) A_\pi(\vec{p},\ev{\vec{p}}'; \vec{X}', x^0_I), \label{3.11}
\end{align}
which is the same as (\ref{3.3}). 
\item[(v)] 
\begin{align}
\int\!\!d^3x \ \vec{x} \ |A_\pi(\vec{x},x^0_I ; \vec{X},x^0_I, \ev{\vec{p}})|^2
 \simeq \vec{X}+\ev{\vec{v}_\pi}(x^0-x^0_I). \label{3.12}
\end{align}
\end{enumerate}

Next, the relevant effective weak interaction, causing $\pi^+ \rightarrow \bar{l}_\sigma +\nu_\sigma$ 
decay  is  written as
\begin{align}
\mathcal{H}_W(x)=-\mathcal{L}_{\text{int}}(x)
 =-[\bar{\nu}_\sigma(x)J_\sigma(x)+\bar{J}_\sigma(x) \nu_\sigma(x)]; \label{3.13}
\end{align} 
the effective forms of $J_\sigma(x)$ and $\bar{J}_\sigma(x)$ have been given by (\ref{1.22}) and (\ref{1.23}).
The spin $1/2$-field $\psi(x)$ is expanded in the interaction
representation as 
\begin{align}
\psi(x)=\sum_{\text{helicity}\ r}\int\!\frac{d^3k}{\sqrt{(2\pi)^3 2E(\vec{k})}}
 \left[ \alpha(kr)e^{ikx}u(kr)+\beta^\dagger(kr)e^{-ikx}v(kr) \right]. \label{3.14}
\end{align}
The concrete forms of $u(kr)$ and $v(kr)$ in Kramers 
representation\cite{r16} as well as other details are given in Refs.\cite{r6} and \cite{r7}.

The expectation value of the flavor neutrino current $j_{(\rho)}^a(x)$, given by (\ref{1.26}), when 
the initial state is $\ket{\Psi_\pi(\ev{\vec{p}}; \vec{x}, x^0_I)}$,
given by (\ref{3.1}), is expressed as, in accordance with (\ref{1.19}), 
\begin{align}
\bra{\Psi_\pi(\ev{\vec{p}}; \vec{X}, x^0_I)}S^{-1}(x^0, x^0_I) 
 :j_{(\rho)}^a(x): S(x^0, x^0_I)\ket{\Psi_\pi(\ev{\vec{p}}; \vec{X}, x^0_I)}. \label{3.15}
\end{align}
This consists of a sum of two parts in the lowest order of $\mathcal{H}_W$ ; 
\begin{align}
&\mathcal{E}(x^0; \ev{\vec{p}}, \vec{X},x^0_I)^a_{\text{(I)}\rho} \nn \\
&\ \equiv \bra{\Psi_\pi(\ev{\vec{p}}; \vec{x}, x^0_I)} \int^{x^0}_{x^0_I}\!\!\int^{x^0}_{x^0_I}d^4zd^4y \mathcal{H}_W(z)  
 :j_{(\rho)}^a(x): \mathcal{H}_W(y) \ket{\Psi_\pi(\ev{\vec{p}}; \vec{x}, x^0_I)}^{\text{con}}, \label{3.16}  \\
&\mathcal{E}(x^0; \ev{\vec{p}}, \vec{X},x^0_I)^a_{\text{(II)}\rho} \nn \\
&\ \equiv \bra{\Psi_\pi(\ev{\vec{p}}; \vec{x}, x^0_I)} i^2 \int^{x^0}_{x^0_I}\!\!\int^{x^0}_{x^0_I}d^4zd^4y \nn \\
&\quad \times \left[ \mathcal{H}_W(y) \mathcal{H}_W(z) :j_{(\rho)}^a(x):+:j_{(\rho)}^a(x):\mathcal{H}_W(z) \mathcal{H}_W(y) \right]
 \ket{\Psi_\pi(\ev{\vec{p}}; \vec{x}, x^0_I)}^{\text{con}}. \label{3.17}
\end{align}
In the following we examine (\ref{3.16}), because this includes a dominant contribution. 

\subsection{Form of $\mathcal{E}^a_{\text{(I)}\rho}$}
As shown in FIG.\ref{fig9}, $\mathcal{E}^a_{\text{(I)}\rho}$ consists of two parts expressed as 
\begin{align}
&\begin{bmatrix}
  \mathcal{E}(\vec{x},x^0; \ev{\vec{p}}, \vec{X},x^0_I)^a_{\text{(I-1)}\rho} \\
  \mathcal{E}(\vec{x},x^0; \ev{\vec{p}}, \vec{X},x^0_I)^a_{\text{(I-2)}\rho} 
 \end{bmatrix} \nn \\
&\ =
\bra{\Psi_\pi(\ev{\vec{p}}; \vec{X}, x^0_I)} \int^{x^0}_{x^0_I}\!\!\int^{x^0}_{x^0_I}d^4zd^4y
 \begin{bmatrix}
  \bar{J}_\sigma(z)\nu_\sigma(z) :j_{(\rho)}^a(x): \bar{\nu}_\sigma(y)J_\sigma(y)\\
  \bar{\nu}_\sigma(z)J_\sigma(z) :j_{(\rho)}^a(x): \bar{J}_\sigma(y)\nu_\sigma(y)
 \end{bmatrix}
\ket{\Psi_\pi(\ev{\vec{p}}; \vec{X}, x^0_I)}^{\text{con}}. \label{3.18}
\end{align}
\begin{figure}[h]
 \begin{center}
  \scalebox{0.9}[0.9]{%WinTpicVersion3.08
\unitlength 0.1in
\begin{picture}( 76.6000, 17.7500)(5,-24.1500)
% ELLIPSE 1 0 3 0
% 4 2610 1215 3430 1655 3430 1655 3660 1785
% 
\special{pn 13}%
\special{ar 2610 1216 820 440  0.7910799 6.2831853}%
\special{ar 2610 1216 820 440  0.0000000 0.7853982}%
% LINE 1 0 3 1
% 2 1010 1215 1790 1215
% 
\special{pn 13}%
\special{pa 1010 1216}%
\special{pa 1790 1216}%
\special{fp}%
% LINE 1 0 3 2
% 2 3410 1215 4190 1215
% 
\special{pn 13}%
\special{pa 3410 1216}%
\special{pa 4190 1216}%
\special{fp}%
% DOT 1 0 3 3
% 2 2610 1655 2610 1665
% 
\special{pn 13}%
\special{sh 1}%
\special{ar 2610 1656 10 10 0  6.28318530717959E+0000}%
\special{sh 1}%
\special{ar 2610 1666 10 10 0  6.28318530717959E+0000}%
% VECTOR 1 0 3 4
% 4 3340 1015 3330 1005 3350 1015 3350 1015
% 
\special{pn 13}%
\special{pa 3340 1016}%
\special{pa 3330 1006}%
\special{fp}%
\special{sh 1}%
\special{pa 3330 1006}%
\special{pa 3364 1066}%
\special{pa 3368 1044}%
\special{pa 3392 1038}%
\special{pa 3330 1006}%
\special{fp}%
\special{pa 3350 1016}%
\special{pa 3350 1016}%
\special{fp}%
% VECTOR 1 0 3 5
% 2 3310 1445 3300 1455
% 
\special{pn 13}%
\special{pa 3310 1446}%
\special{pa 3300 1456}%
\special{fp}%
\special{sh 1}%
\special{pa 3300 1456}%
\special{pa 3362 1422}%
\special{pa 3338 1418}%
\special{pa 3334 1394}%
\special{pa 3300 1456}%
\special{fp}%
% STR 2 0 3 6
% 3 2580 2230 2580 2330 5 0
% {\large (1) $\mathcal{E}^a_{\text{(I-1)}\rho}$}
\put(25.8000,-23.3000){\makebox(0,0){{\large (1) $\mathcal{E}^a_{\text{(I-1)}\rho}$}}}%
% STR 2 0 3 7
% 3 3910 985 3910 1085 5 0
% \large$\pi^+$
\put(39.1000,-10.8500){\makebox(0,0){\large$\pi^+$}}%
% STR 2 0 3 8
% 3 3360 730 3360 830 5 0
% \large$\bar{l}_\sigma$
\put(33.6000,-8.3000){\makebox(0,0){\large$\bar{l}_\sigma$}}%
% STR 2 0 3 9
% 3 3330 1535 3330 1635 5 0
% \large$\nu_\sigma$
\put(33.3000,-16.3500){\makebox(0,0){\large$\nu_\sigma$}}%
% STR 2 0 3 10
% 3 1910 1555 1910 1655 5 0
% \large$\nu_\sigma$
\put(19.1000,-16.5500){\makebox(0,0){\large$\nu_\sigma$}}%
% STR 2 0 3 11
% 3 2610 1765 2610 1865 5 0
% \large$(\alpha^\dagger_{\nu\rho} \alpha_{\nu\rho})$
\put(26.1000,-18.6500){\makebox(0,0){\large$(\alpha^\dagger_{\nu\rho} \alpha_{\nu\rho})$}}%
% STR 2 0 3 12
% 3 1280 985 1280 1085 5 0
% \large$\pi^+$
\put(12.8000,-10.8500){\makebox(0,0){\large$\pi^+$}}%
% STR 2 0 3 13
% 3 5815 1765 5815 1865 5 0
% \large$(\beta^\dagger_{\nu\rho}\beta_{\nu\rho})$
\put(58.1500,-18.6500){\makebox(0,0){\large$(\beta^\dagger_{\nu\rho}\beta_{\nu\rho})$}}%
% STR 2 0 3 14
% 3 5185 625 5185 725 5 0
% \large$\pi^+$
\put(51.8500,-7.2500){\makebox(0,0){\large$\pi^+$}}%
% LINE 1 0 3 15
% 2 6655 1175 5925 1005
% 
\special{pn 13}%
\special{pa 6656 1176}%
\special{pa 5926 1006}%
\special{fp}%
% LINE 1 0 3 16
% 2 5005 1185 6605 785
% 
\special{pn 13}%
\special{pa 5006 1186}%
\special{pa 6606 786}%
\special{fp}%
% VECTOR 1 0 3 17
% 2 6285 1195 6085 1195
% 
\special{pn 13}%
\special{pa 6286 1196}%
\special{pa 6086 1196}%
\special{fp}%
\special{sh 1}%
\special{pa 6086 1196}%
\special{pa 6152 1216}%
\special{pa 6138 1196}%
\special{pa 6152 1176}%
\special{pa 6086 1196}%
\special{fp}%
% LINE 1 0 3 18
% 2 6645 1195 4995 1195
% 
\special{pn 13}%
\special{pa 6646 1196}%
\special{pa 4996 1196}%
\special{fp}%
% ELLIPSE 1 0 3 19
% 4 5825 1205 6645 1665 4915 1205 6845 1205
% 
\special{pn 13}%
\special{ar 5826 1206 820 460  6.2831853 6.2831853}%
\special{ar 5826 1206 820 460  0.0000000 3.1415927}%
% DOT 1 0 3 20
% 2 5815 1655 5815 1665
% 
\special{pn 13}%
\special{sh 1}%
\special{ar 5816 1656 10 10 0  6.28318530717959E+0000}%
\special{sh 1}%
\special{ar 5816 1666 10 10 0  6.28318530717959E+0000}%
% VECTOR 1 0 3 21
% 2 6575 1375 6565 1385
% 
\special{pn 13}%
\special{pa 6576 1376}%
\special{pa 6566 1386}%
\special{fp}%
\special{sh 1}%
\special{pa 6566 1386}%
\special{pa 6626 1352}%
\special{pa 6604 1348}%
\special{pa 6598 1324}%
\special{pa 6566 1386}%
\special{fp}%
% STR 2 0 3 22
% 3 6385 625 6385 725 5 0
% \large$\pi^+$
\put(63.8500,-7.2500){\makebox(0,0){\large$\pi^+$}}%
% STR 2 0 3 23
% 3 6265 1225 6265 1325 5 0
% \large$l_\sigma$
\put(62.6500,-13.2500){\makebox(0,0){\large$l_\sigma$}}%
% STR 2 0 3 24
% 3 6535 1535 6535 1635 5 0
% \large$\bar{\nu}_\sigma$
\put(65.3500,-16.3500){\makebox(0,0){\large$\bar{\nu}_\sigma$}}%
% STR 2 0 3 25
% 3 5115 1555 5115 1655 5 0
% \large$\bar{\nu}_\sigma$
\put(51.1500,-16.5500){\makebox(0,0){\large$\bar{\nu}_\sigma$}}%
% LINE 1 0 3 26
% 2 5780 960 5010 790
% 
\special{pn 13}%
\special{pa 5780 960}%
\special{pa 5010 790}%
\special{fp}%
% STR 2 0 3 27
% 3 5860 2230 5860 2330 5 0
% {\large (2) $\mathcal{E}^a_{\text{(I-2)}\rho}$}
\put(58.6000,-23.3000){\makebox(0,0){{\large (2) $\mathcal{E}^a_{\text{(I-2)}\rho}$}}}%
\end{picture}%
}
 \end{center}
\caption{Diagrams representing \eqref{3.18}}
\label{fig9}
\end{figure}

\noindent
We obtain 
\begin{align}
&\mathcal{E}(\vec{x},x^0; \ev{\vec{p}}, \vec{X},x^0_I)^a_{\text{(I-1)}\rho} \nn \\
&=\int^{x^0}_{x^0_I}\!\!\int^{x^0}_{x^0_I}d^4zd^4y \ 
 \bra{\Psi_\pi(\ev{\vec{p}}; \vec{X}, x^0_I)} if^\ast_{\pi \sigma} \partial_b \phi^\dagger_\pi(x) \nn \\
&\times \bra{0}\bar{l}_\sigma(z)v^b \bra{0}\nu_\sigma(z) :j_{(\rho)}^a(x): 
 \bar{\nu}_\sigma(y)\ket{0} v^d l_\sigma(y)\ket{0}
\cdot if_{\pi \sigma} \partial_d \phi_\pi(x) \ket{\Psi_\pi(\ev{\vec{p}}; \vec{X}, x^0_I)}^{\text{con}} \nn \\
&=\int^{x^0}_{x^0_I}\!\!\int^{x^0}_{x^0_I}d^4zd^4y \int\!\!d^3p' \int\!\!d^3p
 \frac{-|f_{\pi \sigma}|^2}{(2\pi)^3 \sqrt{2E_\pi(\vec{p}') 2E_\pi(\vec{p})}} 
 A_\pi(\vec{p}', \ev{\vec{p}}) A^\ast_\pi(\vec{p}, \ev{\vec{p}}) \nn \\
&\times \int\!\!d^3q \int\!\!d^3k'\int\!\!d^3k (2\pi)^{-9} 
 \sum_{j,l} Z^{1/2}_{\sigma j}{Z^{1/2}_{\rho j}}^\ast Z^{1/2}_{\rho l}{Z^{1/2}_{\sigma l}}^\ast 
 B_{\sigma, jl}^{(1)a}(q ; p', p ; k', k)\nn \\
&\times e^{-i\vec{p}'\cdot(\vec{z}-\vec{X})+iE_\pi(\vec{p}')(z^0-x^0_I)}
 e^{i\vec{p}\cdot(\vec{y}-\vec{X})-iE_\pi(\vec{p})(y^0-x^0_I)} \nn \\
&\times e^{i(q+k'_j)z-i(k_l+q)y}
 e^{i(k_l-k_j)x}, \label{3.19}
\end{align}
where $k_l=(\vec{k},i\omega_l(\vec{k})), \omega_l(\vec{k})=\sqrt{\vec{k}^2+m^2_l}$ ;
\begin{align}
&B_{\sigma, jl}^{(1)a}(q ; p', p ; k', k)\nn \\
&\equiv \sum_{s,r,r'} \bar{v}_\sigma(qs)\sla{p}'(1+\gamma_5)u_j(k'r')\
 (-i)\bar{u}_j(k'r')\gamma^a u_l(kr)\cdot
 \bar{u}_l(kr)\sla{p}(1+\gamma_5)v_\sigma(qs). \label{3.20}
\end{align}
By taking the new parameters
\begin{align}
t=x^0-x^0_I,\quad Z^0=z^0-x^0_I,\quad Y^0=y^0-x^0_I, \label{3.21} 
\end{align}
and adding by hand the factor of $\pi$ lifetime in the same way as (\ref{2.38}), we obtain 
\begin{align}
&\mathcal{E}(x^0; \ev{\vec{p}}, \vec{X},x^0_I)^a_{\text{(I-1)}\rho} \nn \\
&=\int^t_0\!\!\int^t_0 dZ^0 dY^0 \int\!\!d^3p'\!\int\!\!d^3p
 \frac{-|f_{\pi \sigma}|^2}{(2\pi)^3 \sqrt{2E_\pi(\vec{p}') 2E_\pi(\vec{p})}} 
 A_\pi(\vec{p}', \ev{\vec{p}}) A^\ast_\pi(\vec{p}, \ev{\vec{p}}) \nn \\
&\times \int\!\!d^3q \int\!\!d^3k'\int\!\!d^3k (2\pi)^{-9} 
 \sum_{j,l} Z^{1/2}_{\sigma j}{Z^{1/2}_{\rho j}}^\ast Z^{1/2}_{\rho l}{Z^{1/2}_{\sigma l}}^\ast  
 B_{\sigma, jl}^{(1)a}(q ; p', p ; k', k)\nn \\
&\times \int\!\!\int\! d^3z d^3y 
 e^{i\vec{z}\cdot(-\vec{p}'+\vec{q}+\vec{k}')+i\vec{y}\cdot(\vec{p}-\vec{q}-\vec{k})}
 e^{i\vec{X}\cdot(\vec{p}'-\vec{p})+i\vec{x}(-\vec{k}'+\vec{k})} \nn \\
&\times e^{iZ^0( E_\pi(\vec{p}')-E_\sigma(\vec{q})-\omega_j(\vec{k}')+i\Gamma_\pi(\vec{p}')/2 )}
 e^{iY^0(-E_\pi(\vec{p})+E_\sigma(\vec{q})+\omega_l(\vec{k})+i\Gamma_\pi(\vec{p})/2 )} \nn \\
&\times e^{it( \omega_j(\vec{k}')-\omega_l(\vec{k}) )}. \label{3.22}
\end{align}
The factors depending on $Z^0$ and $Y^0$, when we perform the $Z^0$- and $Y^0$-integrations and 
require the conditions
\begin{align}
t \Gamma_\pi(\vec{p}') \gg 1 \ , \ t \Gamma_\pi(\vec{p}) \gg 1, \label{3.23}
\end{align}
lead to the factor
\begin{align}
\left\{ 
 \left[E_\pi(\vec{p}')-E_\sigma(\vec{q})-\omega_j(\vec{k}')+i\frac{\Gamma_\pi(\vec{p}')}{2}\right]
 \left[E_\pi(\vec{p})-E_\sigma(\vec{q})-\omega_l(\vec{k})-i\frac{\Gamma_\pi(\vec{p})}{2} \right]_{
\substack{\vec{q}=\vec{p}'-\vec{k}'\\ =\vec{p}-\vec{k}}}
\right\}^{-1} \label{3.24}
\end{align}
Then we see that possible candidates having a connection with the oscillation in R.H.S. of (\ref{3.22}) 
are 
\begin{align}
\text{i)\ } e^{i(\omega_j(\vec{k}')-\omega_l(\vec{k}))t} \quad 
\text{and\quad ii)\ }
 e^{i(\vec{p}'-\vec{p})\cdot\vec{X}+i(-\vec{k}'+\vec{k})\cdot\vec{x}}
 \rightarrow e^{i(-\vec{k}'+\vec{k})\cdot\vec{L}} \quad \text{with}
 \quad \vL =\vec{x}-\vec{X}. \label{3.25}
\end{align}
We will examine the situation in the next section.

Similar to $\mathcal{E}^a_{\text{(I-1)}\rho}$, we obtain
\begin{align}
&\mathcal{E}(x^0; \ev{\vec{p}}, \vec{X},x^0_I)^a_{\text{(I-2)}\rho} \nn \\
&=\int^t_0\!\!\int^t_0 dZ^0 dY^0 \int\!\!d^3p\int\!\!d^3p'  
\frac{-|f_{\pi \sigma}|^2}{(2\pi)^6 \sqrt{2E_\pi(\vec{p}) 2E_\pi(\vec{p}')}} 
 A_\pi(\vec{p}, \ev{\vec{p}}) A^\ast_\pi(\vec{p}', \ev{\vec{p}}) \nn \\
&\times \sum_{j,l} Z^{1/2}_{\sigma j}{Z^{1/2}_{\rho j}}^\ast Z^{1/2}_{\rho l}{Z^{1/2}_{\sigma l}}^\ast  
 B_{\sigma, lj}^{(2)a}(q ; p, p' ; k, k') 
\cdot e^{i(\vec{k}-\vec{k}')\cdot(\vec{x}-\vec{X})}
 e^{i(-\omega_l(\vec{k})+\omega_j(\vec{k}'))t} \nn \\
&\times e^{iZ^0(-E_\pi(\vec{p}')-E_\sigma(\vec{q})-\omega_j(\vec{k}')+i\Gamma_\pi(\vec{p}')/2)}
 e^{iY^0(E_\pi(\vec{p})+E_\sigma(\vec{q})+\omega_l(\vec{k})+i\Gamma_\pi(\vec{p})/2)}
 \bigg|_{\vec{q}=-\vec{p}'-\vec{k}=-\vec{p}-\vec{k}'} ; \label{3.26}
\intertext{here,}
&B_{\sigma, lj}^{(2)a}(q ; p, p' ; k, k') \nn \\
&\equiv \sum_{s,r,r'} \bar{u}_\sigma(qs)\sla{p}(1+\gamma_5)v_l(kr)
 \cdot i\bar{v}_l(kr)\gamma^a v_j(k'r')\cdot
 \bar{v}_j(k'r')\sla{p}'(1+\gamma_5)u_\sigma(qs). \label{3.27}
\end{align}
The factor corresponding to (\ref{3.24}) is
\begin{align}
\left\{\!
 \left[E_\pi(\vec{p}')+E_\sigma(\vec{q})+\omega_j(\vec{k}')-i\frac{\Gamma_\pi(\vec{p}')}{2}\right]\!
 \left[E_\pi(\vec{p})+E_\sigma(\vec{q})+\omega_l(\vec{k})+i\frac{\Gamma_\pi(\vec{p})}{2} \right]_{
 \substack{\vec{q}=-\vec{p}'-\vec{k}'\\ =-\vec{p}-\vec{k}}} \!
\right\}^{-1}. \label{3.28}
\end{align}
Though detailed numerical analyses are nescessary, we can expect the main contribution in 
$\mathcal{E}^a_{\text{(I)}\rho}$ comes from $\mathcal{E}^a_{\text{(I-1)}\rho}$.  
In the next subsection, we perform further evaluations of $\mathcal{E}^a_{\text{(I-1)}\rho}$.

\subsection{Evaluation of $\mathcal{E}^a_{\text{(I-1)}\rho}$}
With the aim of examining the effect due to the wave packet of $\pi^+$, we perform 
the integrations in R.H.S. of (\ref{3.22}) with respect to $\vec{p}'$ and $\vec{p}$ firstly; 
then the $\vec{z}$ and $\vec{y}$ integrations are carried out.  

\subsubsection{$\vec{p}'$- and $\vec{p}$-integrations}
We consider the $\vec{p}$ integration appearing in (\ref{3.22}); 
\begin{align}
\int\!\!d^3p A^\ast_\pi(\vec{p},\ev{\vec{p}})
 e^{iY^0(-E_\pi(\vec{p})+i\Gamma_\pi(\vec{p})/2)+i(\vec{y}-\vec{X})\cdot\vec{p}}\ F(\vec{p}), \label{3.29}
\end{align}
and take the approximation, similar to (\ref{3.5}),
\begin{align}
e^{iY^0(-E_\pi(\vec{p})+i\Gamma_\pi(\vec{p})/2)} \simeq 
 e^{iY^0(-\ev{E_\pi}-\ev{\vec{v}_\pi}\cdot(\vec{p}-\ev{\vec{p}})+i\ev{\Gamma_\pi}/2)}; \label{3.30}
\end{align}
here, $\ev{\Gamma_\pi} \equiv \Gamma_\pi(\ev{\vec{p}})$.    Also $F(\vec{p})$ in the integrand 
is assumed to be replaced approximately by $F(\ev{\vec{p}})$;
then we have 
\begin{align}
\eqref{3.29} &\simeq F(\ev{\vec{p}}) \int\!\frac{d^3p}{[\sqrt{2\pi}\sigma_\pi]^{3/2}}
 \exp\left[ -\frac{(\vec{p}-\ev{\vec{p}})^2}{4\sigma_\pi^2}
  +i(\vec{p}-\ev{\vec{p}})\cdot(\vec{y}-\vec{X}-Y^0\ev{\vec{v_\pi}}) \right.\nn \\
&\left. \hspace{3.5cm} +i\ev{\vec{p}}\cdot(\vec{y}-\vec{X}) +iY^0(-\ev{E_\pi}+i\frac{\ev{\Gamma_\pi}}{2}) \right] \nn \\
&=F(\ev{\vec{p}})\int\!\frac{d^3p}{[\sqrt{2\pi}\sigma_\pi]^{3/2}}
 \exp\left[ -\frac{1}{4\sigma_\pi^2} (\vec{p}-2i\sigma_\pi^2(\vec{y}-\vec{X}-Y^0\ev{v_\pi}))^2 \right] \nn \\ 
& \hspace{2.5cm} \times
 \exp\left[-\sigma_\pi^2(\vec{y}-\vec{X}-Y^0\ev{\vec{v}_\pi})^2 +
 i\ev{\vec{p}}\cdot(\vec{y}-\vec{X}) +iY^0(-\ev{E_\pi}+i\frac{\ev{\Gamma_\pi}}{2}) \right] \nn \\
&\simeq F(\ev{\vec{p}})(2\sigma_\pi\sqrt{2\pi})^{3/2} \exp\left[ -\sigma_\pi^2(\vec{y}-\vec{X}-Y^0\ev{\vec{v}_\pi})^2 
 +i\ev{\vec{p}}\cdot(\vec{y}-\vec{X}) +iY^0(-\ev{E_\pi}+i\frac{\ev{\Gamma_\pi}}{2}) \right]. \label{3.31}
\end{align}
Thus, (\ref{3.22}) is written as 
\begin{align}
&\mathcal{E}(\vec{x},x^0; \ev{\vec{p}}, \vec{X},x^0_I)^a_{\text{(I-1)}\rho} \nn \\
&\simeq \int\!\!\int^t_0 dZ^0 dY^0 \int\!\!d^3z\!\int\!\!d^3y
 \frac{-|f_{\pi \sigma}|^2}{(2\pi)^3 2\ev{E_\pi}} 
 \int\!\!d^3q\!\int\!\!d^3k'\!\int\!\!d^3k \nn \\
&\times \sum_{j,l} Z^{1/2}_{\sigma j}{Z^{1/2}_{\rho j}}^\ast Z^{1/2}_{\rho l}{Z^{1/2}_{\sigma l}}^\ast  
 B_{\sigma, jl}^{(1)a}(q ; \ev{\vec{p}},\ev{\vec{p}} ; k', k)
 [2\sigma_\pi \sqrt{2\pi}]^3 \nn \\
&\times e^{i\vec{z}\cdot(-\ev{\vec{p}}+\vec{q}+\vec{k}')+i\vec{y}\cdot(\ev{\vec{p}}-\vec{q}-\vec{k})} \nn \\
&\times e^{iZ^0( \ev{E_\pi}-E_\sigma(\vec{q})-\omega_j(\vec{k}')+i\ev{\Gamma_\pi}/2 )}
 e^{iY^0(-\ev{E_\pi}+E_\sigma(\vec{q})+\omega_l(\vec{k})+i\ev{\Gamma_\pi}/2 )} \nn \\
&\times e^{i\vec{x}\cdot(-\vec{k}'+\vec{k})+it( \omega_j(\vec{k}')-\omega_l(\vec{k}) )} \nn\\
&\times \exp\left[ -\sigma_\pi^2(\vec{z}-\vec{X}-Z^0\ev{\vec{v}_\pi})^2
 -\sigma_\pi^2(\vec{y}-\vec{X}-Y^0\ev{\vec{v}_\pi})^2
 \right]. \label{3.32} 
\intertext{We see}
& \vec{y}-\vec{X}-Y^0\ev{\vec{v}_\pi}=0, \qquad \vec{z}-\vec{X}-Z^0\ev{\vec{v}_\pi}=0 \label{3.33}
\end{align}
correspond to the classical relations as shown in Fig.\ref{fig10}.
\begin{figure}[h]
 \begin{center}
  \scalebox{0.9}[0.9]{%WinTpicVersion3.08
\unitlength 0.1in
\begin{picture}(14,14)( 26,-21)
% LINE 1 0 3 0
% 2 3990 1590 2590 1590
% 
\special{pn 13}%
\special{pa 3990 1590}%
\special{pa 2590 1590}%
\special{fp}%
% VECTOR 1 0 3 1
% 2 2590 1590 1590 990
% 
\special{pn 13}%
\special{pa 2590 1590}%
\special{pa 1590 990}%
\special{fp}%
\special{sh 1}%
\special{pa 1590 990}%
\special{pa 1638 1042}%
\special{pa 1636 1018}%
\special{pa 1658 1008}%
\special{pa 1590 990}%
\special{fp}%
% VECTOR 1 0 3 2
% 2 2570 1600 1570 2030
% 
\special{pn 13}%
\special{pa 2570 1600}%
\special{pa 1570 2030}%
\special{fp}%
\special{sh 1}%
\special{pa 1570 2030}%
\special{pa 1640 2022}%
\special{pa 1620 2010}%
\special{pa 1624 1986}%
\special{pa 1570 2030}%
\special{fp}%
% STR 2 0 3 3
% 3 3340 1350 3340 1450 5 0
% \large$\pi$
\put(33.4000,-14.5000){\makebox(0,0){\large$\pi$}}%
% DOT 1 0 3 4
% 1 4010 1590
% 
\special{pn 13}%
\special{sh 1}%
\special{ar 4010 1590 10 10 0  6.28318530717959E+0000}%
% STR 2 0 3 5
% 3 4480 1370 4480 1470 5 0
% \large$(\vec{X},t=0)$
\put(44.8000,-14.7000){\makebox(0,0){\large$(\vec{X},t=0)$}}%
% STR 2 0 3 6
% 3 1880 850 1880 950 5 0
% \large$\bar{l}_\sigma$
\put(18.8000,-9.5000){\makebox(0,0){\large$\bar{l}_\sigma$}}%
% STR 2 0 3 7
% 3 1870 1960 1870 2060 5 0
% \large$\nu_\sigma$
\put(18.7000,-20.6000){\makebox(0,0){\large$\nu_\sigma$}}%
% STR 2 0 3 8
% 3 2710 1340 2710 1440 5 0
% \large$(\vec{y})$
\put(27.1000,-14.4000){\makebox(0,0){\large$(\vec{y})$}}%
% STR 2 0 3 9
% 3 3300 1660 3300 1760 5 0
% \scalebox{3}[1.5]{\LARGE$\underbrace{\ }$}
\put(33.0000,-17.6000){\makebox(0,0){\scalebox{3}[1.5]{\LARGE$\underbrace{\ }$}}}%
% STR 2 0 3 10
% 3 3300 1960 3300 2060 5 0
% \large$\vev{\pi}y^0$
\put(33.0000,-20.6000){\makebox(0,0){\large$\vev{\pi}y^0$}}%
\end{picture}%
}
 \end{center}
\caption{Classical Relation \eqref{3.33}.}
\label{fig10}
\end{figure}

\subsubsection{$\vec{z}$- and $\vec{y}$-integrations}
By employing 
\begin{align}
-\sigma_\pi^2(\vec{y}-\vec{X}- & Y^0\ev{\vec{v}_\pi})^2+i\vec{y}\cdot(\ev{\vec{p}}-\vec{q}-\vec{k}) \nn \\
&= -\sigma_\pi^2[(\vec{y}-\vec{X}-Y^0\ev{\vec{v}_\pi})-\frac{i}{2\sigma^2_\pi}(\ev{\vec{p}}-\vec{q}-\vec{k})]^2 \nn \\
&\qquad -\frac{1}{4\sigma_\pi^2}(\ev{\vec{p}}-\vec{q}-\vec{k}))^2+i(Y^0\ev{\vec{v}_\pi}+\vec{X})(\ev{\vec{p}}-\vec{q}-\vec{k}),
\label{3.34}
\end{align}
we obtain from (\ref{3.32})
\begin{align}
&\mathcal{E}(\vec{x},x^0; \ev{\vec{p}}, \vec{X},x^0_I)^a_{\text{(I-1)}\rho} \nn \\
&\simeq \int\!\!\int^t_0 dZ^0 dY^0 
 \frac{-|f_{\pi \sigma}|^2}{(2\pi)^3 2\ev{E_\pi}} 
 \int\!\!d^3q\!\int\!\!d^3k'\!\int\!\!d^3k 
 \sum_{j,l} Z^{1/2}_{\sigma j}{Z^{1/2}_{\rho j}}^\ast Z^{1/2}_{\rho l}{Z^{1/2}_{\sigma l}}^\ast  \nn \\
&\times B_{\sigma, jl}^{(1)a}(q ; \ev{\vec{p}},\ev{\vec{p}} ; k', k)
 [2\sigma_\pi \sqrt{2\pi}]^3 \left[\frac{\sqrt{\pi}}{\sigma_\pi}\right]^6 \nn \\
&\times \exp \left[-\frac{1}{4\sigma_\pi^2}(\ev{\vec{p}}-\vec{q}-\vec{k}))^2 -\frac{1}{4\sigma_\pi^2}(\ev{\vec{p}}-\vec{q}-\vec{k}'))^2 \right] \nn \\
&\times e^{iZ^0( \ev{E_\pi}-E_\sigma(\vec{q})-\omega_j(\vec{k}')+i\ev{\Gamma_\pi}/2 +\ev{\vec{v}_\pi}\cdot(-\ev{\vec{p}}+\vec{q}+\vec{k}'))}
 e^{iY^0(-\ev{E_\pi}+E_\sigma(\vec{q})+\omega_l(\vec{k})+i\ev{\Gamma_\pi}/2 +\ev{\vec{v}_\pi}\cdot(\ev{\vec{p}}-\vec{q}-\vec{k}) )} \nn \\
&\times e^{i(\vec{x}-\vec{X})\cdot(-\vec{k}'+\vec{k})+it( \omega_j(\vec{k}')-\omega_l(\vec{k}) )}. \label{3.35} 
\end{align}

\subsubsection{$\vec{k}$- and $\vec{k}'$-integrations}
Similarly we can perform $\vec{k}$-  and $\vec{k}'$-integrations approximately.  For a fixed $\vec{K}$, 
which is defined by
\begin{align}
\vec{K} \equiv \ev{\vec{p}}-\vec{q} \label{3.36}
\end{align}
the main contributions to the $\vec{k}$ and $\vec{k}'$-integrations come from the regions 
\begin{align}
\vec{k} \simeq \vec{K}, \qquad \vec{k}' \simeq \vec{K}. \label{3.37}
\end{align}
Then, we are allowed to employ the relation
\begin{align}
\omega_l(\vec{k}) \simeq \omega_l(\vec{K})+\del{\omega_l}{\vec{k}}\bigg|_\vec{K}\cdot(\vec{k}-\vec{K})
 =\omega_l(\vec{K})+\vec{v}_l(\vec{K})\cdot(\vec{k}-\vec{K}), \label{3.38}
\end{align}
and that for $\omega_j(\vec{k}')$. Then we rewrite the exponent of (\ref{3.35}) as 
\begin{align}
&-\frac{1}{4\sigma_\pi^2}(\vec{k}-\vec{K})^2 -i\omega_l(\vec{k})(t-Y^0)
 -iY^0\ev{\vec{v}_\pi}\cdot(\vec{k}-\vec{K})+i\vec{k}\cdot(\vec{x}-\vec{X})\nn \\
&\simeq -\frac{1}{4\sigma_\pi^2}(\vec{k}-\vec{K})^2 +i(\vec{k}-\vec{K})\cdot[\vec{L}-\vec{v}_l(t-Y^0)-Y^0\ev{\vec{v}_\pi}] \nn \\
&\quad -i\omega_l(\vec{K})(t-Y^0)+i\vec{K}\cdot\vec{L} \nn \\
&= -\frac{1}{4\sigma^2_\pi}\left[(\vec{k}-\vec{K})-2i\sigma^2_\pi\vec{F}_l(\vec{K}; \vec{L},t,Y^0)\right]^2
 -\sigma^2_\pi\vec{F}^2_l(\vec{K}; \vec{L},t,Y^0) \nn \\
&\quad -i\omega_l(\vec{K})(t-Y^0)+i\vec{K}\cdot\vec{L}, \label{3.39}
\intertext{where $\vec{L}=\vec{x}-\vec{X}$ and} 
&\vec{F}_l(\vec{K}; \vec{L},t,Y^0)=\vec{L}-\vec{v}_l(\vec{K})(t-Y^0)-Y^0\ev{\vec{v}_\pi}. \label{3.40}
\end{align}
By employing (\ref{3.39}), we obtain from (\ref{3.35}) 
\begin{align}
&\mathcal{E}(\vec{x},x^0; \ev{\vec{p}}, \vec{X},x^0_I)^a_{\text{(I-1)}\rho} \nn \\
&\simeq \int^t_0\!\!\int^t_0 dZ^0 dY^0 
 \frac{-|f_{\pi \sigma}|^2}{(2\pi)^{12} 2\ev{E_\pi}} 
 \int\!\!d^3q 
 \sum_{j,l} {Z^{1/2}_{\rho j}}^\ast Z^{1/2}_{\rho l}Z^{1/2}_{\sigma
 j}{Z^{1/2}_{\sigma l}}^\ast \nn \\
&\times B_{\sigma, jl}^{(1)a}(q ; \ev{\vec{p}},\ev{\vec{p}} ; \vec{K},\vec{K})
 [2\sigma_\pi \sqrt{2\pi}]^3
 \left[\frac{\sqrt{\pi}}{\sigma_\pi}\right]^6 
 (2\sigma_\pi \sqrt{\pi})^6 \nn \\
&\times e^{iZ^0( \ev{E_\pi}-E_\sigma(\vec{q})+i\ev{\Gamma_\pi}/2 )}
 e^{iY^0(-\ev{E_\pi}+E_\sigma(\vec{q})+i\ev{\Gamma_\pi}/2 )} \nn \\
&\times \exp \left[ -\sigma^2_\pi \vec{F}_j(\vec{K;\vec{L},t,Z^0})^2 
 -\sigma^2_\pi \vec{F}_l(\vec{K;\vec{L},t,Y^0})^2 \right] \nn \\
&\times e^{i\omega_j(\vec{K})(t-Z^0)-i\omega_l(\vec{K})(t-Y^0)}
\bigg|_{\vec{K}=\ev{\vec{p}}-\vec{q}}. \label{3.41}
\end{align}

\begin{figure}[h]
 \begin{center}
  \scalebox{1}[1]{%WinTpicVersion3.08
\unitlength 0.1in
\begin{picture}( 57.3000, 11.4500)( 15.1000,-22.3000)
% VECTOR 1 0 3 0
% 2 5800 1400 2600 1400
% 
\special{pn 13}%
\special{pa 5800 1400}%
\special{pa 2600 1400}%
\special{fp}%
\special{sh 1}%
\special{pa 2600 1400}%
\special{pa 2668 1420}%
\special{pa 2654 1400}%
\special{pa 2668 1380}%
\special{pa 2600 1400}%
\special{fp}%
% DOT 1 0 3 1
% 3 5810 1400 2570 1400 2570 1400
% 
\special{pn 13}%
\special{sh 1}%
\special{ar 5810 1400 10 10 0  6.28318530717959E+0000}%
\special{sh 1}%
\special{ar 2570 1400 10 10 0  6.28318530717959E+0000}%
\special{sh 1}%
\special{ar 2570 1400 10 10 0  6.28318530717959E+0000}%
% VECTOR 1 0 3 2
% 2 5795 1945 4795 1945
% 
\special{pn 13}%
\special{pa 5796 1946}%
\special{pa 4796 1946}%
\special{fp}%
\special{sh 1}%
\special{pa 4796 1946}%
\special{pa 4862 1966}%
\special{pa 4848 1946}%
\special{pa 4862 1926}%
\special{pa 4796 1946}%
\special{fp}%
% VECTOR 1 1 3 3
% 2 4795 1945 2595 1945
% 
\special{pn 13}%
\special{pa 4796 1946}%
\special{pa 2596 1946}%
\special{da 0.070}%
\special{sh 1}%
\special{pa 2596 1946}%
\special{pa 2662 1966}%
\special{pa 2648 1946}%
\special{pa 2662 1926}%
\special{pa 2596 1946}%
\special{fp}%
% STR 2 0 3 4
% 3 4210 1070 4210 1170 5 0
% \large$\vec{L}(=\vec{x}-\vec{X})$
\put(42.1000,-11.7000){\makebox(0,0){\large$\vec{L}(=\vec{x}-\vec{X})$}}%
% STR 2 0 3 5
% 3 6295 1525 6295 1625 5 0
% \large$(\vec{X},t=0)$
\put(62.9500,-16.2500){\makebox(0,0){\large$(\vec{X},t=0)$}}%
% DOT 1 0 3 6
% 3 5805 1945 2565 1945 2565 1945
% 
\special{pn 13}%
\special{sh 1}%
\special{ar 5806 1946 10 10 0  6.28318530717959E+0000}%
\special{sh 1}%
\special{ar 2566 1946 10 10 0  6.28318530717959E+0000}%
\special{sh 1}%
\special{ar 2566 1946 10 10 0  6.28318530717959E+0000}%
% STR 2 0 3 7
% 3 2365 1525 2365 1625 5 0
% \large$(\vec{x},t)$
\put(23.6500,-16.2500){\makebox(0,0){\large$(\vec{x},t)$}}%
% STR 2 0 3 8
% 3 3685 1685 3685 1785 5 0
% \large$\nu_l$
\put(36.8500,-17.8500){\makebox(0,0){\large$\nu_l$}}%
% STR 2 0 3 9
% 3 5315 1695 5315 1795 5 0
% \large$\pi^+$
\put(53.1500,-17.9500){\makebox(0,0){\large$\pi^+$}}%
% STR 2 0 3 10
% 3 4825 1525 4825 1625 5 0
% \large$(\vec{y})$
\put(48.2500,-16.2500){\makebox(0,0){\large$(\vec{y})$}}%
% STR 2 0 3 11
% 3 5285 2045 5285 2145 5 0
% \large$\vev{\pi}y^0$
\put(52.8500,-21.4500){\makebox(0,0){\large$\vev{\pi}y^0$}}%
% STR 2 0 3 12
% 3 3425 2045 3425 2145 5 0
% \large$\vec{v}_l(\vec{K})\cdot (t-y^0)$
\put(34.2500,-21.4500){\makebox(0,0){\large$\vec{v}_l(\vec{K})\cdot (t-y^0)$}}%
\end{picture}%
}
 \end{center}
\caption{Classical Relations}
\label{fig11}
\end{figure}
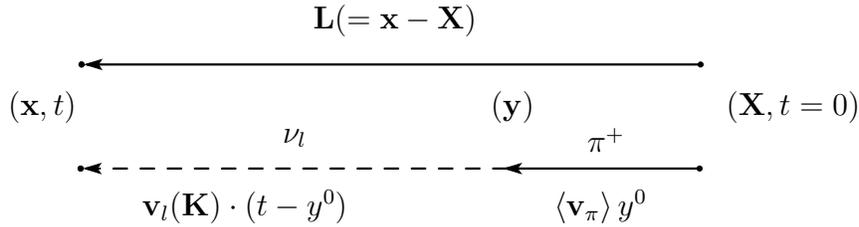

In R.H.S. of \eqref{3.41}, the $\vL$-dependent oscillation factor in
\eqref{3.35}, $\exp[i(\vec{k}-\vec{k}')\cdot\vL]$, disappears. The
$\vec{L}$-dependence is expected to be recovered due to $\vec{F}_j^2$-
and $\vec{F}_l^2$-terms through the time integration in \eqref{3.41}.
In order to see this possibility qualitatively, we evaluate roughly the
$Y^0$- and $Z^0$- integration in \eqref{3.41} under the classical
relations (as shown in FIG.\ref{fig11})
\begin{align}
 \vec{F}_l(\vec{K}; \vec{L}, t, Y^0)=\vec{F}_j(\vec{K}; \vec{L}, t,
 Z^0)=0\ ; \label{3.42}
\end{align}

Under the parallet condition, where the relevant momenta, $\ev{\vec{p}}$
and $\ev{\vec{K}}=\ev{\vec{p}}-\vec{q}$, are parallel to $\vL$, the
solutions $Y^0_c$ and $Z^0_c$ of \eqref{3.42} are given by 
\begin{align}
 Y^0_c&=\frac{v_l t-L}{v_l-\vev{\pi}}\quad ; \quad Z^0_c=Y^0_c(v_l \rightarrow
 v_j), \label{3.43}
\intertext{or}
t-Y^0_c&=\frac{L-\vev{\pi}t}{v_l-\vev{\pi}}=\frac{L-Y^0_c\vev{\pi}}{v_l},
 \qquad t-Z^0_c=t-Y^0_c(v_l \rightarrow v_j). \label{3.44}
\end{align}
By assuming $0<\{Y^0_c,Z^0_c\}<t$, and by taking the rough evaluation of
integrations 
\begin{align}
 \int^t_0\!dZ^0\int^t_0\!dY^0\cdots \simeq
 \int^t_0\!dZ^0\int^t_0\!dY^0\delta(Z^0-Z^0_c)\delta(Y^0-Y^0_c), \nn
\end{align}
we obtain
\begin{align}
 \eqref{3.41} &\simeq
 \frac{-|f_{\pi\sigma}|^2}{\sqrt{2\pi}^9}(2\sigma_\pi)^3
 \sum_{j,l}Z^{1/2}_{\sigma j}{Z^{1/2}_{\rho j}}^\ast Z^{1/2}_{\rho
 l}{Z^{1/2}_{\sigma l}}^\ast \int\!\!d^3q
 B^{(1)a}_{\sigma,jl}(q;\ev{\vec{p}},\ev{\vec{p}};\vec{K},\vec{K}) \nn \\
 &\times
 \exp\big[(\mathcal{P}_{ji})_{\text{clas}}\big]\bigg|_{\vec{K}=\ev{\vec{p}}-\vec{q}}, \label{3.45}
\intertext{where}
(\mathcal{P}_{ji})_{\text{clas}}&\equiv
 i(\omega_j(\vec{K})-\omega_l(\vec{K}))t-Z^0_c(i\delta
 E_j+\frac{\plt}{2})+Y^0_c(i\delta E_l-\frac{\plt}{2}), \label{3.46}
\end{align}
with $\delta E_j=E_\sigma(\vec{q})+\omega_j(\vec{K})-\ev{E_\pi}$. We
obtain easily
\begin{align}
 Im(\mathcal{P}_ji)_{\text{clas}}=(L-\vev{\pi}t)\frac{v_j-v_l}{(v_j-\vev{\pi})(v_l-\vev{\pi})}
 \left[ -\frac{v_j+v_l-\vev{\pi}}{v_j v_l}K+\ev{E_\pi}-E_\sigma
 \right]. \label{3.47}
\end{align}
This unfamiliar form suggests a more precise evaluation of the
integrations in \eqref{3.41} is necessary, which we will consider as
follows

\subsubsection{$Z^0$- and $Y^0$-integrations}
For simplicity, we write $\vec{F}_j(\vec{K};\vec{L},t,Z^0)$ as
$\vec{F}_j(Z^0)$, and define $\Zd$ by
\begin{align}
\left[\frac{d}{dZ^0}\left(\vec{F}_j(Z^0)\right)^2\right]_{\Zd}=0. \label{3.48}
\end{align}
Then we obtain
\begin{align}
\Zd&=\frac{1}{(\vec{v}_j-\vev{\pi})^2} 
 (\vec{v}_j-\vev{\pi})\cdot(\vec{v}_jt-\vL) \hspace{2cm} \nn \\
&=2\sg_{\pi j}\sg_\pi(\vec{v}_j-\vev{\pi})\cdot(\vec{v}_jt-\vL),
 \label{3.49} \\
\sg_\pi[\vec{F}_j(Z^0)]^2&=\frac{1}{2\sg_{\pi j}}(Z^0-\Zd)^2 +
 \sg_\pi[\vec{F}_j(\Zd)]^2, \label{3.50}
\intertext{where}
\sg_\pi(\vec{v}_j-\vev{\pi})^2&=\frac{1}{2\sg_{\pi j}}. \label{3.51}
\end{align}
\eqref{3.41} is rewritten as 
\begin{align}
&\mathcal{E}(\vL,t;\ev{\vec{p}})^a_{\text{(I-1)}\rho}
 \equiv\mathcal{E}(\vec{x},x^0,\ev{\vec{p}},\vec{X},x^0_I)^a_{\text{(I-1)}\rho} \nn \\
&\simeq \frac{-|f_{\pi \sigma}|^2}{2\ev{E_\pi}}
 \frac{(2\sigma_\pi)^3}{(\sqrt{2\pi})^9} 
 \int\!\!d^3q 
 \sum_{j,i} Z^{1/2}_{\sigma j}{Z^{1/2}_{\rho j}}^\ast Z^{1/2}_{\rho i}{Z^{1/2}_{\sigma i}}^\ast 
 B_{\sigma, ji}^{(1)a}(q ; \ev{\vec{p}},\ev{\vec{p}} ; \vec{K},\vec{K})
 \nn \\
&\times \exp\!\bigg[i(\omega_j(\vec{K})-\omega_i(\vec{K}))t
 -\sigma^2_\pi \vec{F}_j(\vec{K;\vec{L},t,\Zd})^2 
 -\sigma^2_\pi \vec{F}_i(\vec{K;\vec{L},t,\Yd})^2 \nn \\
&\hspace{1cm}
 -\Zd(i\delta E_j+\frac{\plt}{2}) 
 +\Yd(i\delta E_i-\frac{\plt}{2}) \bigg] \nn \\
&\times M_j(t;\Zd)\cdot M_i(t;\Yd)^\ast, \label{3.52}
\end{align}
where
\begin{align}
M_j(t;\Zd) \equiv 
 \int^t_0\!dZ^0 \exp\left[-\frac{1}{2\sg_{\pi
 j}}(Z^0-\Zd)^2-(Z^0-\Zd)(i\delta E_j+\frac{\plt}{2}) \right]. \label{3.53}
\end{align}
The integration form of \eqref{3.53} is expressed as
\begin{align}
M_j(t,\Zd) &=\sqrt{2 \sg_{\pi j}}
\exp\left[\frac{1}{2}\sg_{\pi
 j}(i\delta E_j+\frac{\plt}{2})^2
 \right]\ J_j(u_j(t);\alpha_j,\beta_j),\label{3.54}
\intertext{where}
\alpha_j&=\frac{\sigma_{\pi j}\delta E_j}{\sqrt{2}},\qquad 
\beta_j=\frac{1}{\sqrt{2}\sigma_{\pi j}}(-\Zd+\frac{\sg_{\pi j}}{2}\plt),\nn \\
J_j(u_j(t);\alpha_j,\beta_j)&=\int^{u_j(t)}_0 du_j 
\exp[-(u_j+i\alpha+\beta_j)^2],\label{3.55}
\end{align}
with $u_j=Z^0/(\sqrt{2}\sigma_{\pi j})$ and $u_j(t)=t/(\sqrt{2}\sigma_{\pi j})$.
We define the two functions as
\begin{align}
f_c(y;\alpha,\beta)&=\int^y_0 du \exp[\alpha^2-(u+\beta)^2]\cos(2\alpha(u+\beta)),\nn \\
f_s(y;\alpha,\beta)&=\int^y_0 du \exp[\alpha^2-(u+\beta)^2]\sin(2\alpha(u+\beta)); \label{3.56}
\end{align}
then, \eqref{3.54} is expressed as
\begin{align}
M_j(t,\Zd)&=\sqrt{2\sg_{\pi j}}\exp\left[\frac{1}{2}\sg_{\pi j}(i\delta E_j+\frac{\plt}{2})^2\right]
 |J_j(u_j(t);\alpha_j,\beta_j)| \exp[-i\Theta(u_j(t);\alpha_j,\beta_j)],\label{3.57}
\intertext{where}
\Theta(y;\alpha,\beta)&\equiv 
\arctan\left[\frac{f_s(y;\alpha,\beta)}{f_c(y;\alpha,\beta)}\right].\label{3.58}
\end{align}
Thus, \eqref{3.52} is expressed as
\begin{align}
\mathcal{E}(\vec{L},t,\ev{\vec{p}})^a_{(I-1)}&\simeq
 -\frac{|f_{\pi \sigma}|}{2\ev{E_\pi}}\frac{(2\sigma_\pi)^3}{(\sqrt{2\pi})^9}
 \int dq^3 \sum_{j,i}Z^{1/2}_{\sigma j}{Z^{1/2}_{\rho j}}^\ast Z^{1/2}_{\rho i}{Z^{1/2}_{\sigma i}}^\ast \nn\\
&\times B^{(1)a}_{\sigma,j i}(q;\ev{\vec{p}},\ev{\vec{p}};\vec{K},\vec{K})\
 (2\sigma_{\pi j}\sigma_{\pi i})\ \exp\left[\mathcal{P}_{ji}+
 \frac{1}{2}\sg_{\pi j}(i\delta E_j+\frac{\plt}{2})^2
 +\frac{1}{2}\sg_{\pi i}(-i\delta E_i+\frac{\plt}{2})^2\right] \nn\\
&\times |J_j(u_j(t);\alpha_j,\beta_j)||J_i(u_i(t);\alpha_i,\beta_i)|\
 \exp\bigg[-i\big(\Theta(u_j(t;\alpha_j,\beta_j))-\Theta(j\rightarrow
 i)\big)\bigg]\bigg|_{\vec{K}=\ev{\vec{p}}-\vec{q}},
 \label{3.59}
\intertext{where}
\mathcal{P}_{ji}&\equiv i(\omega_j(\vec{K})-\omega_i(\vec{K}))t
 -\sg_\pi\big[\f_j(\Zd)\big]^2-\sg_\pi\big[\f_i(\Yd)\big]^2\nn\\
&\qquad -\Zd(i\delta E_j+\frac{\plt}{2})-\Yd(-i\delta E_i+\frac{\plt}{2}),\label{3.60}\\
\big[\f_j(\Zd)\big]^2&=\frac{1}{(\vv_j-\vev{\pi})^2}
 \big[(\vv_j-\vev{\pi})^2(\vv_jt-\vL)^2-\{(\vv_j-\vev{\pi})\cdot(\vv_jt-\vL)\}^2\big] \label{3.61}
\end{align}
with $\vv_j=\vec{K}/\sqrt{\vec{K}^2+m_j^2}, \vec{K}=\ev{\vec{p}}-\vec{q}$.

Though the $t$- and $\vL$-dependent part \eqref{3.59} is complicated, it becomes
a little simpler when the parallel condition is assumed. From 
$ [\vec{F}_j(\Zd)^2+\vec{F}_i(\Yd)^2]_{\text{para}}=0$, which leads to the classical 
relation
\begin{align}
[\Zd]_{\text{para}}=\frac{(v_jt-L)}{v_j-\ev{v_\pi}},\label{3.62}
\end{align}
we obtain $\im[\mathcal{P}_{ji}]_{\text{para}}$, 
\begin{align}
\im[\mathcal{P}_{ji}]_{\text{para}}=(L-\ev{v_\pi}t)\frac{v_j-v_i}{(v_j-\ev{v_\pi})(v_i-\ev{v_\pi})}
 \bigg[-\frac{v_j+v_i-\ev{v_\pi}}{v_jv_i}K+\ev{E_\pi}-E_\sigma \bigg],\label{3.63}
\end{align}
which is the same as \eqref{3.47}.

We can examine the qualitative structure of \eqref{3.59} under the condition of neutrino 
high energies in comparison with $m_j$'s. This is to be explained in the last section Sec.V.

%% section 4 %%
\section{EXPECTATION VALUE OF NEUTRINO CURRENT WITH AN ADDITIONAL OBSERVED PARTICLE}
As pointed out in Sec.III, it is worthwhile for us to examine the structure of 
the expectation value of the flavor current, where the charged lepton observation is 
taken into consideration.
\subsection{Boson model;  the plane wave case}

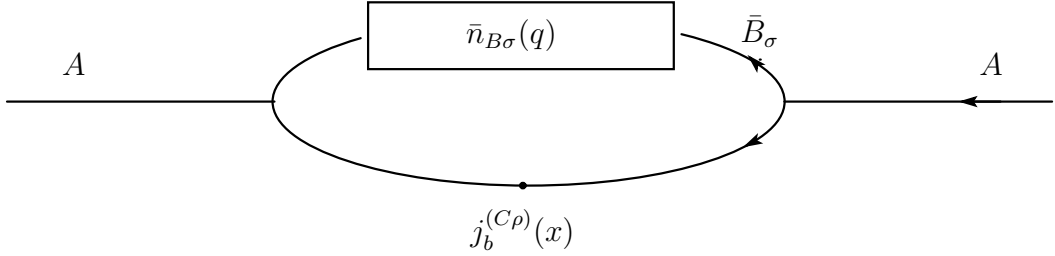
\begin{figure}[h]
 \begin{center}
  \scalebox{1}[1]{%WinTpicVersion3.08
\unitlength 0.1in
\begin{picture}( 56.1000, 12.7500)(  7.4500,-20.3)
% ELLIPSE 1 0 3 0
% 4 3530 1400 2190 1840 2340 950 4460 990
% 
\special{pn 13}%
\special{ar 3530 1400 1340 440  5.3524223 6.2831853}%
\special{ar 3530 1400 1340 440  0.0000000 3.9971878}%
% BOX 1 0 3 1
% 2 2690 880 4290 1230
% 
\special{pn 13}%
\special{pa 2690 880}%
\special{pa 4290 880}%
\special{pa 4290 1230}%
\special{pa 2690 1230}%
\special{pa 2690 880}%
\special{fp}%
% STR 2 0 3 2
% 3 3440 950 3440 1050 5 0
% \large$\bar{n}_{B\sigma}(q)$
\put(34.4000,-10.5000){\makebox(0,0){\large$\bar{n}_{B\sigma}(q)$}}%
% LINE 1 0 3 3
% 2 2200 1400 800 1400
% 
\special{pn 13}%
\special{pa 2200 1400}%
\special{pa 800 1400}%
\special{fp}%
% LINE 1 0 3 4
% 2 6270 1400 4870 1400
% 
\special{pn 13}%
\special{pa 6270 1400}%
\special{pa 4870 1400}%
\special{fp}%
% DOT 0 0 3 5
% 1 3500 1840
% 
\special{pn 20}%
\special{sh 1}%
\special{ar 3500 1840 10 10 0  6.28318530717959E+0000}%
% VECTOR 1 0 3 6
% 2 6000 1400 5800 1400
% 
\special{pn 13}%
\special{pa 6000 1400}%
\special{pa 5800 1400}%
\special{fp}%
\special{sh 1}%
\special{pa 5800 1400}%
\special{pa 5868 1420}%
\special{pa 5854 1400}%
\special{pa 5868 1380}%
\special{pa 5800 1400}%
\special{fp}%
% STR 2 0 3 7
% 3 5950 1100 5950 1200 5 0
% \large$A$
\put(59.5000,-12.0000){\makebox(0,0){\large$A$}}%
% STR 2 0 3 8
% 3 1150 1100 1150 1200 5 0
% \large$A$
\put(11.5000,-12.0000){\makebox(0,0){\large$A$}}%
% VECTOR 1 0 3 9
% 4 4710 1190 4690 1170 4740 1180 4740 1180
% 
\special{pn 13}%
\special{pa 4710 1190}%
\special{pa 4690 1170}%
\special{fp}%
\special{sh 1}%
\special{pa 4690 1170}%
\special{pa 4724 1232}%
\special{pa 4728 1208}%
\special{pa 4752 1204}%
\special{pa 4690 1170}%
\special{fp}%
\special{pa 4740 1180}%
\special{pa 4740 1180}%
\special{fp}%
% VECTOR 1 0 3 10
% 2 4690 1620 4670 1630
% 
\special{pn 13}%
\special{pa 4690 1620}%
\special{pa 4670 1630}%
\special{fp}%
\special{sh 1}%
\special{pa 4670 1630}%
\special{pa 4740 1618}%
\special{pa 4718 1606}%
\special{pa 4722 1582}%
\special{pa 4670 1630}%
\special{fp}%
% STR 2 0 3 11
% 3 3500 1970 3500 2070 5 0
% \large$j_b^{(C\rho)}(x)$
\put(35.0000,-20.7000){\makebox(0,0){\large$j_b^{(C\rho)}(x)$}}%
% STR 2 0 3 12
% 3 4740 930 4740 1030 5 0
% \large$\bar{B}_\sigma$
\put(47.4000,-10.3000){\makebox(0,0){\large$\bar{B}_\sigma$}}%
\end{picture}%
}
 \end{center}
\caption{Diagram reprenting \eqref{4.7}}
\label{fig12}
\end{figure}

We first consider the boson model in the plane wave case with the aim of seeing the 
way how to take into account the observation of an additional particle.    We examine the 
same model as that in Sec.IIB ;
\begin{align}
\mathcal{H}_{\text{int}}&=-f_{ABC\sigma} \phi^\dagger_{C\sigma}(x)
 \phi_{B\sigma}(x) \phi_A(x)+H.c. \nn \\
&=\mathcal{H}_1(x)+\mathcal{H}_2(x), \label{4.1} \\
\mathcal{H}_1(x)&=\phi^\dagger_{C\sigma}(x)J_\sigma(x),\quad
 \mathcal{H}_2(x)=\mathcal{H}^\dagger_1(x)=J^\dagger_\sigma(x)\phi_{C\sigma}(x),
 \label{4.2} \\
j^{(C\rho)}_b(x)&=i\phi^\dagger_{C\rho}(x)\olra{\partial}_b\phi_{C\rho}(x). \label{4.3}
\end{align}
From the expectation value (\ref{2.28})，we pick up the part，
corresponding to FIG.\ref{fig7}(1), i.e.
\begin{align}
 \mathcal{E}(A(x^0_I);C_\rho\text{-cur(x)})^{\text{(I-1)}}_b &=
 \int^{x^0}_{x^0_I}\!\!d^4z\!\int^{x^0}_{x^0_I}\!\!d^4y \
 \bra{0}\alpha_A(\vec{p})H_2(z) :j^{(C\rho)}_b(x):
 H_1(x)\alpha^\dagger_A(\vec{p}) \ket{0}^{\text{con}} \nn \\
&=\int^{x^0}_{x^0_I}\!\!d^4z\!\int^{x^0}_{x^0_I}\!\!d^4y \
 \bra{0}\alpha_A(\vec{p}) J^\dagger_\sigma(z)\cdot 1 \cdot J_\sigma(y)
 \alpha_A^\dagger(\vec{p}) \ket{0} \nn \\
&\qquad \times \bra{0} \phi_{C\sigma}(z) :j^{(C\rho)}_b(x):
 \phi^\dagger_{C\sigma}(y) \ket{0}^{\text{con}}. \label{4.4}
\end{align}
The first part in the last integrand includes the propagator
$\bra{0}\phi^\dagger_{B\sigma}(z) \phi_{B\sigma}(y)\ket{0}$.  
This can be effectively set equal to
\begin{align}
\bra{0}\phi^\dagger_{B\sigma}(z) \phi_{B\sigma}(y)\ket{0}&=
 \bra{0}\phi^\dagger_{B\sigma}(z) \int\!\frac{d^3q}{2E_B(\vec{q})}
 \beta^\dagger_{B\sigma}(\vec{q}) \ket{0}
 \bra{0} \beta_{B\sigma}(\vec{q})\phi_{B\sigma}(y)\ket{0} \nn \\
&=\bra{0}\phi^\dagger_{B\sigma}(z)
 \left[-\!\int\!\!\bar{n}_{B\sigma}(q)d^3q\right]
 \phi_{B\sigma}(y)\ket{0}^{\text{con}} \label{4.5}
\end{align}
with
$-\bar{n}_{B\sigma}(q)=\beta^\dagger_{B\sigma}(\vec{q})\beta_{B\sigma}(\vec{q})
/(2E_B(\vec{q}))$，as seen from
\begin{align}
\bra{0}\beta_{B\sigma}(\vec{Q}')
 \left[-\!\int\!\!\bar{n}_{B\sigma}(q)d^3q\right]
 \beta^\dagger_{B\sigma}(\vec{Q})\ket{0}^{\text{con}}&= 
 \int\!\!\frac{d^3q}{2E_B(\vec{q})}2E_B(\vec{Q}')\delta(\vec{Q}'-\vec{q})
 2E_B(\vec{Q})\delta(\vec{Q}-\vec{q}) \nn \\
&=\bra{0}\beta_{B\sigma}(\vec{Q}')\beta^\dagger_{B\sigma}(\vec{Q})\ket{0}. \label{4.6}
\end{align}
Thus，we can regard the expectation value of the flavor current with the 
additional particle observation is given by 
\begin{align}
\mathcal{E}(A(x^0_I);C_\rho\text{-Cur}(x),B_\sigma(\vec{q})\text{-obs})^{(I-1)}_b
 &\equiv \int^{x^0}_{x^0_I}\!\int^{x^0}_{x^0_I}\!d^4zd^4y
 \bra{0}\alpha_A(\vec{p})J^\dagger_\sigma(z)(-\bar{n}_{B\sigma}(q))
 J_\sigma(y)\alpha^\dagger_A(\vec{p})\ket{0}^{\text{con}} \nn \\
&\times \bra{0}\phi_{C\sigma}(z) :j^{(C\rho)}_b(x):
 \phi^\dagger_{C\sigma}(y)\ket{0}^{\text{con}}. \label{4.7}
\end{align}
(See FIG.\ref{fig12}.) 

The concrete form of (\ref{4.7}) is given by 
\begin{align}
&\mathcal{E}(A(x^0_I);C_\rho\text{-Cur}(x),B_\sigma(\vec{q})\text{-obs})^{(I-1)}_b
 \nn \\
&=\int^{x^0}_{x^0_I}\!\int^{x^0}_{x^0_I}d^4zd^4y|f_{ABC\sigma}|^2\frac{1}{(2\pi)^6
 2E_B({\vec{q}})}
 \times \sum_{j,l} Z^{1/2}_{\sigma j} {Z^{1/2}_{\rho j}}^\ast 
 Z^{1/2}_{\rho j}{Z^{1/2}_{\sigma l}}^\ast e^{iz(-p+q)+iy(-q+p)} \nn \\
&\times \int\!\!\int\!\!d^3k'\!d^3k
 \frac{i^2(k_b'+k_b)e^{i(k-k')x+ik'z-iky}}{(2\pi)^6
 2\omega_j(\vec{k}')2\omega_l(\vec{k})} \nn\\
&=\int^t_0\!\!\int^t_0\!dz^0dy^0 \frac{|f_{ABC\sigma}|^2}{(2\pi)^6
 2E_B(\vec{q})} \ 
 \sum_{j,l} Z^{1/2}_{\sigma j} {Z^{1/2}_{\rho j}}^\ast 
 Z^{1/2}_{\rho j}{Z^{1/2}_{\sigma l}}^\ast \
 e^{i(\omega_j(\vec{k})-\omega_l(\vec{k}))t}
 \frac{i^2}{2\omega_j(\vec{k})2\omega_l(\vec{k})} \nn \\
&\times 
 \begin{bmatrix}
  2\vec{k} \\
  i(\omega_j(\vec{k})+\omega_l(\vec{k}))
 \end{bmatrix} \ 
 e^{i(E_A(\vec{p})-E_B(\vec{q})-\omega_j(\vec{k}))z^0
  -i(E_A(\vec{p})-E_B(\vec{q})-\omega_l(\vec{k}))y^0}\bigg|_{\vec{k}=\vec{p}-\vec{q}}, 
 \label{4.8}
\end{align}
where $t = x^0-x^0_I$. We see (\ref{4.8}) leads certainly to 
\begin{align}
\int\!\!d^3x\!\int\!\!d^3k \ 
 i\mathcal{E}(A(x^0_I);C_\rho\text{-Cur}(x),B_\sigma(\vec{q})\text{-obs})^{(I-1)}_4
=\ev{N_{C\rho}(x^0)^{(1)}}_{A(\vec{p}),x^0_I}, \label{4.9}
\end{align}
given by (\ref{2.35}). 

Due to the plane-wave calculation, the $\vec{x}$-dependence disappears in (\ref{4.8}); 
thus, there is no spacial coordinates specifying the observation. 
In the next subsection we examine the model calculation by employing the
wave packets for the external particles 
( i.e. for $\pi^+$ and the "observed" $B_\sigma$-particle). 

\subsection{Boson model; the wave-packet case}
First we have to take out the part, which is suitable for the present purpose, of 
\begin{align}
 \bra{\Psi_A(\ev{\vec{p}};\vec{X}_A,x^0_I)}\int^{x^0}_{x^0_I}\!\!\int^{x^0}_{x^0_I}\!d^4zd^4y 
 \mathcal{H}_{\text{int}}(z) :j^{(C\rho)}_b(x): \mathcal{H}_{\text{int}} 
 \ket{\Psi_A(\ev{\vec{p}};\vec{X}_A,x^0_I)}^{\text{con}}. \label{4.10}
\end{align}
Here, we choose the wave-packet state in the same way as (\ref{3.1}); 
\begin{align}
\ket{\Psi_A(\ev{\vec{p}};\vec{X}_A,x^0_I)}&=
 \int\!\!d^3p A_A^\ast(\vec{p},\ev{\vec{p}};\vec{X}_A,x^0_I)
 \frac{1}{\sqrt{2E_A(\vec{p})}} \alpha^\dagger_A(\vec{P}) \ket{0},
 \label{4.11} \\
A_A^\ast(\vec{p},\ev{\vec{p}};\vec{X}_A,x^0_I)&=\frac{1}{(\sqrt{2\pi}\sigma_A)^{3/2}}
 \exp\left[ -\frac{(\vec{p}-\ev{\vec{p}})^2}{4\sigma_A^2}
 -i\vec{p}\cdot\vec{X}_A +iE_A(\vec{p})x^0_I\right]. \label{4.12}
\end{align}
This state is normalized in the way as noted by (\ref{3.3}). (\ref{4.11}) is regarded 
as the state obtained through substituting the wave-packet state for 
the plane-wave state
$\frac{1}{\sqrt{2E_A(\ev{\vec{p}})}}\alpha^\dagger_A(\vec{p})\ket{0}$.
In the same way, the wave-packet of $\bar{B}_\sigma$ incorporated in (\ref{4.5}) through the substitution 
\begin{align}
\frac{1}{2E_B(\ev{\vec{q}})}\beta^\dagger_{B\sigma}(\ev{\vec{q}})\ket{0}
 \bra{0}\beta_{B\sigma}(\ev{\vec{q}}) \longrightarrow
 \ket{\Psi_B(\ev{\vec{q}}; \vec{X}_B,x^0_B)}\bra{\Psi_B(\ev{\vec{q}};
 \vec{X}_B,x^0_B)} \label{4.13}
\intertext{with}
\ket{\Psi_B(\ev{\vec{q}}; \vec{X}_B,x^0_B)}=\int\!\!d^3q
 A^\ast_B(\vec{q},\ev{\vec{q}};\vec{X}_B,x^0) \frac{\beta^\dagger_{B\sigma}(\vec{q})}{\sqrt{2E_B(\vec{q})}}\ket{0},\label{4.14}
\end{align}
where $A^\ast_B$ is defined similarly to (\ref{4.12}).

Thus, we are led to examine 
\begin{align}
&\mathcal{E}[A(\ev{\vec{p}};\vec{X}_A,x^0_I);C_\rho\text{-Cur}(x),
\bar{B}_\sigma(\ev{\vec{q}};\vec{X}_B,x^0_B)]^{(I-1)}_b \nn \\
&\equiv \int^{x^0}_{x^0_I}\!\!\int^{x^0}_{x^0_I}\!d^4zd^4y |f_{ABC\sigma}|^2
 \bra{\Psi_A(\ev{\vec{p}};\vec{X}_A,x^0_I)}
 \phi^\dagger_A(z)\phi_A(y)\ket{\Psi_A(\ev{\vec{p}};\vec{X}_A,x^0_I)}^{\text{con}} \nn \\
&\times\bra{0}\phi^\dagger_{B\sigma}(z)\ket{\Psi_B(\ev{\vec{q}};\vec{X}_B,x^0_B)}
 \bra{\Psi_B(\ev{\vec{q}};\vec{X}_B,x^0_B)}\phi_{B\sigma}(y)\ket{0} \nn \\
&\times\bra{0}\phi_{C\sigma}(z) :j_b^{(C\rho)}(x):
 \phi^\dagger_{C\sigma}(y)\ket{0}^{\text{con}}. \label{4.15} 
\end{align}
We add a remark on the choice of (\ref{4.13}). As noted above the state 
normalization is changed from
$\bra{0}\alpha_A(\vec{p}')\alpha^\dagger(\vec{p})\ket{0}= 
2E_A(\vec{p})\delta(\vec{p}'-\vec{p})$ to 
\begin{align}
\bracket{\Psi_A(\ev{\vec{p}'};\vec{X}_A,x^0_I)}{\Psi_A(\ev{\vec{p}};\vec{X}_A,x^0_I)} 
=\exp\left[ -\frac{(\ev{\vec{p}'}-\ev{\vec{p}})^2}{8\sigma_A^2} \right]: \label{4.16}
\end{align}

In the same way, we adopt (\ref{4.13}) which satisfies 
\begin{align}
&\bra{0}\beta_{B\sigma}(\vec{Q}') 
\int\!d^3\ev{q} \ket{\Psi_{B\sigma}(\ev{\vec{q}};\vec{X}_B,x^0_B)}
\bra{\Psi_{B\sigma}(\ev{\vec{q}};\vec{X}_B,x^0_B)}
\beta^\dagger_{B\sigma}(\vec{Q})\ket{0} \nn \\
&=\int\!d^3\ev{q}\sqrt{2E_B(\vec{Q}')2E_B(\vec{Q})}
A^\ast_B(\vec{Q}',\ev{\vec{q}})A_B(\vec{Q},\ev{\vec{q}}) \nn \\
&=\sqrt{2E_B(\vec{Q}')2E_B(\vec{Q})}
\exp\left[-\frac{(\ev{\vec{Q}'}-\ev{\vec{Q}})^2}{8\sigma_B^2}\right], \label{4.17}
\end{align}
instead of adopting
\begin{align}
\int\!d^3q A^\ast_B(\vec{q},\ev{\vec{q}})A_B(\vec{q},\ev{\vec{q}})\
\frac{1}{2E_B(\vec{q})}\
\beta^\dagger_{B\sigma}(\vec{q})\ket{0}\bra{0}\beta_{B\sigma}(\vec{q}). \label{4.18}
\end{align}
(\ref{4.18}) satisfies 
\begin{align}
\bra{0}\beta_{B\sigma}(\vec{Q}')\left(\int\!d^3\ev{q}\eqref{4.18}\right)\beta^\dagger_{B\sigma}(\vec{Q})\ket{0}
&=2E_B(\vec{Q})\delta(\vec{Q}'-\vec{Q}), \label{4.19}
\end{align}
which corresponds to the plane-wave normalization.

We obtain from (\ref{4.15}) 
\begin{align}
&\mathcal{E}[A(\ev{\vec{p}};\vec{X}_A,x^0_I);C_\rho\text{-Cur}(x),
\bar{B}_\sigma(\ev{\vec{q}};\vec{X}_B,x^0_B)]^{(I-1)}_b \nn \\
&= \int^{x^0}_{x^0_I}\!\!\int^{x^0}_{x^0_I}\!d^4zd^4y |f_{ABC\sigma}|^2
 \int\!d^3p'\!\int\!d^3p\!\int\!d^3q'\!\int\!d^3q\!\int\!d^3k'\!\int\!d^3k
 \frac{1}{(2\pi)^6\sqrt{2E_A(\vec{p}')2E_A(\vec{p})2E_B(\vec{q}')2E_B(\vec{q})}}\nn \\
&\times\sum_{j,i}Z^{1/2}_{\sigma j}{Z^{1/2}_{\rho j}}^\ast 
 Z^{1/2}_{\rho i} {Z^{1/2}_{\sigma i}}^\ast
 A_A(\vec{p}',\ev{\vec{p}})A_A^\ast(\vec{p},\ev{\vec{p}})
 A_B^\ast(\vec{q}',\ev{\vec{q}})A_B(\vec{q},\ev{\vec{q}}) \nn \\
&\times\frac{i^2(k'_j+k_i)_b}{(2\pi)^6 2\omega_j(\vec{k}')2\omega_i(\vec{k})} \
 e^{i(-\vec{p}'+\vec{q}'+\vec{k}')\cdot\vec{z}+i(\vec{p}-\vec{q}-\vec{k})\cdot\vec{y}}
 e^{i(\vec{p}'-\vec{p})\cdot\vec{X}_A-i(\vec{q}'-\vec{q})\cdot\vec{X}_B 
 +i(-\vec{k}'+\vec{k})\cdot\vec{x}} \nn \\
&\times\exp\left[i(E_A(\vec{p}')-E_B(\vec{q}')-\omega_j(\vec{k}'))(z^0-x^0_I)
 +i(-E_A(\vec{p})+E_B(\vec{q})+\omega_i(\vec{k}))(y^0-x^0_I) \right. \nn \\
&\hspace{2cm}\left. +i(\omega_j(\vec{k}')-\omega_i(\vec{k}))(x^0-x^0_I)
 +i(E_B(\vec{q}')-E_B(\vec{q}))(x^0_B-x^0_I) \right]. \label{4.20}
\end{align}

We perform the integrations with respect to $\vec{p}', \vec{p},
\vec{q}'$ and $\vec{q}$ by employing the approximate relations as 
already used in Sec.III (e.g.(\ref{3.29}) ). and obtain 
\begin{align}
&\mathcal{E}[A(\ev{\vec{p}};\vec{X}_A,x^0_I);C_\rho\text{-Cur}(x),
\bar{B}_\sigma(\ev{\vec{q}};\vec{X}_B,x^0_B)]^{(I-1)}_b \nn \\
&\simeq\int^t_0\!\!\int^t_0\!dZ^0dY^0\!\int\!d^3z\!\int\!d^3y\!\int\!d^3k'\!\int\!d^3k\ 
 \frac{(2\sqrt{2\pi}\sigma_A)^3(2\sqrt{2\pi}\sigma_B)^3}{(2\pi)^6
 2\ev{E_A} 2\ev{E_B}}\ |f_{ABC\sigma}|^2 \nn \\
&\times\sum_{j,i}Z^{1/2}_{\sigma j}{Z^{1/2}_{\rho j}}^\ast 
 Z^{1/2}_{\rho i} {Z^{1/2}_{\sigma i}}^\ast\ 
 \frac{i^2(k'_j+k_i)_b}{(2\pi)^6 2\omega_j(\vec{k}')2\omega_i(\vec{k})} \
 e^{i(-\vec{k}'+\vec{k})\cdot\vec{x}+i(\omega_j(\vec{k}')-\omega_i(\vec{k}))t} \nn \\
&\times\exp\left[i\ev{\vec{p}}\cdot(\vec{y}-\vec{X}_A)+(-i\ev{E_A}-\ev{\Gamma_A}/2)Y^0
 +i\ev{\vec{p}}\cdot(-\vec{z}+\vec{X}_A)+(i\ev{E_A}-\ev{\Gamma_A}/2)Z^0
 \right. \nn \\
&\hspace{1cm}+i\ev{\vec{q}}\cdot(-\vec{y}+\vec{X}_B)+i\ev{E_B}(Y^0-X^0_B)
 +i\ev{\vec{q}}\cdot(\vec{z}-\vec{X}_B)-i\ev{E_B}(Z^0-X^0_B) \nn \\
&\hspace{1cm}\left. -h(\vec{y},Y^0)-h(\vec{z},Z^0)
 +i(-\omega_j(\vec{k}')Z^0+\omega_i(\vec{k})Y^0
 +i(\vec{k}'\cdot\vec{z}-\vec{k}\cdot\vec{y}))\right], \label{4.21}
\end{align}
where the weak-decay width of A-particle is added by
hand; $t=x^0-x^0_I,Y^0=y^0-x^0_I$, $Z^0=z^0-x^0_I, X^0_B=x^0_B-x^0_I$;
\begin{align}
h(\vec{y},Y^0)=\sigma^2_A(\vec{y}-\vec{X}_A-\ev{\vec{v}_A}Y^0)^2
 +\sigma^2_B(\vec{y}-\vec{X}_B-\ev{\vec{v}_B}(Y^0-X^0_B))^2, \label{4.22}
\end{align}

With the aim of execute the $\vec{z}$- and $\vec{y}$-integrations, 
we first examine the structure of 
\begin{align}
\exp\left[-h(\vec{z},Z^0)+i(-\ev{\vec{p}}+\ev{\vec{q}}+\vec{k}')\cdot\vec{z}\right],\label{4.23}
\end{align}
which is included in the integrand of (\ref{4.21}). The concrete form of
$\vl{z}$, defined by 
\begin{align}
\left.\del{h(\vec{z},Z^0)}{\vec{z}}\right|_{\vl{z}}=0, \label{4.24}
\end{align}
is given by 
\begin{align}
\vl{z}=\frac{1}{\sigma_A^2+\sigma_B^2}
 \left[ \sigma^2_A(\vec{X}_A+\ev{\vec{v}_A}Z^0)
 +\sigma^2_B(\vec{X}_B-\ev{\vec{v}_B}(X^0_B-Z^0)) \right]; \label{4.25}
\end{align}
then, we obtain
\begin{align}
h(\vl{z},Z^0)=\frac{\sigma_A^2\sigma_B^2}{\sigma_A^2+\sigma_B^2}
 [\vec{F}^{AB}(Z^0,X^0_B)]^2, \label{4.26}
\intertext{with}
\vec{F}^{AB}(Z^0,X^0_B)
\equiv\vec{X}_B-\ev{\vec{v}_B}(X^0_B-Z^0)-\ev{\vec{v}_A}Z^0-\vec{X}_A, \label{4.27}
\intertext{and}
h(\vec{z},Z^0)=(\sigma_A^2+\sigma_B^2)(\vec{z}-\vl{z})^2+h(\vl{z},Z^0). \label{4.28}
\end{align}
We see $\exp[-h(\vec{z},Z^0)]$ is maximum at $\vec{z}=\vl{z}$, and 
\begin{align}
h(\vl{z},Z^0)=0 \leftrightarrow \vec{F}^{AB}(Z^0,X^0_B)=0. \label{4.29}
\end{align}
$\vec{F}^{AB}(Z^0,X^0_B)$ leads to the relation among classical
trajectories.
%as shown in FIG.\ref{fig13}. 
%\begin{figure}[h]
% \begin{center}
%  \scalebox{1}[1]{\input{fig13.tex}}
% \end{center}
%\caption{}
%\label{fig13}
%\end{figure}

By noting 
\begin{align}
&\exp\left[-(\sigma_A^2+\sigma_B^2)(\vec{z}-\vl{z})^2 
 +i(-\ev{\vec{p}}+\ev{\vec{q}}+\vec{k})\cdot\vec{z} \right] \nn \\
&=\exp\left[ -(\sigma_A^2+\sigma_B^2) \left\{\vec{z}-\vl{z} 
 -\frac{i}{2(\sigma_A^2+\sigma_B^2)}(-\ev{\vec{p}}+\ev{\vec{q}}+\vec{k})
 \right\}^2 \right. \nn \\ 
&\hspace{1cm} 
 \left.-\frac{(-\ev{\vec{p}}+\ev{\vec{q}}+\vec{k})^2}{4(\sigma_A^2+\sigma_B^2)}
 +i(-\ev{\vec{p}}+\ev{\vec{q}}+\vec{k})\cdot\vl{z} \right], \label{4.30}
\end{align}
we perform the $\vec{z}$ and $\vec{y}$ integrations of (\ref{4.21}) and obtain 
\begin{align}
&\mathcal{E}[A(\ev{\vec{p}};\vec{X}_A,x^0_I);C_\rho\text{-Cur}(x),
\bar{B}_\sigma(\ev{\vec{q}};\vec{X}_B,x^0_B)]^{(I-1)}_b \nn \\
&\simeq\int^t_0\!\!\int^t_0\!dZ^0dY^0\!\int\!d^3k'\!\int\!d^3k\ 
 |f_{ABC\sigma}|^2 \ 
 \frac{(2\sqrt{2\pi}\sigma_A)^3(2\sqrt{2\pi}\sigma_B)^3}{(2\pi)^62\ev{E_A} 2\ev{E_B}}\ 
 \left[ \frac{\pi}{\sigma_A^2+\sigma_B^2} \right]^3 \nn \\
&\times\sum_{j,i}Z^{1/2}_{\sigma j}{Z^{1/2}_{\rho j}}^\ast 
 Z^{1/2}_{\rho i} {Z^{1/2}_{\sigma i}}^\ast\
 \frac{i^2(k'_j+k_i)_b}{(2\pi)^6 2\omega_j(\vec{k}')2\omega_i(\vec{k})} \
 e^{i(-\vec{k}'+\vec{k})\cdot\vec{x}+i(\omega_j(\vec{k}')-\omega_i(\vec{k}))t} \nn \\
&\times\exp\bigg[-h(\vl{z},Z^0)+i(-\ev{\vec{p}}+\ev{\vec{q}}+\vec{k}')\cdot\vl{z}
 -h(\vl{y},Y^0)+i(\ev{\vec{p}}-\ev{\vec{q}}-\vec{k})\cdot\vl{y} \nn \\
&\hspace{1cm}-\frac{1}{4(\sigma_A^2+\sigma_B^2)}
 \left\{(-\ev{\vec{p}}+\ev{\vec{q}}+\vec{k}')^2+(-\ev{\vec{p}}+\ev{\vec{q}}+\vec{k})^2
 \right\} \nn \\
&\hspace{1cm}+iZ^0(\ev{E_A}-\ev{E_B}-\omega_j(\vec{k}'))
 +iY^0(-\ev{E_A}+\ev{E_B}+\omega_i(\vec{k}))
 -\frac{1}{2}\ev{\Gamma_A}(Z^0+Y^0) \bigg]. \label{4.31}
\end{align}
      
Next we proceed to integrate with respect to $\vec{k}'$ and $\vec{k}$.
By employing the definitions
\begin{align}
\ev{\vec{k}}\equiv\ev{\vec{p}}-\ev{\vec{q}},\ \ 
\ev{\omega_j}\equiv\omega_j(\ev{\vec{k}}),\ \ 
\ev{\vec{v_j}}\equiv\vec{v}_j(\ev{\vec{k}}), \label{4.32}
\end{align}
and 
\begin{align}
&\exp\bigg[-h(\vl{z},Z^0)+i(-\ev{\vec{k}}+\vec{k}')\cdot\vl{z}
 +iZ^0\big(\ev{E_A}-\ev{E_B}-\ev{\omega}_j
 -\ev{\vec{v}_j}\cdot(\vec{k}'-\ev{\vec{k}})\big) \nn \\
&\hspace{1cm}-i\vec{k}'\cdot\vec{x}
 +i\big(\ev{\omega_j}+\ev{\vec{v}_j}\cdot(\vec{k}'-\ev{\vec{k}}) \big)t
 -\frac{1}{4(\sigma_A^2+\sigma_B^2)}(-\ev{\vec{k}}+\vec{k}')^2 \bigg]
 \nn \\
&=\exp\bigg[-h(\vl{z},Z^0)
 -\frac{1}{4(\sigma_A^2+\sigma_B^2)}
 \left\{\vec{k}'-\ev{\vec{k}}+2i(\sigma_A^2+\sigma_B^2)\cdot
 (-\vl{z}+Z^0\ev{\vec{v}_j}+\vec{x}-\ev{\vec{v}_j}t)\right\}^2 \nn \\
&\hspace{1cm}-(\sigma_A^2+\sigma_B^2)
\left\{-\vl{z}+Z^0\ev{\vec{v}_j}+\vec{x}-\ev{\vec{v}_j}t\right\}^2
+iZ^0(\ev{E_A}-\ev{E_B}-\ev{\omega_j}) \nn \\
&\hspace{1cm}-i\ev{\vec{k}}\cdot\vec{x}+i\ev{\omega_j}t \bigg], \label{4.33}
\end{align}
we obtain from (\ref{4.31}) 
\begin{align}
&\mathcal{E}[A(\ev{\vec{p}};\vec{X}_A,x^0_I);C_\rho\text{-Cur}(x),
\bar{B}_\sigma(\ev{\vec{q}};\vec{X}_B,x^0_B)]^{(I-1)}_b \nn \\
&\simeq\sum_{j,i}Z^{1/2}_{\sigma j}{Z^{1/2}_{\rho j}}^\ast 
 Z^{1/2}_{\rho i} {Z^{1/2}_{\sigma i}}^\ast
 \int^t_0\!\!\int^t_0\!dZ^0dY^0\ |f_{ABC\sigma}|^2 \ 
 \frac{(2\sqrt{2\pi}\sigma_A)^3(2\sqrt{2\pi}\sigma_B)^3}{(2\pi)^62\ev{E_A} 2\ev{E_B}}\ 
 \left[ \frac{\pi}{\sigma_A^2+\sigma_B^2} \right]^3 \nn \\
&\times[2\sqrt{\pi(\sigma_A^2+\sigma_B^2)}]^6\
 \times e^{i(\ev{\omega_j}-\ev{\omega_i})t} \nn \\
&\times\exp\bigg[-G_j(Z^0,\vec{X}_A; X^0_B,\vec{X}_B; t,\vec{x})
-G_i(Y^0,\vec{X}_A; X^0_B,\vec{X}_B; t,\vec{x}) \nn \\
&\hspace{1cm}+iZ^0(\ev{E_A}-\ev{E_B}-\ev{\omega_j})
 +iY^0(-\ev{E_A}+\ev{E_B}+\ev{\omega_i})
 -\frac{1}{2}\ev{\Gamma_A}(Z^0+Y^0)\bigg] \nn \\
&\times \frac{i^2}{(2\pi)^6 2\ev{\omega_j} 2\ev{\omega_i}}
 \begin{bmatrix}
  2\ev{\vec{k}} \\
  i(\ev{\omega_j}+\ev{\omega_i})
 \end{bmatrix}, \label{4.34}
\end{align}
where
\begin{align}
G_j(Z^0,\vec{X}_A; X^0_B,\vec{X}_B; t,\vec{x})\equiv
 h(\vl{z},Z^0)+(\sigma_A^2+\sigma_B^2)\{\vec{x}-\vl{z}-\ev{\vec{v}_j}(t-Z^0)\}.
 \label{4.35}
\end{align}

In order to see the properties of the integrand, it seems convenient for us to define 
\begin{align}
\vec{F}^{AC}_j(Z^0)&\equiv
 \vec{x}-\ev{\vec{v}_A}Z^0-\ev{\vec{v}_j}(t-Z^0)-\vec{X}_A,\label{4.36}\\
\vec{F}^{BC}_j(Z^0,X^0_B)&\equiv
 \vec{F}^{AB}(Z^0,X^0_B)-\vec{F}^{AC}_j(Z^0)\nn \\
&=\vec{X}_B-\ev{\vec{v}_B}(X^0_B-Z^0)+\ev{\vec{v}_j}(t-Z^0)-\vec{x},\label{4.37}
\end{align}
where $\vec{F}^{AB}(Z^0,X^0_B)$ is given by (\ref{4.27}). 
The factor $\vec{x}-\vl{z}-\ev{\vec{v}_j}(t-Z^0)$ included in $G_j$, 
given by (\ref{4.35}), is expressed as 
\begin{align}
&\vec{x}-\vl{z}-\ev{\vec{v}_j}(t-Z^0)=
 \vec{F}^{AC}_j(Z^0)+\ev{\vec{v}_A}Z^0+\vec{X}_A \nn \\
&\hspace{1cm}-\frac{1}{\sigma_A^2+\sigma_B^2}
 \bigg[\sigma_A^2(\vec{X}_A+\ev{\vec{v}_A}Z^0)
 +\sigma_B^2\{\vec{X}_B-\ev{\vec{v}_B}(X^0_B-Z^0)\}\bigg] \quad
 \text{due to \eqref{4.25}} \nn\\
&=\vec{F}^{AC}_j(Z^0)-\frac{\sigma_B^2}{\sigma_A^2+\sigma_B^2}
 \vec{F}^{AB}(Z^0,X^0_B)\nn\\
&=\frac{1}{\sigma_A^2+\sigma_B^2}
 \bigg[\sigma_A^2\vec{F}^{AC}_j(Z^0)-\sigma_B^2\vec{F}^{BC}_j(Z^0,X^0_B)\bigg];\label{4.38}
\end{align}
thus we obtain from (\ref{4.26}) and (\ref{4.35}) 
\begin{align}
&G_j(Z^0,\vec{X}_A; X^0_B,\vec{X}_B; t,\vec{x}) \nn \\
&=\frac{1}{\sigma_A^2+\sigma_B^2}
 \bigg[\sigma_A^2\sigma_B^2(\vec{F}^{AB}(Z^0,X^0_B))^2
 +\{\sigma_A^2\vec{F}^{AC}_j(Z^0)-\sigma_B^2\vec{F}^{BC}_j(Z^0,X^0_B)\}^2\bigg]\nn\\
&=\sigma_A^2\big(\vec{F}^{AC}_j(Z^0)\big)
 +\sigma_B^2\big(\vec{F}^{BC}_j(Z^0,X^0_B)\big)^2. \label{4.39}
\end{align}

Here we give a remark on the possibility of the minimum of $G_j$ to
be $0$. The condition $G_j=0$ for $Z^0=Z^0_{\text{clas}}$ is 
\begin{align}
\vec{F}^{AC}_j(Z^0_{\text{clas}})=0,\ \ \text{and} \ \
 \vec{F}^{BC}_j(Z^0_{\text{clas}})=0,\ \ \text{[due to \eqref{4.39}]} \label{4.40}
\end{align}
leading respectively to 
\begin{align}
\Zc&=\frac{1}{(\vev{j}-\vev{A})^2}
 (\vev{j}-\vev{A})\cdot(-\vec{x}+\vec{X}_A+\vev{j}t), \label{4.41} \\
\Zc&=\frac{1}{(\vev{j}-\vev{B})^2}
 (\vev{j}-\vev{A})\cdot(-\vec{x}+\vec{X}_B-\vev{B}X^0_B+\vev{j}t), \label{4.42}
\end{align}
together with the componets of $-\vec{x}+\xa+\vev{j}t$ and 
$-\vec{x}+\xb-\vev{B}\tb+\vev{j}t$ perpendicular respectively to 
$\vev{j}-\vev{A}$ and $\vev{j}-\vev{B}$ to be equal
$0$. Eqs.(\ref{4.40}) means $\vec{F}^{AB}(\Zc,\tb)=0$ [due to
\eqref{4.37}], which leads to
\begin{align}
\Zc=\frac{1}{(\vev{A}-\vev{B})^2}
 (\vev{A}-\vev{B})\cdot(\xb-\vev{B}\tb-\xa), \label{4.43}
\end{align}
together with the component of $\xb-\vev{B}\tb-\xa$ perpendicular to 
$\vev{A}-\vev{B}$ to be $0$.     These $\Zc$'s give three kinds of the 
classical relations. 
%such as shown by FIG.XIII.
We see, however, these
$\Zc$'s should not be taken to be equal especially due to the $j$-independence 
of (\ref{4.43}).

It is necessary for us to consider how to perform the $Y^0$ - and $Z^0$-integrations 
over a finite time-interval; this is to be considered in the next subsection，where
the minimum of $G_j$ is examined.

\subsection{Time integrations}
We examine in detail the structures of the integrand in (\ref{4.34}).
\subsubsection{The minimum point $\udl{Z}^0$ of $G_j$}     
We write for simplicity $G_j$ as $G_j(Z^0)$ and rewite it 
in the form 
\begin{align}
G_j(Z^0)=\alpha_j(Z^0-\udl{Z}^0)^2+G_j(\udl{Z}^0). \label{4.44}
\end{align}
$\udl{Z}^0$ is determined by 
\begin{align}
&\left.\frac{dG_j(Z^0)}{dZ^0}\right|_{\udl{Z}^0}=0
 =2\bigg[\sg_A(\vev{j}-\vev{A})\cdot\vec{F}^{AC}_j(\Zd) 
 -\sg_B(\vev{j}-\vev{B})\cdot\vev{F}^{BC}_j(\Zd,\tb) \bigg] \nn 
\intertext{due to \eqref{4.39}}
&=2\bigg[\big\{\sg_A(\vev{j}-\vev{A})^2+\sg_B(\vev{j}-\vev{B})^2\big\}\Zd\nn \\
&\hspace{0.5cm}+\sg_A(\vev{j}-\vev{A})\cdot\vec{F}^{AC}_j(Z^0=0)
 -\sg_B(\vev{j}-\vev{B})\cdot\vec{F}^{BC}_j(0,\tb)  \bigg] \label{4.45}
\intertext{which leads to}
&\Zd=2\sg_j\bigg[-\sg_A(\vev{j}-\vev{A})\cdot\vec{F}^{AC}_j(0)
 +\sg_B(\vev{j}-\vev{B})\cdot\vec{F}^{BC}_j(0,\tb)\bigg]; \label{4.46}
\end{align}
here
\begin{align}
&\frac{1}{\sg_j}\equiv\frac{1}{\sg_{jA}}+\frac{1}{\sg_{jB}}, \nn
\intertext{with}
&\frac{1}{2\sg_{jA}}\equiv\sg_A(\vev{j}-\vev{A})^2,\ \ \ 
\frac{1}{2\sg_{jB}}\equiv\sg_B(\vev{j}-\vev{B})^2. \label{4.47}
\end{align}
By noting $\alpha_j$ to be given by
$\frac{1}{2}(d^2G_j/d(Z^0)^2)=1/(2\sg_j)$, 
we have 
\begin{align}
G_j(Z^0)&=\frac{1}{2\sg_j}(Z^0-\Zd)^2+G_j(\Zd)\nn \\
&=\frac{1}{2\sg_j}(Z^0-\Zd)^2+\sg_A\big(\vec{F}^{AC}_j(\Zd)\big)^2
 +\sg_B\big(\vec{F}^{BC}_j(\Zd,\tb)\big)^2. \label{4.48}
\end{align}

\subsubsection{Another expression of $G_j(Z^0)$}
We write the points $Z^0$'s giving the minima of $(\vec{F}^{AC}_j)^2$ and
$(\vec{F}^{BC}_j)^2$ as $T^{AC}_j$ and $T^{BC}_j$, respectively ; 
\begin{align}
\fac_j(\Tac_j)\cdot(\vev{j}-\vev{A})=0,\ \ 
\fbc_j(\Tbc_j,\tb)\cdot(\vev{j}-\vev{B})=0. \label{4.49}
\end{align}
We obtain 
\begin{align}
\Tac_j&=\frac{1}{(\vev{j}-\vev{A})^2}(\vev{j}-\vev{A})\cdot(-\vec{x}+\xa+\vev{j}t)\nn\\
&=2\sg_A\sg_{jA}(\vev{j}-\vev{A})\cdot(-\fac_j(Z^0=0)), \label{4.50}\\
\Tbc_j&=\frac{1}{(\vev{j}-\vev{B})^2}(\vev{j}-\vev{B})
 \cdot(-\vec{x}+\vev{j}t-\vev{B}\tb+\xb)\nn\\
&=2\sg_B\sg_{jB}(\vev{j}-\vev{B})\cdot\fbc_j(Z^0=0,\tb); \label{4.51}
\end{align}
then (\ref{4.46}) leads to 
\begin{align}
\Zd&=\sg_j\bigg[\frac{1}{\sg_{jA}}\Tac_j+\frac{1}{\sg_{jB}}\Tbc_j\bigg]\nn\\
&=\frac{1}{\sg_{jA}+\sg_{jB}}\bigg[\sg_{jB}\Tac_j+\sg_{jA}\Tbc_j \bigg].\label{4.52}
\end{align}
By noting 
\begin{align}
\frac{1}{2}\frac{\partial^2}{\partial (Z^0)^2}
 \begin{bmatrix}
  (\fac_j)^2 \\
  (\fbc_j)^2
 \end{bmatrix}
=
 \begin{bmatrix}
  \frac{1}{2\sg_A\sg_{jA}} \\
  \frac{1}{2\sg_B\sg_{jB}}
 \end{bmatrix}, \label{4.53}
\end{align}
we have from (\ref{4.39})
\begin{align}
G_j(Z^0)&=\frac{1}{2\sg_{jA}}(Z^0-\Tac_j)^2+\sg_A[\fac_j(\Tac_j)]^2
 +\frac{1}{2\sg_{jB}}(Z^0-\Tbc_j)^2+\sg_B[\fbc_j(\Tbc_j,\tb)]^2. \label{4.54}
\end{align}
It is easy for us to confirm the equality 
\begin{align}
&\frac{1}{2\sg_j}(Z^0-\Zd)^2+\frac{1}{2(\sg_{jA}+\sg_{jB})}(\Tac_j-\Tbc_j)^2
 \nn\\
&=\frac{1}{2\sg_{jA}}(Z^0-\Tac_j)^2+\frac{1}{2\sg_{jB}}(Z^0-\Tbc_j)^2; \label{4.55}
\end{align}
because, due to (\ref{4.52}) we have 
\begin{align}
\text{L.H.S. of \eqref{4.55}}&=\frac{1}{2\sg_j(\sg_{jA}+\sg_{jB})^2}
 \bigg[\sg_{jB}(Z^0-\Tac_j)+\sg_{jA}(Z^0-\Tbc_j)\bigg]^2 \nn\\
&\hspace{0.5cm}+\frac{1}{2(\sg_{jA}+\sg_{jB})}
 \bigg[-(Z^0-\Tac_j)+(Z^0-\Tbc_j)\bigg]^2 \nn\\
&=\text{R.H.S. of \eqref{4.55}}. \label{4.56}
\end{align}
Thus we obtain 
\begin{align}
G_j(\Zd)=\sg_A[\fac_j(\Tac_j)]^2+\sg_B[\fbc_j(\Tbc_j,\tb)]^2
 +\frac{(\Tac_j-\Tbc_j)^2}{2(\sg_{jA}+\sg_{jB})}. \label{4.57}
\end{align}
\subsubsection{The expression of $\mathcal{E}_b^{(I-1)}$}
(\ref{4.34}) is expressed as 
\begin{align}
&\mathcal{E}[A(\ev{\vec{p}};\vec{X}_A,x^0_I);C_\rho\text{-Cur}(x),
\bar{B}_\sigma(\ev{\vec{q}};\vec{X}_B,x^0_B)]^{(I-1)}_b \nn \\
&\simeq\sum_{j,i}Z^{1/2}_{\sigma j}{Z^{1/2}_{\rho j}}^\ast 
 Z^{1/2}_{\rho i} {Z^{1/2}_{\sigma i}}^\ast\ |f_{ABC\sigma}|^2 \ 
 (\sg_A\sg_B)^{3/2} \left[\frac{2}{\pi}\right]^3 
 \frac{i^2}{2\ev{E_A}2\ev{E_B}2\ev{\omega_j}2\ev{\omega_i}} 
 \begin{bmatrix}
  2\ev{\vec{k}} \\
  i(\ev{\omega_j}+\ev{\omega_i})
 \end{bmatrix}\nn \\
&\times\exp\bigg[i(\ev{\omega_j}-\ev{\omega_i})t
 -G_j(\udl{Z}^0,\vec{X}_A; X^0_B,\vec{X}_B; t,\vec{x})
 -G_i(\udl{Y}^0,\vec{X}_A; X^0_B,\vec{X}_B; t,\vec{x}) \nn \\
&\hspace{1cm}-\udl{Z}^0(i\Delta E_j+\frac{\ev{\Gamma_A}}{2})
 +\udl{Y}^0(i\Delta E_i+\frac{\ev{\Gamma_A}}{2}) \bigg] \nn \\
&\times I_j(\vec{x},t; \xa,\xb,\tb)\ I_i^\ast(\vec{x},t; \xa,\xb,\tb),\label{4.58}
\end{align}
\begin{align}
I_j(\vec{x},t; \xa,\xb,\tb)&\equiv
 \int^{\tb}_0\!dZ^0\exp\bigg[-\frac{(Z^0-\Zd)^2}{2\sg_j}
 -(Z^0-\Zd)(i\Delta E_j+\frac{\ev{\Gamma_A}}{2})\bigg]; \label{4.59}
\intertext{here, we}
\Delta E_j&\equiv \ev{E_B}+\ev{\omega_j}-\ev{E_A}. \label{4.60}
\end{align}
and required a physical condition on the time ordering 
\begin{align}
 x^0 \ge x^0_B \ge (y^0,z^0), \quad \text{i.e.}\quad t\ge \tb >(Y^0,Z^0) \label{4.61}
\end{align}

In the same way as $M_j(t;\Zd)$ given by \eqref{3.54}, we can write $I_j$ as
\begin{align}
I_j=\sqrt{2\sg_j}\exp\left[\frac{1}{2}\sg_j(i\Delta E_j+\frac{\life}{2})^2\right]\
  J_j(u_j(\tb);\tilde{\alpha}_j,\tilde{\beta}_j),\label{4.62}
\end{align}
where $\tilde{\alpha}_j=\sigma_j\Delta E_j/\sqrt{2},\
\tilde{\beta}_j=(-\Zd+\frac{1}{2}\sg_j\ev{\Gamma_A})/(\sqrt{2}
\sigma_j)$ and $u_j(\tb)=t/(\sqrt{2}\sigma_j)$.
Then, \eqref{4.58} is expressed as 
\begin{align}
&\mathcal{E}[A(\ev{\vec{p}};\vec{X}_A,x^0_I);C_\rho\text{-Cur}(x),
\bar{B}_\sigma(\ev{\vec{q}};\vec{X}_B,x^0_B)]^{(I-1)}_b \nn \\
&=\sum_{j,i}Z^{1/2}_{\sigma j}{Z^{1/2}_{\rho j}}^\ast 
 Z^{1/2}_{\rho i} {Z^{1/2}_{\sigma i}}^\ast\ |f_{ABC\sigma}|^2 \ 
 (\sg_A\sg_B)^{3/2} \left[\frac{2}{\pi}\right]^3 
 \frac{i^2}{2\ev{E_A}2\ev{E_B}2\ev{\omega_j}2\ev{\omega_i}} \nn \\
&\times
 \begin{bmatrix}
  2\ev{\vec{k}} \\
  i(\ev{\omega_j}+\ev{\omega_i})
 \end{bmatrix}\ 
 \exp\bigg[\mathcal{Q}_{ji}
 +\frac{1}{2}\sg_j(i\Delta E_j+\frac{\life}{2})^2
 +\frac{1}{2}\sg_i(-i\Delta E_i+\frac{\life}{2})^2\bigg]\nn\\
&\times \big|J_j(u_j(\tb);\tilde{\alpha}_j,\tilde{\beta}_j)\big|\
 \big|J_i(u_i(\tb);\tilde{\alpha}_i,\tilde{\beta}_i)\big| 
 \exp\bigg[i(\Theta(u_j(\tb);\tilde{\alpha}_j,\tilde{\beta}_j)-\Theta(j\rightarrow i))\bigg],\label{4.63}
\intertext{where}
\mathcal{Q}_{ji}&=i(\ev{\omega_j}-\ev{\omega_i})t 
 -G_j(\udl{Z}^0)-G_i(\udl{Y}^0) 
 -\Zd(i\Delta E_j+\frac{\life}{2}) 
 +\Yd(i\Delta E_i-\frac{\life}{2}).\label{4.64}
\end{align}

We see \eqref{4.64} has structures similar to those of \eqref{3.59}, and 
examine in Sec.V its qualitatives of $t$- and $L$-dependence under the 
condition of neutrino high energies.
Here we add a remark on $G_j(\Zd)$ as follows. From \eqref{4.48} and
\eqref{4.57}, $G_j(\Zd)$ is given by
\begin{align}
G_j(\Zd)&=\sg_A\big(\fac_j(\Zd)\big)^2+\sg_B\big(\fbc_j(\Zd,\tb)\big)^2\nn \\
 &=\sg_A\big(\fac_j(\Tac_j)\big)^2+\sg_B\big(\fbc_j(\Tbc_j,\tb)\big)^2
 +\frac{(\Tac_j-\Tbc_j)^2}{2(\sg_{jA}+\sg_{jB})}.\label{4.65}
\end{align}
As noted in the subsection B of Sec.IV, we can not take two $\Zdc$'s
(given by \eqref{4.41} and \eqref{4.42}) to be equal. This means
$\Tac_j$ and $\Tbc_j$, given by \eqref{4.50} and \eqref{4.51}, are not
equal to each other, since the expressions of \eqref{4.50} and
\eqref{4.51} are equal respectively to \eqref{4.41} and
\eqref{4.42}. Thus, due to \eqref{4.52}, we have
\begin{align}
\Tac_j > \Zd > \Tbc_j \quad \text{or} \quad \Tac_j < \Zd < \Tbc_j. \label{4.66}
\end{align}
Under the parallel condition we obtain
\begin{align}
\tacp{j}&=\frac{\vev{j}t-L}{\vev{j}-\vev{A}},\qquad L=x-X_A,\label{4.67}\\
\tbcp{j}&=\frac{1}{\vev{j}-\vev{B}}(\vev{j}t-L+L_{BA}-\vev{B}\tb), 
 \qquad L_{BA}=X_B-X_A,\label{4.68}\\
\Zdp&=\frac{1}{\sg_{jA}+\sg_{jB}}\big[\sg_{jB}\tacp{j}+\sg_{jA}\tbcp{j}\big];\label{4.69}
\intertext{the minimum of $G_j(\Zd)$ is given by}
G_j(\Zdp)&=\frac{1}{2(\sg_{jA}+\sg_{jB})}\big[\tacp{j}-\tbcp{j}\big]^2.\label{4.70}
\end{align}
%In FIG.XIV, the relation among the classical trajectories given by
%$\fac_j(\tacp{j})=0$ and $\fbc_j(\tbcp{j,\tb})=0$ are shown in the case 
%of $\tbcp{j}>\tacp{j}$.

%% section 5  %%

\section{DISCUSSIONS AND FINAL REMARKS}
We have investigated the structures of the expectation values of the
flavor current with respect to an one-particle state which plays a role
of the neutrino source. The present approach is certainly simple one to
the neutrino oscillation without recourse to any one-particle state of
flavor neutrino. In Sec.III, the expectation value of the flavor
neutrino current with respect to one $\pi^+$ wave-packet state was given
as \eqref{3.59}. In Sec.IV, we have examined in the boson model the same
type of the expectation value with an additional ``observed'' charged
particle, and shown that the main part of the expectation value of
\eqref{4.63} has the structures similar to those of \eqref{3.59} derived
in Sec.III.

The expectation values \eqref{3.59} and \eqref{4.63} have somewhat
complicated structure in comparison with the standard oscillation
formula; therefore, in order to make clear characteristic features
included in \eqref{3.59} and \eqref{4.63}, we are necessary to perform
numerical calculations, which will be done in a subsequent paper
\cite{r18}. Here we give some remarks in the following.

It is worthwhile for us to examine the features of contributions from
high-energy neutrinos to \eqref{3.59} and \eqref{4.63}.

First we considered the structures of the integration of the integrand
of \eqref{3.59} in the case of $m^2_j/k^2 \ll 1$ for all j's and
$1-\vev{\pi} \simeq \mathcal{O}(1)$.

As to the phase parts, we obtain from \eqref{3.63}
\begin{align}
\im(\mathcal{P}_{ji})_{\text{para}}&\simeq\frac{\Delta m^2_{ji}}{2k}\
 \frac{L-t\ev{v_\pi}}{1-\ev{v_\pi}}\left[1+\frac{\ev{\delta
 E}}{k(1-\ev{v_\pi})}\right] ;\label{5.1}
\intertext{with $\ev{\delta E}=E_\sigma+K-\ev{E_\pi}$, and also}
-\Theta(u_j(t);\alpha_j,\beta_j)+\Theta(u_i(t);\alpha_i,\beta_i)&
 \simeq-\frac{dv_j}{dm^2_j}\left[\del{\Theta(u_j(t);\alpha_j,\beta_j)}{v_j}\right]_{m_j=0}
 m^2_j+(j\rightarrow i).\label{5.2}
\end{align}
By using th j-independent quantity
\begin{align}
\left[\del{\Theta}{v}\right]_0 &\equiv
 \left[\del{\Theta(u_j(t);\alpha_j,\beta_j)}{v_j}\right]_{m_j=0} \nn\\
&=-\frac{1}{|J_0(u(t);\alpha,\beta)|}e^{\alpha^2}
 \left[
 e^{-\beta^2}\sqrt{(\alpha')^2+(\beta')^2}
 \sin(2\alpha\beta+\Theta-\varphi^1)\right. \nn \\
&\hspace{3.5cm}\left.+e^{-(u(t)+\beta)^2}\sqrt{(\alpha')^2+\big((u(t)+\beta)'\big)^2}
 \sin(2\alpha(u(t)+\beta)+\Theta-\varphi^2)
 \right],\label{5.3}
\intertext{where primes ($'$) denote the derivative with respect to
 neutrino velocity, $v$, and}
\varphi^1&=\arctan\bigg(\frac{\alpha'}{\beta'}\bigg),
 \qquad\varphi^2=\arctan\bigg(\frac{\alpha'}{(u(t)+\beta)'}\bigg),\nn
\end{align}
we obtain
\begin{align}
-\Theta(u_j(t);\alpha_j,\beta_j)+\Theta(u_i(t);\alpha_i,\beta_i)&
 \simeq\frac{\Delta m^2_{ji}}{2k^2}\left[\del{\Theta}{v}\right]_0,\label{5.4}
\end{align}
This is the additional phase coming from the finite-time integrals in
\eqref{3.52}. One can see from \eqref{5.3} that this additional phase 
will vanish when $\beta$ and $u(t)+\beta$ become much 
larger than 1 as well as $\alpha$. This result can be understood
because, since 
$\beta$'s represent the ratio of the distances $(\Zd,t-\Zd)$ between the peak of
gaussian $(\Zd)$ and both limits of integral region $(0,t)$ to the
gaussian width $(\sigma_{\pi j})$,
$\beta$'s $\gg1$ means that the gaussian is so sharp that one
can regard the both limits as infinite. Then the finite-time integral is
reduced to the real function. However, the
situation will be different when $\beta$'s are smaller than 1 and 
$\alpha$. Numerical calculations are expected to make a behavior of this
phase clearer. 
%$\left[\del{\Theta}{v}\right]$ is not a simple
%function of $L$ and $t$, since $\beta_j$ depends on $\Zd$ and dependence
%of $\{u_j(t),\alpha_j,\beta_j\}$ on $v_j$ is not simple. So 

As mentioned before, the condition
\begin{align}
\frac{m^2_j}{k^2}\ \frac{1}{1-\ev{v_\pi}} \ll 1, \label{5.5}
\end{align}
is necessary for deriving \eqref{5.1}. By taking $\ev{v_\pi}\sim0$ and
$\ev{\delta E}/k \sim0$, we obtain
\begin{align}
\im(\mathcal{P}_{ji})_{\text{para}} \simeq \frac{\Delta m^2_{ji}}{2k}L, \label{5.6}
\end{align}
which is the same as the standard one, apart from
$\vec{K}=\vec{K}(\vec{q})=\ev{\vec{p}}-\vec{q}$. It will be instructive
to evaluate typical $\ev{v_\pi}$-values a shown in Table \ref{table1}(a); by the
way, rough values of $K$ by assuming $|\delta E_j|/\ev{E_\pi}\lesssim
\mathcal{O}(1/10)$ are given in Table \ref{table1}(b).
\begin{center}
\begin{table}[h]
\begin{tabular}{|c|c|c|c|c|c|} \hline
 & $\ev{E}$ (MeV)& 200 & 300 & 500 & 1000 \\ \hline \hline
\raisebox{-1.5ex}[0pt][0pt]{(a)} &$\ev{p}$ (MeV)& 142.8 & 265.3 & 480.0 & 990.1 \\ \cline{2-6} 
 & $\ev{v_\pi}/c$ & 0.71 & 0.88 & 0.96 & 0.99 \\ \hline \hline
(b) & $k$ (MeV)$^\ast$ & 74 & 122 & 211 & 427 \\ \hline
\multicolumn{6}{c}{$\ast$ Values of $\frac{1}{2}(m^2_\pi-m^2_\mu)/(\ev{E_\pi}-\ev{p})$}
\end{tabular}
\caption{(a) and (b)}
\label{table1}
\end{table}
\end{center}
It is possible for us to extract information on the coherence length
from $|J_j(u_j(t);\alpha_j,\beta_j)|\cdot|J_i(u_i(t);\alpha_i,\beta_i)|$
as follows as $\text{Re}\mathcal{P}_{ji}$. This is one of the remaining
tasks.

Next we give a remark on the structures of \eqref{4.64}. In this case,
$\ev{\vec{k}}(=\ev{\vec{p}}-\ev{\vec{q}})$ is fixed when $\ev{\vec{p}}$
and $\ev{\vec{q}}$ are given. We have the phase parts in \eqref{4.63}
expressed as 
\begin{align}
\im(\mathcal{Q}_{ji})_{\text{para}}
 =\big[(\ev{\omega_j}-\ev{\omega_i})t-\Zd\Delta E_j+\Yd\Delta
 E_i\big]_{\text{para}}, \label{5.7} \\ 
[\Theta(u_j(t);\tilde{\alpha}_j,\tilde{\beta}_j)-\Theta(j\rightarrow
 i)]_{\text{para}}. \label{5.8}
\end{align}
In comparison with the case \eqref{4.64}, some complication appears due
to the difference between $(Z^0)_{\text{para}}$ given by \eqref{4.69}
and $\Zdp$ given by \eqref{3.62}. This difference arises from
\eqref{4.66}, i.e. from $\Tac_j \neq \Tbc_j$, which leads to nonvanishing
$[G_j(\Zd)+G_i(\Yd)]_{\text{para}}$ in
$\text{Re}[\mathcal{Q}_{ji}]_{\text{para}}$ and also to the additional
$v_j$-dependence of $\Zd$ through $\sg_{jA}$ and $\sg_{jB}$.

\end{document}